\def\nh{{n_{\rm H}}}
\def\nh2{{n(\rm H_2)}}
\def\h2{${\rm H_2}$}

\def\3cm{\rm {cm^{-3}}}
\def\2cm{\rm {cm^{-2}}}
\def\s-1{\rm {s^{-1}}}

\def\mum {\hbox{$\mu$m}}
\def\kms {\hbox{${\rm km\,s}^{-1}$}}
\def\Kkms {\hbox{${\rm K}\,{\rm km}\,{\rm s}^{-1}$}}

\def\ergcmssr{{\rm {erg}~{{cm^{-2}}~s^{-1}~sr^{-1}}}}

\def\twco{\hbox{$^{12}$CO}}
\def\thco{\hbox{$^{13}$CO}}

\def\ci{\hbox{\rm {[C {\scriptsize I}]}}}
\def\cii{\hbox{\rm {[C {\scriptsize II}]}}}
\def\nii{\hbox{\rm {[N {\scriptsize II}]}}}
\def\hii{\hbox{\rm {H {\scriptsize II}}}}
\def\c18o{\hbox{C$^{18}$O}}
\def\cilo{\hbox{\rm {[C {\scriptsize I}]}~$^3P_1\to{^3P_0}$}}

\def\ciitrans{\hbox{\rm {[C {\scriptsize II}]}~$^3P_{3/2}\to{^3P_{1/2}}$}}
\def\niilo{\hbox{\rm {[N {\scriptsize II}]}~$^3P_1\to{^3P_0}$}}
\def\niiup{\hbox{\rm {[N {\scriptsize II}]}~$^3P_2\to{^3P_1}$}}
\documentclass{aa}
\usepackage[varg]{txfonts}
\usepackage{natbib}

\usepackage{amsmath}
\bibpunct{(}{)}{;}{a}{}{,} % to follow the A&A style
\usepackage{multirow,tabularx}
\usepackage{graphicx}
\usepackage{textcomp}
\usepackage{ulem}
\usepackage{censor}
\usepackage{epsfig}
\usepackage{diagbox}
\usepackage{array}
\usepackage{txfonts}
\usepackage{psfrag}
\usepackage{units}
\usepackage{textgreek}
\usepackage{hyperref}
\usepackage{relsize}  %to adjust font size in URL footnotes

\begin{document}
\title{[CII]~158{\textmu}m and  [NII]~205{\textmu}m emission from IC~342}
\subtitle{Disentangling the emission from ionized and photo-dissociated regions}
\author{M. R\"ollig\inst{1} \and 
        R.~Simon\inst{1} \and
        R.~G\"usten\inst{2} \and
        J.~Stutzki\inst{1} \and
        F.P.~Israel\inst{3} \and 
        	K.~Jacobs\inst{1} 
}
\offprints{M. R\"ollig}
\institute{
 I. Physikalisches Institut der Universit\"at zu K\"oln, Z\"ulpicher Stra\ss e 77, 50937 K\"oln, Germany\\
 \email{roellig@ph1.uni-koeln.de} % changed VO
\and
 Max-Planck-Institut f\"ur Radioastronomie, Auf dem H\"ugel 69, 53121 Bonn, Germany 
\and
Leiden Observatory, Leiden University, PO Box 9513, NL 2300 RA Leiden, The Netherlands
}
\date{Received  / Accepted  }
\titlerunning{{\cii}~158{\textmu}m and {\nii}~205{\textmu}m emission from IC~342}
%\authorrunning{ }

%
\abstract
  {
  Atomic fine-structure line emission is a major cooling process in the interstellar medium (ISM). In particular the \cii\ 158{\textmu}m line is one of the dominant cooling lines in photon-dominated regions (PDRs). However, it is not confined to PDRs but can also originate from the ionized gas closely surrounding young massive stars. The proportion of the \cii\ emission from \hii\ regions relative to that from PDRs can vary significantly. 
  }
  { 
  We investigate the question of how much of the \cii\ emission in the  nucleus of the nearby spiral galaxy IC~342 is contributed by PDRs and by the ionized gas. We examine the spatial variations of starburst/PDR activity and study the correlation of the \cii\ line with the  \nii\ 205{\textmu}m emission line coming exclusively from the \hii\ regions.
  }
  { 
  We present small maps of \cii\ 158{\textmu}m and \nii\ 205{\textmu}m lines recently observed with the GREAT receiver on board SOFIA.\thanks{\cii\ and \nii\ spectra are available in electronic form (in GILDAS/CLASS format at the CDS via anonymous ftp to \url{cdsarc.u-strasbg.fr} (130.79.128.5) or via \url{http://cdsweb.u-strasbg.fr/cgi-bin/qcat?J/A+A/}}   
  We present different methods to utilize the superior spatial and spectral resolution of our new data to infer information on how the gas kinematics in the nuclear region influence the observed line profiles. In particular we present a super-resolution method to derive how unresolved, kinematically correlated structures in the beam contribute to the observed line shapes.
  }
  {
   We find that the emission coming from the ionized gas shows a kinematic component in addition to the general Doppler signature of the molecular gas. We interpret this as the signature of two bi-polar lobes of ionized gas expanding out of the galactic plane. We then show how this requires an adaptation of our understanding of the geometrical structure of the nucleus of IC~342. Examining the starburst activity we find ratios $I({\cii})/I(\twco (1-0))$ between 400 and 1800 in energy units. 
   Applying predictions from numerical models of \hii\ and PDR regions to derive the contribution from the ionized phase to the total \cii\ emission we find that 35-90\% of the observed \cii\ intensity stems from the ionized gas if both phases contribute. 
   Averaged over the central few hundred parsec we find for the \cii\ contribution a {\hii}-to-PDR ratio of 70:30.
  }
  {
   The ionized gas in the center of IC~342 contributes more strongly to the overall \cii\ emission than is commonly observed on larger scales and than is predicted. Kinematic analysis shows that the majority of the \cii\ emission is related to the strong but embedded star formation in the nuclear molecular ring and only marginally emitted from the expanding bi-polar lobes of ionized gas.
  }
  

\maketitle

\section{Introduction}

The \cii\ 158{\textmu}m emission line is one of the strongest cooling lines in the interstellar medium (ISM) as long as most of the carbon exists as C$^+$. This is true for the ionized phase, e.g. in \hii\ regions, as well as in the outer regions of molecular clouds, in so-called photo-dissociation regions (PDR). For PDRs this is particularly interesting because this line, owing to its not too high optical depths, traces almost the entire carbon content of a molecular cloud. Spatially, the \cii\ emission of a PDR originates from parts that are CO-dark. Consequently, it should also trace the fraction of molecular hydrogen gas that is  spatially not coexistent with CO and therefore complements the standard CO-\h2\ correlation in regions of high UV flux. The \cii\ line also carries important information on the energetic state of the cloud.
Unfortunately, it is almost impossible to observe a PDR without picking up contributions from the accompanying \hii\ region. Discriminating between \cii\ emission coming from the \hii\ region and coming from the PDR is not easy, but crucial because a {\hii}-pollution of the \cii\ signal could significantly affect the conclusion of any emission line analysis. \citet{abel06b} presented numerical calculations of models of ionized and PDR gas showing that up to 50\% of a  detected \cii\ line intensity can come from the \hii\ region. 
One suggestion to clean a \cii\ signal from \hii\ contamination is to compare it with emission lines that are exclusively produced in the ionized gas, such as \nii\ emission lines. Atomic nitrogen has an ionization potential of 14.53 eV, prohibiting N$^+$ production below the Lyman edge. \nii\ emission is therefore only produced in the \hii\ region and because of the comparable excitation conditions and critical densities of C$^+$ and N$^+$ it should be an excellent tracer of \cii\ emitted from the \hii\ region. 

\begin{figure*}[ht]
\centering
\begin{minipage}{8.5cm}
\centering
\includegraphics[width=8.5cm]{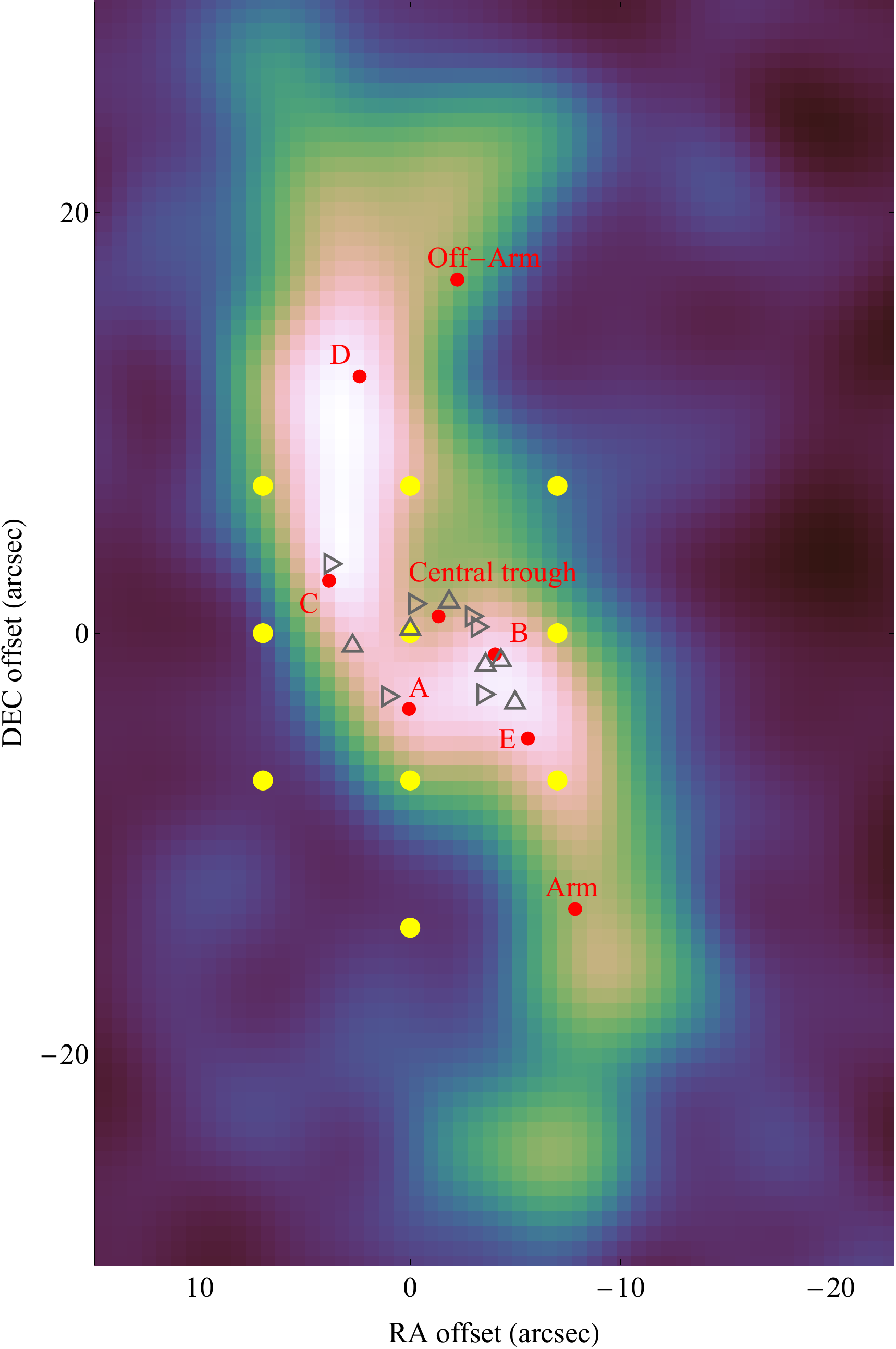}
\end{minipage}
\begin{minipage}{8.5cm}
\centering
\includegraphics[width=8.5cm]{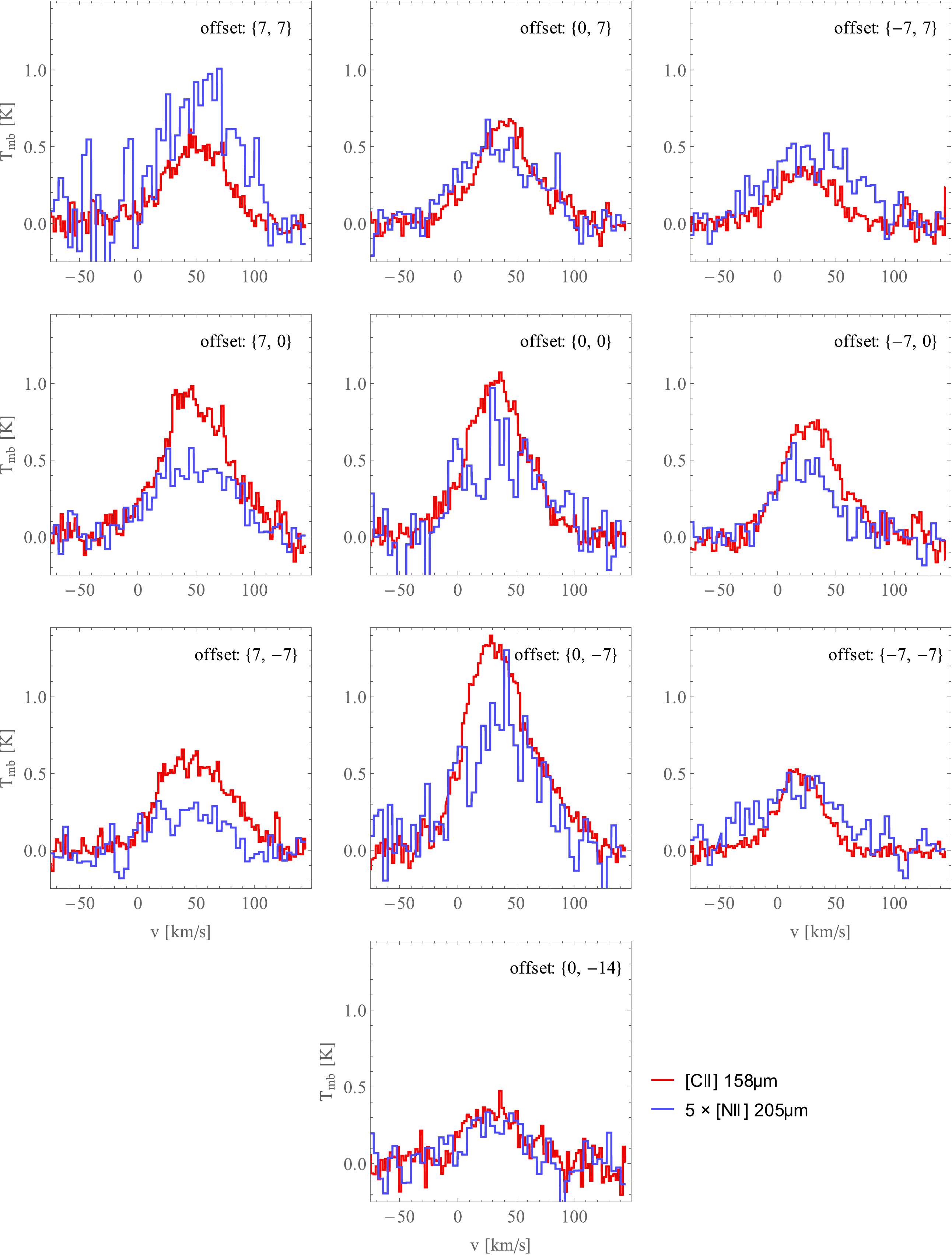}
\end{minipage}
\caption{Line-integrated map of the \twco(1-0) transition from the BIMA-SONG sample. The yellow points show the observed GREAT positions  in IC~342. The (0,0) position corresponds to (RA, DEC) (J2000) (03:46:48.5 68:05:47). The red points indicate named GMCs and other structures according to Table~3 in \citet{meier2001}.
Triangles facing up and right indicate \hii\ regions and super nova remnants, respectively, as given by \citet{tsai2006}.  The \cii\ and \nii\ spectra are shown on the right side. The \nii\ intensity scale is multiplied by a factor of 5.}\label{figObs}
\end{figure*}

\object{IC 342}, a face-on spiral galaxy at a distance of $3.9\pm0.1$~Mpc \citep{tikhonov2010}, has a nuclear region with active star formation. \citet{downes1992} showed that five giant molecular clouds (GMC) with masses of $\sim 10^6$~M$_\odot$ are surrounding a young central star cluster in a ring of dense molecular gas. Two molecular arms of a mini-spiral originate from the molecular ring, north and south of the galaxy center (see also Fig.~\ref{figObs}, left panel). The nuclear star cluster illuminates the molecular ring with intense far-ultraviolet (FUV) radiation  producing photo-dissociation regions (PDRs) on the inner side facing the central cluster. The nucleus of IC~342 shows a great similarity to the center of our Galaxy. In particular, the spatial size of the central GMCs as well as the infrared luminosity of the central few hundred pc of IC~342 are comparable to the Milky Way. In an earlier paper \citep{roellig2012} we presented early Stratospheric Observatory For Infrared Astronomy (SOFIA, \citet{young2012}) observations of the \cii~158~\mum\ fine-structure transition of C$^+$  at 1900.536900 GHz and the \twco(11-10) transition at 1496.922909 GHz at two GMCs in the central molecular ring of IC~342. Using KOSMA-$\tau$ PDR model calculations \citep{stoerzer1996, roellig2006, roellig2013dust} we were able to distinguish between a strong PDR/star-burst emission in the southern GMC E and the much more quiescent conditions in the cooler and denser GMC C in the northern arm, confirming the findings of \citet{meier2005}.

However, the relative contribution of the diffuse material to the overall \cii\ emission was unknown. With the upgraded spectral capabilities of the German REceiver
for Astronomy at Terahertz frequencies (GREAT\footnote{GREAT is a development by the MPI f\"ur Radioastronomie and the  KOSMA/Universit\"at zu K\"oln  in cooperation
with the MPI f\"ur Sonnensystemforschung and the DLR Institut
f\"ur Planetenforschung.}, \citet{GREAT}) receiver on SOFIA, we are for the first time able to investigate the \nii\ 205{\textmu}m and \cii\ 158{\textmu}m fine-structure emission of the ionized material from \hii\ and PDRs simultaneously. The aim of this paper  is to study how the relative contributions from these two ISM phases vary spatially. 
 
\section{Observations}
We used the dual-channel receiver GREAT on SOFIA to perform pointed observations close to the nucleus of IC~342. We used the L1/L2 GREAT configuration with the L1 channel tuned to the \nii\ 205{\textmu}m fine-structure line \niilo\ ($\nu=1461.13$ GHz) and the L2 channel tuned to the \cii\ 158 {\textmu}m fine-structure line \ciitrans\  ($\nu=1900.5369$ GHz). For the rest of this paper \nii\ will always refer to the 205{\textmu}m line. The observations were done in dual beam-switch mode (chop rate 1 Hz; the chop throw was  100{\arcsec} both for \cii\ and {\nii}, at an angle of 20 deg counterclockwise from the R.A. axis. ) toward selected positions centered around the nucleus of IC~342 on a half-beam sampled 7\arcsec grid.
We do not see any signs of self-chopping in our data, but the 100\arcsec chop throw does not exclude possible weak contamination of the two off-source positions. We made sure not to chop onto the spiral arms, but hardware limitations did not allow us to chop out of the galaxy.
The center positions for all observations is RA,DEC (J2000) 03:46:48.5 68:05:47 (offset: 0\arcsec,0\arcsec). We observed a rectangular 3x3 grid centered around the (0\arcsec,0\arcsec) position plus an additional pointing at (0\arcsec,-14\arcsec). The observations took place in February 2014 during three flights.  In total we present data for ten positions. The total observing time per position is between 2.5 and 7.5 minutes on-source, $T_\mathrm{sys}(\mathrm{SSB})$ varied between about 1300~K and 1400~K for \nii\ and between 1900~K and 3200~K for the \cii\ line depending on the date of observations. 
\citet{GREAT} described the overall pointing accuracy as a combination of the accuracy of the boresight determination (within 1-2\arcsec) and the stability during flight, controlled with the optical guide cameras to 3-5\arcsec. However, since then the pointing accuracy has  improved considerably. Pointing instabilities due to drifts no longer occur and the total pointing accuracy (boresight determination and optical camera) is now below 1\arcsec. 
Therefore, it is unlikely that a systematic pointing error may have contaminated the observed line profiles.
  
We used a fast Fourier transform spectrometer (XFFTS, \citet{klein2012}) with 32768 channels. The XFFTS provides 2.5~GHz bandwidth and about 88.5~kHz spectral resolution. The data were converted to line brightness temperature $T_{B}=\eta_{f}\times T_{A}^{*}/\eta_{c}$ applying a  beam efficiency $\eta_c\approx0.67 (\mathrm{L1})$ and $0.65(\mathrm{L2})$ and a forward efficiency ($\eta_f$) of 0.97. Baselines were corrected with polynomials up to the fourth order. The reduction of these calibrated data were made with the GILDAS\footnote{\url{http://www.iram.fr/IRAMFR/GILDAS}} package CLASS90. The data analysis and most of the figures in this paper were made using Mathematica \footnote{Wolfram Research, Inc., Mathematica, Version 10.0, Champaign, IL (2014),\url{http://www.wolfram.com} }
In this paper we use integrated line intensities in units of energy, $[I]=$~erg~s$^{-1}$~cm$^{-2}$~sr$^{-1}$ , and temperature, $\left[\int T dv\right]=$~{\Kkms}, as is common in the literature. The conversion between the two is achieved with the following formula: 
$\int I\, d\nu \,\;\mathrm{erg\,s}^{-1}\,\mathrm{cm}^{-2}\,\mathrm{sr}^{-1}=\tfrac{2 k {\nu}^3}{c^3}\int T dv\;$~\Kkms. 
When discussing intensity ratios we will specify the underlying intensity units.

%-------------------------------------------
%  Section DATA OVERVIEW
%-------------------------------------------
 
\section{Data overview}
  \begin{table*}[!pt]
      \caption[]{Line parameters derived from Gaussian fits. The complementary data was smoothed to the \cii\ beam resolution when possible. For details see Sect.~\ref{app1}.}
      \label{tab:1}
         \centering
         \scriptsize
         \begin{tabular*}{1.0\textwidth}{@{\extracolsep{\fill}} l l l l l l l l l l}
           \hline\hline
	    \noalign{\smallskip}

&$^{12}$CO(1-0)&$^{12}$CO(2-1)&$^{12}$CO(3-2)&$^{12}$CO(4-3)&$^{13}$CO(2-1)&$^{13}$CO(3-2)&\ci&\cii&\nii \\
             \noalign{\smallskip}
             \hline
             \noalign{\smallskip}
            \multicolumn{10}{c} {(7\arcsec,7\arcsec)} \\
            \noalign{\smallskip}
            \hline
            \noalign{\smallskip}
T$_\mathrm{pk}$~(K)&4.97$\pm$0.12&2.51$\pm$0.06&1.76$\pm$0.02&1.72$\pm$0.07&-&-&0.45$\pm$0.05&0.51$\pm$0.02&0.17$\pm$0.01\\
v$_0$ (\kms)&47.21$\pm$0.59&30.78$\pm$0.65&30.79$\pm$0.31&40.77$\pm$1.23&-&-&43.94$\pm$2.30&49.99$\pm$0.93&54.14$\pm$2.77\\
FWHM (\kms)&49.17$\pm$1.40&56.72$\pm$1.53&54.37$\pm$0.74&57.45$\pm$2.89&-&-&45.82$\pm$5.41&59.46$\pm$2.18&72.31$\pm$6.53\\
            \noalign{\smallskip}
            \hline
            \noalign{\smallskip}

            \multicolumn{10}{c} {(0\arcsec,7\arcsec)} \\
            \noalign{\smallskip}
            \hline
            \noalign{\smallskip}
T$_\mathrm{pk}$~(K)&6.31$\pm$0.09&2.51$\pm$0.06&1.76$\pm$0.02&2.44$\pm$0.07&-&0.30$\pm$0.09&0.72$\pm$0.04&0.62$\pm$0.02&0.10$\pm$0.01\\
v$_0$ (\kms)&40.76$\pm$0.36&30.78$\pm$0.65&30.79$\pm$0.31&39.82$\pm$0.75&-&30.57$\pm$6.35&36.46$\pm$1.57&40.25$\pm$0.7&32.82$\pm$2.36\\
FWHM (\kms)&50.21$\pm$0.84&56.72$\pm$1.53&54.37$\pm$0.74&55.09$\pm$1.77&-&44.54$\pm$14.94&53.48$\pm$3.69&58.71$\pm$1.64&77.46$\pm$5.56\\
            \noalign{\smallskip}
            \hline
            \noalign{\smallskip}

            \multicolumn{10}{c} {(-7\arcsec,7\arcsec)} \\
            \noalign{\smallskip}
            \hline
            \noalign{\smallskip}
T$_\mathrm{pk}$~(K)&3.85$\pm$0.08&2.51$\pm$0.06&1.76$\pm$0.02&2.29$\pm$0.07&-&0.30$\pm$0.09&0.66$\pm$0.05&0.34$\pm$0.01&0.09$\pm$0.01\\
v$_0$ (\kms)&29.75$\pm$0.56&30.78$\pm$0.65&30.79$\pm$0.31&31.10$\pm$0.79&-&30.57$\pm$6.34&29.67$\pm$1.70&23.05$\pm$1.20&29.36$\pm$2.38\\
FWHM (\kms)&53.05$\pm$1.32&56.72$\pm$1.53&54.37$\pm$0.74&52.34$\pm$1.85&-&44.54$\pm$14.94&50.38$\pm$4.00&55.79$\pm$2.81&87.52$\pm$5.61\\
            \noalign{\smallskip}
            \hline
            \noalign{\smallskip}

            \multicolumn{10}{c} {(7\arcsec,0\arcsec)} \\
            \noalign{\smallskip}
            \hline
            \noalign{\smallskip}
T$_\mathrm{pk}$~(K)&3.76$\pm$0.09&2.51$\pm$0.06&1.76$\pm$0.02&1.14$\pm$0.07&0.20$\pm$0.04&-&0.24$\pm$0.05&0.90$\pm$0.02&0.10$\pm$0.01\\
v$_0$ (\kms)&40.69$\pm$0.55&30.78$\pm$0.65&30.79$\pm$0.31&38.14$\pm$1.64&25.67$\pm$11.14&-&38.66$\pm$4.04&48.80$\pm$0.69&47.37$\pm$2.04\\
FWHM (\kms)&49.02$\pm$1.30&56.72$\pm$1.53&54.37$\pm$0.74&51.49$\pm$3.87&112.70$\pm$26.50&-&41.24$\pm$9.51&68.50$\pm$1.63&78.62$\pm$4.82\\
            \noalign{\smallskip}
            \hline
            \noalign{\smallskip}

            \multicolumn{10}{c} {(0\arcsec,0\arcsec)} \\
            \noalign{\smallskip}
            \hline
            \noalign{\smallskip}
T$_\mathrm{pk}$~(K)&6.11$\pm$0.09&2.51$\pm$0.06&1.76$\pm$0.02&1.56$\pm$0.06&0.46$\pm$0.03&0.31$\pm$0.02&0.57$\pm$0.03&0.98$\pm$0.01&0.12$\pm$0.01\\
v$_0$ (\kms)&32.86$\pm$0.34&30.78$\pm$0.65&30.79$\pm$0.31&33.21$\pm$0.91&24.03$\pm$1.73&30.19$\pm$1.20&32.86$\pm$1.48&33.28$\pm$0.44&38.09$\pm$3.86\\
FWHM (\kms)&49.95$\pm$0.81&56.72$\pm$1.53&54.37$\pm$0.74&51.38$\pm$2.15&47.83$\pm$4.07&45.84$\pm$2.84&52.76$\pm$3.48&60.08$\pm$1.03&77.69$\pm$9.09\\
            \noalign{\smallskip}
            \hline
            \noalign{\smallskip}

            \multicolumn{10}{c} {(-7\arcsec,0\arcsec)} \\
            \noalign{\smallskip}
            \hline
            \noalign{\smallskip}
T$_\mathrm{pk}$~(K)&5.50$\pm$0.15&2.54$\pm$0.07&2.32$\pm$0.03&2.13$\pm$0.08&0.51$\pm$0.04&-&0.65$\pm$0.04&0.74$\pm$0.02&0.10$\pm$0.01\\
v$_0$ (\kms)&22.50$\pm$0.61&21.95$\pm$0.68&22.58$\pm$0.31&22.09$\pm$0.91&24.03$\pm$1.81&-&27.20$\pm$1.55&28.51$\pm$0.63&20.87$\pm$1.52\\
FWHM (\kms)&47.27$\pm$1.45&52.07$\pm$1.60&51.25$\pm$0.74&50.35$\pm$2.15&48.31$\pm$4.27&-&-53.47$\pm$3.65&56.02$\pm$1.48&53.55$\pm$3.57\\
            \noalign{\smallskip}
            \hline
            \noalign{\smallskip}

            \multicolumn{10}{c} {(+7\arcsec,-7\arcsec)} \\
            \noalign{\smallskip}
            \hline
            \noalign{\smallskip}
T$_\mathrm{pk}$~(K)&2.41$\pm$0.06&2.51$\pm$0.06&1.76$\pm$0.02&1.19$\pm$0.18&0.20$\pm$0.04&-&0.10$\pm$0.06&0.60$\pm$0.01&0.05$\pm$0.01\\
v$_0$ (\kms)&33.31$\pm$0.53&30.78$\pm$0.65&30.79$\pm$0.31&30.87$\pm$2.53&25.67$\pm$11.14&-&33.00$\pm$11.63&48.49$\pm$0.85&42.19$\pm$3.41\\
FWHM (\kms)&47.10$\pm$1.25&56.72$\pm$1.53&54.37$\pm$0.74&33.68$\pm$5.95&112.70$\pm$26.50&-&40.00$\pm$27.40&71.77$\pm$1.99&62.43$\pm$8.03\\
            \noalign{\smallskip}
            \hline
            \noalign{\smallskip}

            \multicolumn{10}{c} {(0\arcsec,-7\arcsec)} \\
            \noalign{\smallskip}
            \hline
            \noalign{\smallskip}
T$_\mathrm{pk}$~(K)&4.84$\pm$0.06&2.51$\pm$0.06&1.76$\pm$0.02&1.44$\pm$0.14&-&-&0.39$\pm$0.05&1.31$\pm$0.02&0.17$\pm$0.01\\
v$_0$ (\kms)&25.75$\pm$0.30&30.78$\pm$0.65&30.79$\pm$0.31&26.85$\pm$1.75&-&-&21.09$\pm$2.72&34.37$\pm$0.56&36.40$\pm$2.36\\
FWHM (\kms)&46.35$\pm$0.72&56.72$\pm$1.53&54.37$\pm$0.74&36.90$\pm$4.11&-&-&40.94$\pm$6.41&66.81$\pm$1.31&73.99$\pm$5.57\\
            \noalign{\smallskip}
            \hline
            \noalign{\smallskip}
            \multicolumn{10}{c} {(-7\arcsec,-7\arcsec)} \\
            \noalign{\smallskip}
            \hline
            \noalign{\smallskip}
T$_\mathrm{pk}$~(K)&5.94$\pm$0.13&2.54$\pm$0.07&2.32$\pm$0.03&1.62$\pm$0.08&0.51$\pm$0.04&0.41$\pm$0.07&0.58$\pm$0.05&0.51$\pm$0.01&0.08$\pm$0.01\\
v$_0$ (\kms)&17.34$\pm$0.46&21.95$\pm$0.68&22.58$\pm$0.31&18.75$\pm$1.06&24.03$\pm$1.81&20.20$\pm$2.20&19.05$\pm$1.91&18.86$\pm$0.44&17.27$\pm$3.21\\
FWHM (\kms)&42.61$\pm$1.08&52.07$\pm$1.60&51.25$\pm$0.74&44.55$\pm$2.50&48.31$\pm$4.27&26.74$\pm$5.17&46.72$\pm$4.49&44.30$\pm$1.04&86.56$\pm$7.56\\
            \noalign{\smallskip}
            \hline
            \noalign{\smallskip}

            \multicolumn{10}{c} {(0\arcsec,-14\arcsec)} \\
            \noalign{\smallskip}
            \hline
            \noalign{\smallskip}
T$_\mathrm{pk}$~(K)&3.10$\pm$0.07&2.51$\pm$0.06&1.76$\pm$0.02&1.15$\pm$0.11&0.16$\pm$0.06&-&0.35$\pm$0.06&0.35$\pm$0.02&0.06$\pm$0.01\\
v$_0$ (\kms)&21.46$\pm$0.53&30.78$\pm$0.65&30.79$\pm$0.31&22.47$\pm$2.14&22.46$\pm$7.95&-&13.56$\pm$2.89&30.31$\pm$1.71&30.27$\pm$3.72\\
FWHM (\kms)&46.39$\pm$1.25&56.72$\pm$1.53&54.37$\pm$0.74&45.30$\pm$5.04&40.00$\pm$18.73&-&34.19$\pm$6.81&56.92$\pm$4.04&54.82$\pm$8.77\\
             \noalign{\smallskip}
             \hline	    	    
   \noalign{\smallskip}
          \end{tabular*}
\end{table*}

All \cii\ and \nii\ spectra  were smoothed to a spectral resolution of 2 and 4~{\kms}, respectively. The baseline noise RMS is between 21 and 62 mK for \nii\ and between 42 and 98 mK for {\cii}. We present the \cii\ and \nii\ in their native spatial resolution of 14\arcsec and 18.3{\arcsec}, respectively, in Fig.~\ref{figObs}. The 3x3 grid covers the nucleus of IC~342, while the (0\arcsec,-14\arcsec) position covers a position off the southern mini-spiral arm. 

The \cii\ emission is strongest at  (0\arcsec,-7\arcsec) and weakest at the positions (-7\arcsec,7\arcsec) and (0\arcsec,-14\arcsec) off the spiral arm. The positions (0\arcsec,0\arcsec) and (+7\arcsec,0\arcsec) are about 30\% weaker than the strongest  \cii\ position. The line shape is Gaussian to a good degree.  
 Overall the \cii\ emission follows the \twco(1-0) emission showing a correlation between molecular gas and PDR.

The \nii\ emission is weaker than {\cii}, between 1/3 and 1/10 at the peak level, and has a lower signal-to-noise ratio than the stronger \cii\ signal.  Comparing the \cii\ and \nii\ spectra in Fig.~\ref{figObs} we note that \nii\ shows a slightly broader line width than {\cii}. This is not surprising given the very different physical conditions in  \hii\ regions compared to PDRs/GMCs. Generally, the line centers of \nii\ are in good agreement with \cii\ with a recognizable shift of $\sim$8 ~{\kms} at (0\arcsec,7\arcsec) and (-7\arcsec,0\arcsec).

\begin{figure}
\resizebox{\hsize}{!}{\includegraphics{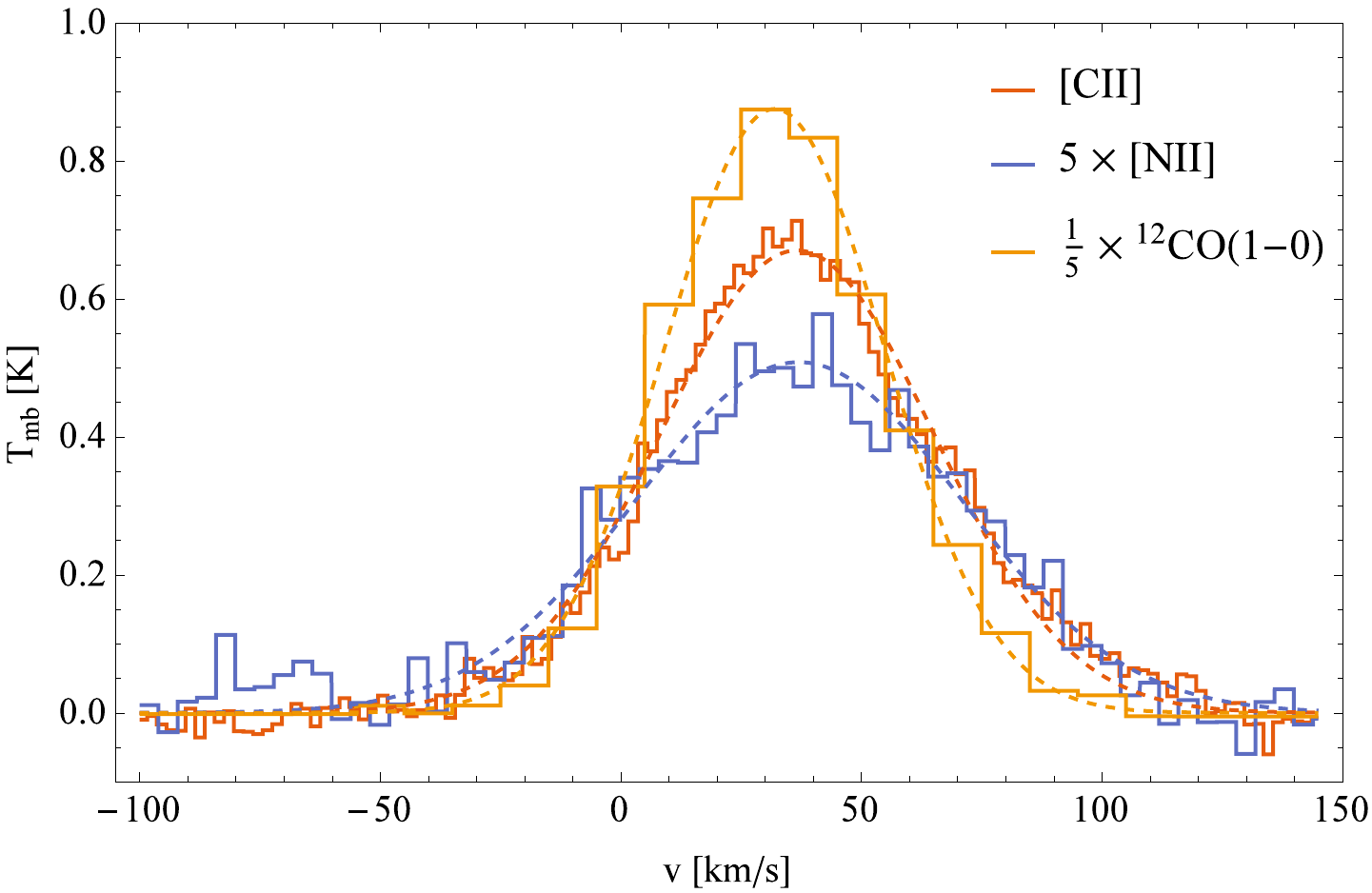}}
\caption{Sum spectra of \cii\ and \nii\ and \twco\ (1-0)  averaged across the central 3x3 grid. Individual positions were equally weighted. The \nii\ intensity scale is multiplied by a factor of 5 and CO is divided by a factor of 5. The dashed lines show the result of Gaussian fits to the lines.}
\label{figAverage}
\end{figure}

To characterize the overall emission of the nucleus of IC~342, we averaged all spectra from the central 3x3 grid of observed positions. Each spectrum was equally weighted. Fig.~\ref{figAverage} shows the resulting averaged spectra of \cii\ and \nii\ . We fitted Gaussian line profiles to both average spectra. For \cii\ the peak intensity is $670\pm 7$~mK, for \nii\ we find $102\pm 3$~mK. Both lines have similar central velocities v$_0$ of $36.6\pm 0.3$~\kms\ ({\cii}) and  $37.0\pm 1.0$~\kms\ ({\nii}). The average \cii\ line is narrower (FWHM) than the \nii\ line: $66.6\pm 0.8$~\kms\ vs.  $79.8\pm 2.4$~\kms . To summarize, the averaged \nii\ emission is slightly redshifted with respect to the \cii\ line and shows a line profile that is about 13~{\kms} broader than {\cii}, otherwise both average spectra are well represented by Gaussians (dashed lines in Fig.~\ref{figAverage}). 

Following the approach in \citet{roellig2012} we also compare the SOFIA data with complementary emission line data of CO and atomic carbon. In the Appendix in Fig.~\ref{figAppendix} we show for each position an overlay of the fine-structure lines presented in this paper with the available data. A direct comparison is complicated by the different spatial resolution and spatial sampling of the various lines \citep[see][for details of whether and how the data was smoothed and/or re-sampled]{roellig2012}.

Generally speaking, the agreement between \cii\ and the molecular gas is strongest on the molecular ring and the spiral arms. Positions away from the spiral arms, e.g. (7\arcsec,-7\arcsec) show a significant difference in line width and central velocity indicating a different kinematic origin.
We note that the various lines in Fig.~\ref{figAppendix} show
significantly different line profiles at some positions. This occurs because the shown data only partly cover the positions observed with SOFIA. When re-sampling was not possible we chose the nearest neighbor spectrum.  An exception is the \twco\ (1-0) data with a beam size and spatial sampling superior to SOFIA data. Hence, in this paper, we do not perform a detailed comparison of the \cii\ and \nii\ lines with all the additional emission lines, but select the \twco\ (1-0) BIMA-SONG data\footnote{\url{http://ned.ipac.caltech.edu/level5/March02/SONG/SONG.html}}\citep{bimasong} as kinematic reference.  

%-------------------------------------------
%  Section KINEMATICS
%-------------------------------------------
      
\section{Kinematics in the nucleus of IC~342\label{sectKinematics}}

\citet{meier2005} combined multi-line millimeter observations to derive an overall scenario of the structure and dynamics of the nucleus of IC~342. In response to the central barred gravitational potential, the molecular gas forms a mini-spiral with trailing arms (in their scenario), which ends in a circumnuclear ring hosting several GMCs. Gas flows along the spiral arms onto the nuclear ring, triggering star formation at the rate of $\sim 0.1 \mathrm{M}_\odot \mathrm{yr}^{-1}$. Recently, \citet{rabidoux2014} used thermal and nonthermal 33~GHz luminosities to derive  star formation rates of 0.4-0.6~$\mathrm{M}_\odot \mathrm{yr}^{-1}$ within the central 23\arcsec. The volume inside the ring is dominated by the  massive central nuclear star cluster. Its intense radiation gives rise to an expanding bi-conical outflow of hot, ionized gas, similar to the Fermi Bubbles observed in the Milky Way \citep{su2010}.

\begin{figure*}[ht]
\centering
\begin{minipage}{8.5cm}
\flushleft
\includegraphics[width=8.2cm]{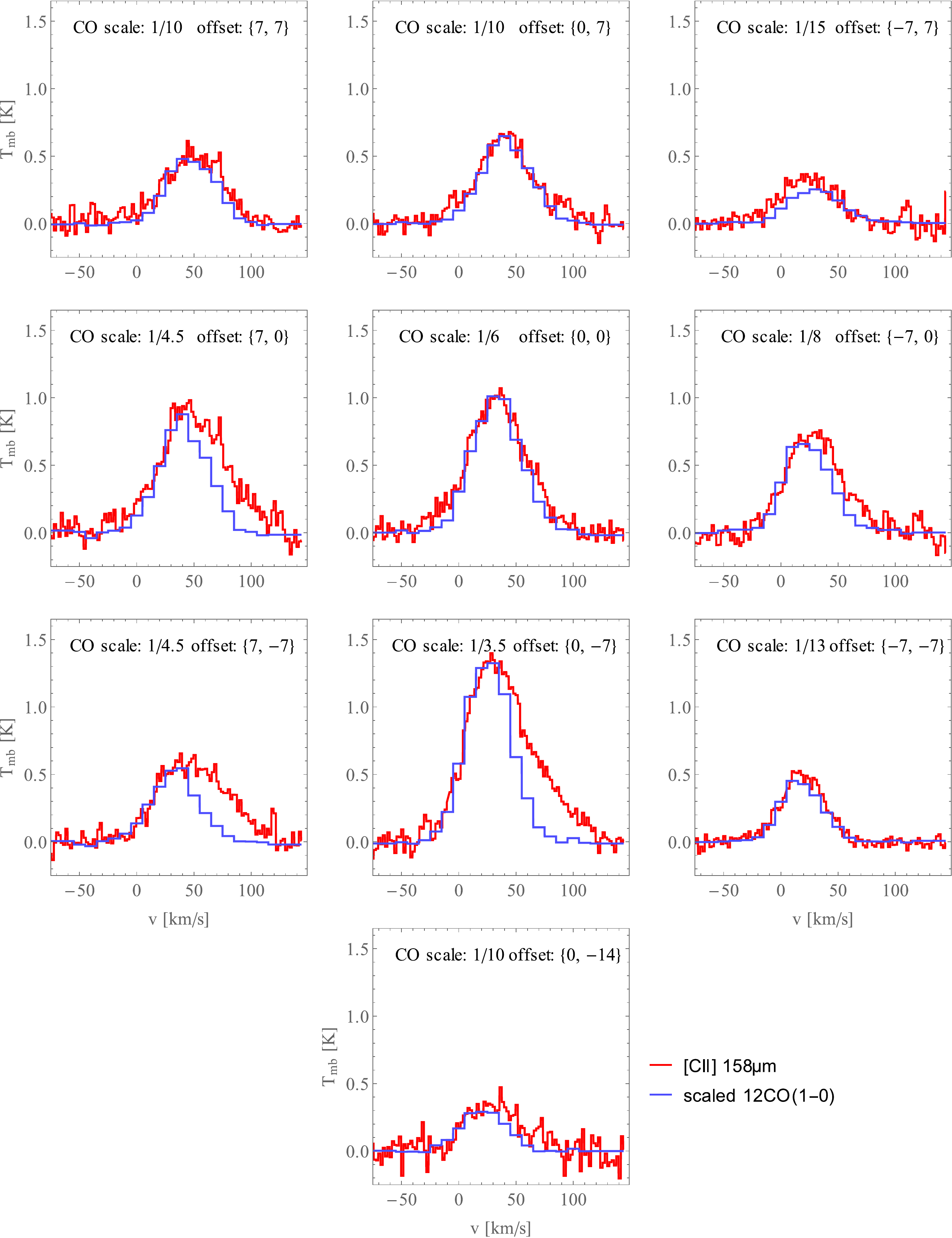}
\end{minipage}
\begin{minipage}{8.5cm}
\flushright
\includegraphics[width=8.2cm]{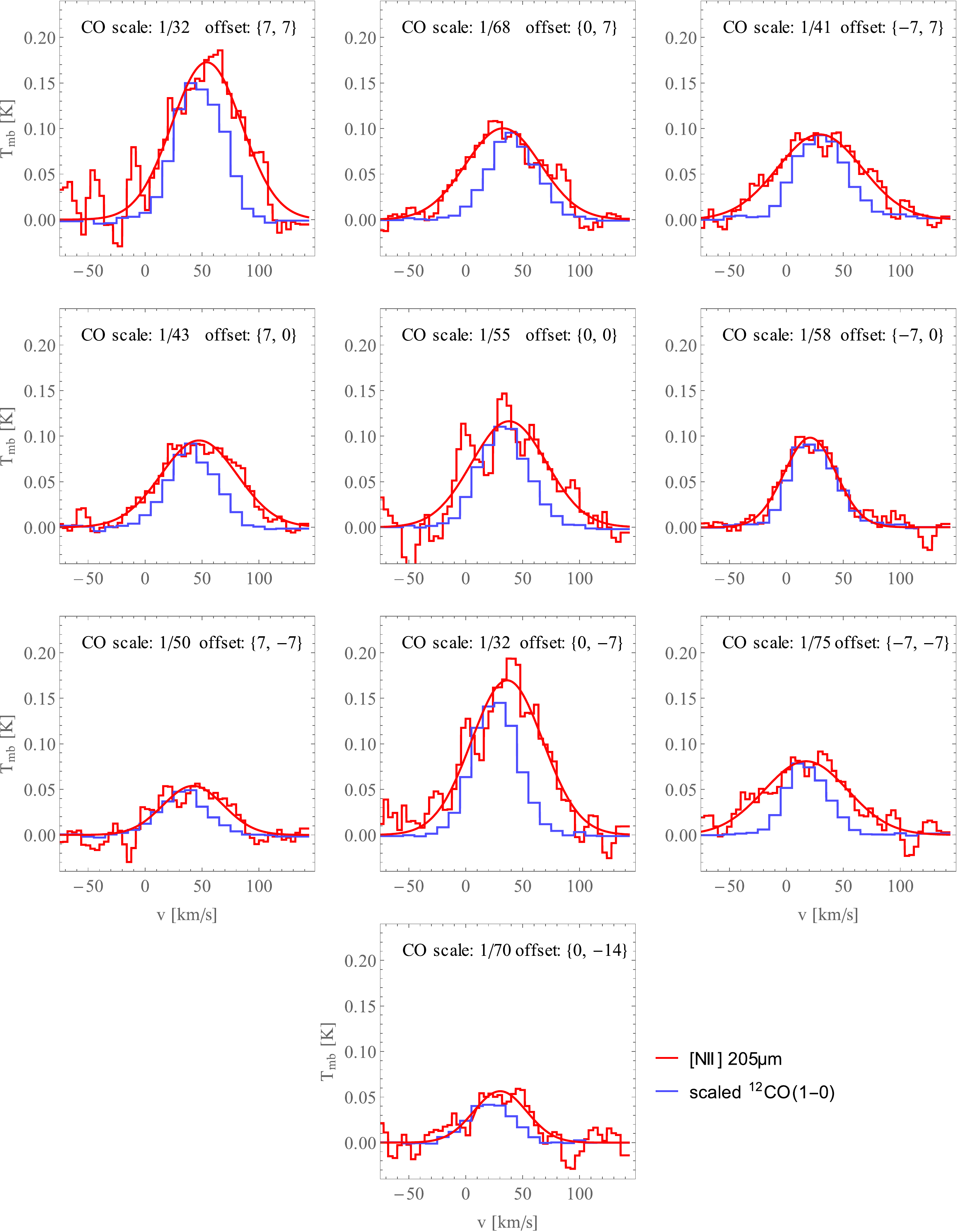}
\end{minipage}
\caption{Comparison of the SOFIA fine-structure spectra with the corresponding \twco\ (1-0) line profiles scaled down to match either the line wings or peak strength of the \cii\ (left panel) and \nii\ (right panel) profiles such that the scaled CO emission is weaker than the \cii\ and {\nii} at all velocities. For easier comparison, we smoothed the \nii\ spectra with a moving average across three spectral channels and also show the Gaussian fits to the lines. The CO scaling factors are given at each position individually.}\label{figObs2}
\end{figure*}

To visualize the kinematic differences between the SOFIA data and molecular gas we show in Fig.~\ref{figObs2} a comparison between the \cii\ and \nii\ line profiles (left and right panel, respectively) with scaled down \twco\ (1-0) line profiles. The scaling was done such that the downscaled CO emission is never stronger than the SOFIA line profiles. This allows us to immediately identify  C$^+$ and N$^+$ gas with different kinematics than the molecular gas. We note that the \cii\ line profiles show a good agreement with the CO along the mini-spiral (diagonally from top left to bottom right). The same is not true for the \nii\ line profiles where we see a significant difference in line shapes compared to the CO. The lower left quadrant of the \cii\ data (positions (+7\arcsec,0\arcsec), (+7\arcsec,-7\arcsec), and (0\arcsec,-7\arcsec)) shows a significant redshifted part that is not visible in the CO data. The same redshifted gas is also visible in {\nii}. The topright position shows additional blueshifted C$^+$ gas, while the right position does not show any blueshifted material, but a weak redshifted contribution. These two positions look different in {\nii}. The topright shows a significantly broader line centered at the \twco\ (1-0) peak, and the right position does not show any component that is kinematically different to the CO. 
\begin{figure}
\centering
\resizebox{\hsize}{!}{\includegraphics{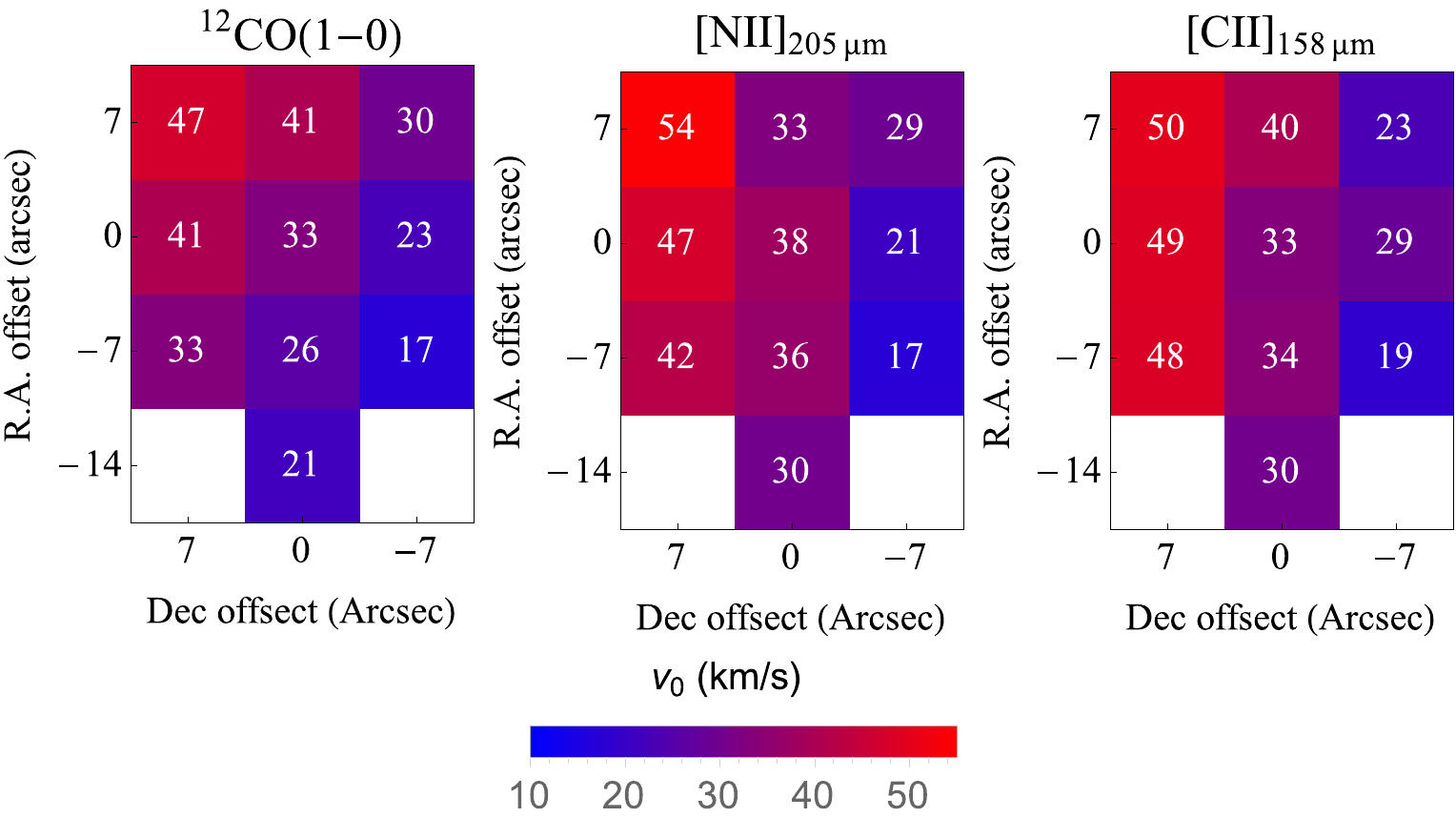}}
\caption{Comparison of the Gaussian line center velocities of \twco\ (1-0), {\cii}, and {\nii}. The white numbers are the corresponding rounded velocities. The spatial resolution is 14\arcsec and 18\arcsec for \cii\ and {\nii}, respectively. }
\label{figVelo1}
\end{figure} 
\begin{figure*}
\centering
\includegraphics[width=8.5cm]{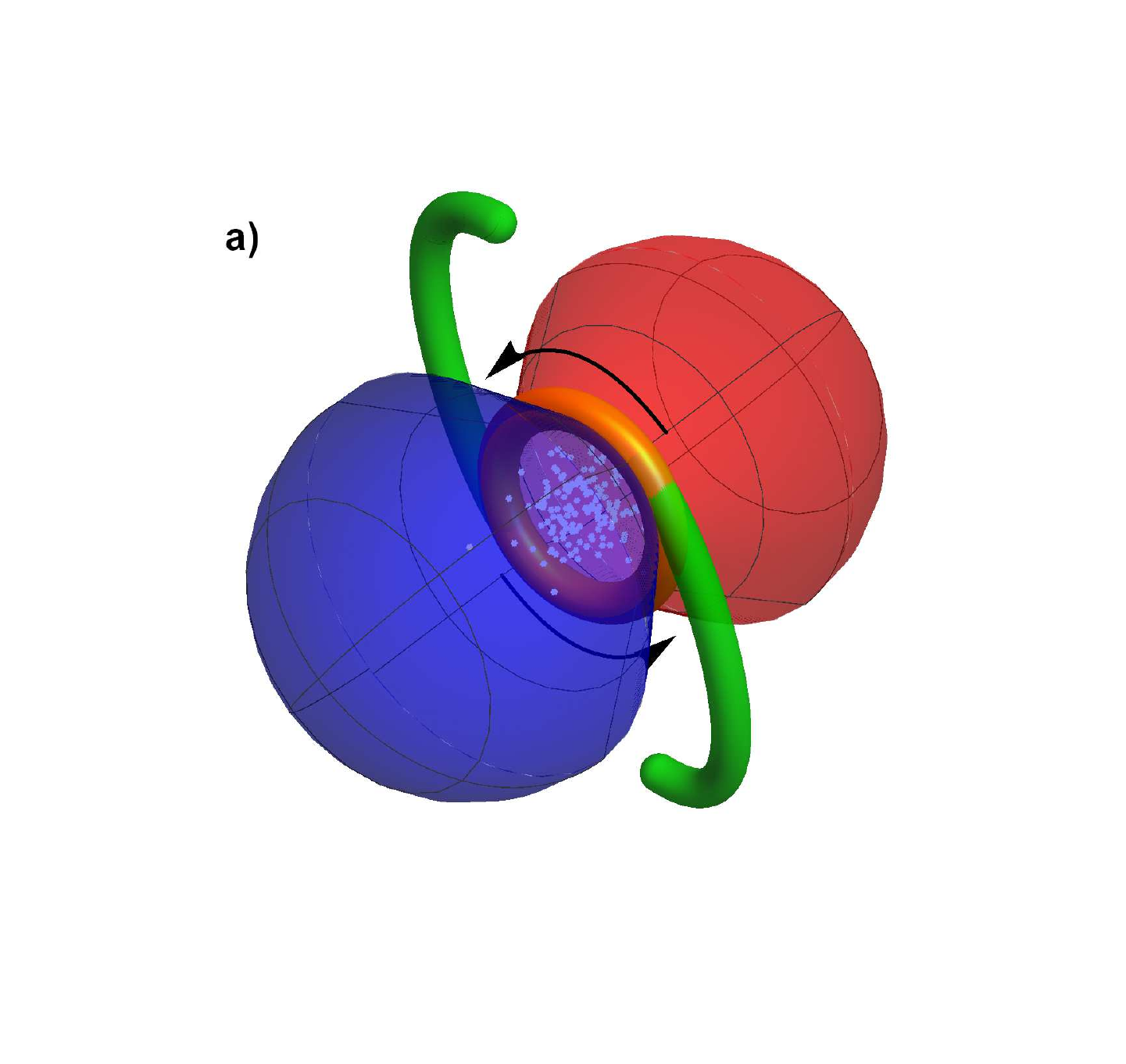}
\hfill
\includegraphics[width=8.5cm]{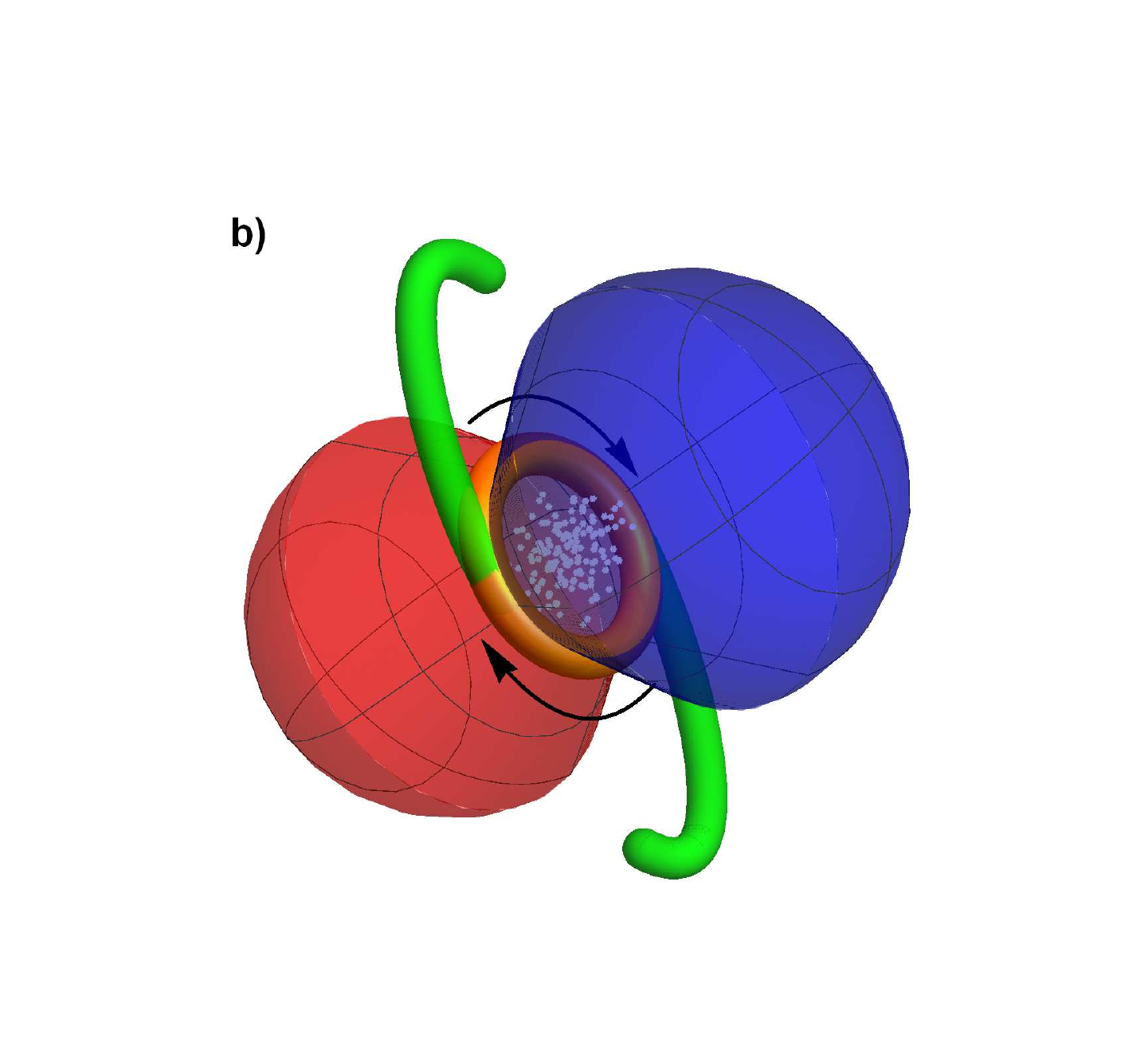}
\caption{Geometrical and kinematic scenarios as explained in Sect.~\ref{sectKinematics} (originally suggested by \citet{meier2005} and slightly modified in this work). The red and blue lobe structures indicate the bi-conic expanding \hii\ gas mentioned in the text. Blue gas is moving toward the observer, red gas is moving away. The mini-spiral (green) and ring (yellow) are surrounding the central nuclear star cluster (white stars), which gives rise to the expanding \hii\ regions as well as to the intense PDR emission from the gas in the ring/spiral facing the cluster. The black arrows indicate a potential  direction of motion consistent with the velocity structure observed in \twco\ (1-0). \textbf{a)} Standard geometry as proposed by \citet{meier2005} with arms in a trailing configuration. \textbf{b)} Proposed new geometry, spiral/arm plane flipped by 90\textdegree\ and a leading arm pattern.  (The observer is located directly above the plane of paper.)}
\label{figGeo}
\end{figure*}

In Fig.~\ref{figVelo1} we compare the Gaussian line center velocities of  \twco\ (1-0), \cii\ , and \nii\ for all ten SOFIA positions. \twco\ (1-0) shows a clear velocity gradient from the southwest to the northeast, a signature of spiral/ring rotational kinematics. \footnote{The spatial resolution of the data is 14\arcsec and 18\arcsec for \cii\ and {\nii}, respectively. Accordingly, at each of our positions we pick up emission from the neighboring pixel.}  It is consistent with a general nuclear rotation of an inclined spiral (the rotation axis is oriented from the SE to the NW) together with  the aforementioned gas flow along the spiral arms. Which of the two motions is dominating is unclear. However, when observing only tracers that are arranged spatially within the rotating plane, thus following the nuclear rotation and gas flow along the arms, one suffers from a degeneracy of inclination angle of the rotating plane and the direction of rotation. To actually distinguish between the two possible configurations it is necessary to observe a physically associated motion perpendicular to the plane, i.e., hot, ionized gas in the scenario of an expanding outflow due to the central star cluster.

Given the scenario above, the two lobes of expanding \hii\ regions should emit significantly in \cii\ and \nii\ and allow us to derive a more detailed geometric model of the nuclear region of IC~342. First of all, we note that the general rotation signature is also visible in \cii\ and \nii\ (Figs.~\ref{figObs2} and \ref{figVelo1})  as of course some of the ionized gas will be associated with and follow the motion of the bulk of the gas in the central region. Superimposed on the general kinematics, we also note some deviations originating from the very different physical conditions in the emitting regions. 

The \nii\ velocities show a stronger redshift to the south and to the southeast and a blue-shift in the north (middle panel in Fig.~\ref{figVelo1}). The stronger redshift of \nii\ in the SE is in agreement with the general kinematic scenario presented by \citet{meier2005}. However, in their geometry (compare Fig. 10 in \citet{meier2005}), the southeastern lobe is moving toward the observer and should therefore be blueshifted. The SOFIA \nii\ data clearly shows the opposite behavior. This requires us to modify the geometrical model by flipping it by 90\textdegree  around an axis along the spiral arms (see Fig.~\ref{figGeo}). Now, the SE lobe is facing away from the observer, leading to the observed redshift in the \nii\ emission while the NW lobe is expected to be blueshifted, just as it is at the (-7\arcsec,7\arcsec) position.   

The \cii\ velocities in the south and the southeast are also redshifted with respect to \twco\ (1-0). The C$^+$ gas, moving away from the observer is clearly visible as additional redshifted gas in the \cii\ at the SE.

We present two possible interpretations of  the velocity information of \twco\ (1-0), {\nii}, and \cii\ : \textbf{1)} The radius of the ring is about 4\arcsec \citep[][and references therein]{montero2006}. Accordingly, within our central 3x3 grid of 7\arcsec spaced observations we pick up information from both the ring and the S-shaped mini-spiral. If the kinematic signature of the molecular ring that revolves around the central cluster is weaker than the gas inflow along the spiral arms, then the velocity pattern in \twco\ (1-0) is consequently also dominated by the spiral arm gas. In order to produce the observed Doppler shifts this requires a large inclination angle (edge-on) of the spiral-arm plane relative to the observer (see Fig.~\ref{figGeo}, panel a) and a slow ring rotation velocity. \textbf{2)} If the velocity signature is mainly produced by the general rotational motion of the gas in the ring, then the projected rotation direction needs to be clockwise. We note that both scenarios lead to a configuration where the mini-spiral is moving in a leading-arm pattern. With the angular resolution of the data at hand a distinction between these two possibilities is not obvious.

The redshifted component observed to the SE of our small maps is not visible in any other molecular line data. It is therefore unlikely that this redshift is the result of shocks. It is clearly not associated with any denser material and we contribute it to the ionized gas moving in a wide-angle outflow/lobe. The (-7\arcsec,7\arcsec) position shows a blueshifted component in \cii\ as well as a much broader line profile in {\nii}. The position (-7\arcsec,0\arcsec) does not show any strong kinematic deviations from the CO gas in either \cii\ or \nii\ . At (0\arcsec,7\arcsec) the \nii\ line is visibly blueshifted compared to the CO line. The blueshifted gas in the NW could be emission from the approaching lobe of ionized gas. An alternative interpretation could be a shock-related origin. \citet{montero2006} presented  detections of hot NH$_3$ with an emission peak close to our (-7\arcsec,0\arcsec) position. \citet{usero2006} also detected strong SiO emission, a typical shock tracer. We probably see a combination of outflow/expansion of the \hii\ gas in the NW together with some shock-related motions. It is unclear why the nearby portion of the bi-polar outflow is much less pronounced than the far-side, redshifted lobe. A possible cause is an asymmetry in the ring/spiral structure hindering the gas flow in our direction.    
 
\subsection{Discussion}
The current standard geometry of IC~342 is based on a number of previous observations. Generally these observations fall into two different categories: in-plane and off-plane. In-plane tracers are situated in the plane of the mini-spiral/ring and take part in the general rotational dynamic. Examples are molecular line observations, such as \twco (1-0). Examples of an off-plane structure are the proposed bi-polar lobes of expanding ionized gas, traced by H\textalpha\  emission, moving perpendicular to the ring/arm plane. Generally, if only  in-plane Doppler data is available it is not possible to distinguish between, for example a counterclockwise rotation with trailing spiral arms and a clockwise rotation with leading arms in a plane that is 90\textdegree\ flipped. Only the combination with additional data can resolve this degeneracy.

Earlier studies combined in-plane CO emission line maps with assumed off-plane H\textalpha\  emission \citep{meier2001,meier2005}. The apparent lack of H\textalpha\ emission at the southern arm of the mini-spiral is interpreted as extinction due to the foreground arm  
while the 3mm continuum emission (predominantly thermal free-free emission) from optically thin \hii\ regions remains unattenuated. This indicates strong star formation activity at the SW portion of the ring shielded by a significant column of foreground material. Assuming that the emission is from the cluster facing side of the ring and both are situated in the same plane, this would suggest an orientation similar to the left panel in Fig.~\ref{figGeo}. If the same scenario were true for the second peak in the 3mm map at the NE part of the arm one would expect to observe unextincted H\textalpha\ emission. A comparison with Fig.~1 in \citet{meier2005} also shows an H\textalpha\ deficit visible as a significantly darker lane following the molecular arm. Following this line of reasoning, both scenarios in Fig.~\ref{figGeo} are possible.

Instead we argue that observed H\textalpha\ and 3mm continuum emission trace two different regimes. The H\textalpha\ emission shown by \citet{meier2001,meier2005} is predominantly emitted by the ionized gas in the cavity and the  expanding lobes, while the 3mm continuum stems from the current but embedded star formation activity in the molecular ring triggered by the inflowing gas.
 In our flipped geometry scenario the ring would account for the foreground extinction visible as significantly darker lane following the northern arm. The \hii\ emission from the NW is much weaker, most likely due to a strong asymmetry between the two lobes. We note that neither the standard nor the flipped geometry explains the absence of ring emission in the northern ring quadrant. Most likely the ring is broken up or fragmented.

The flipped geometry proposed here would imply a leading spiral arm structure within the \textit{inner} Inner Lindblad Resonance (iILR) then transitioning into a trailing arm outside of the \textit{outer} Inner Lindblad Resonance (oILR). The possibility of such a configuration has been shown in numerical computations assuming a weak barred potential \citep{wada1994, pinolferrer2012}. It is important to remember that leading/trailing arms are just transient, rotating patterns not subject to shear, etc. Fundamentally there is no reason to disregard such a configuration.

The flipped geometry is problematic  in the sense that it proposes a tilt between the plane of rotation of the mini-spiral within the ILR and the global plane of the galaxy. The study of three-dimensional orbits in a tri-axial potential is just at its beginning. Three-dimensional N-body simulations of orbits in rotating potentials show the existence of complex three-dimensional orbits with various, sometimes interchanging tilt angles \citep{pfenniger1984,pfenniger1991}. In a recent work, \citet{portail2015} showed N-body simulations of orbits in the Galactic bulge, demonstrating the existence of bent or tilted orbits in barred discs. We conclude that a possible tilt angle of the mini-spiral/ring plane in IC~342 is neither supported nor prohibited by present theoretical models. A significantly different inclination compared to the general orientation of the plane of IC~342 can therefore not be ruled out a priori.
\citet{meidt2009} derived the detailed velocity structure of IC~342 from CO and HI intensities up to a galactocentric distance of 15 kpc. Their first moment map shows indications of a warped outer disk most likely due to tidal interaction with a close companion galaxy.  The velocity pattern of the inner disk 
seems to show some asymmetry very close to the nucleus that could indicate a changing tilt angle, but  the spatial resolution of the data is not sufficient to support or discard this scenario. \citet{schinnerer2003} showed that the CO gas in the central 300~pc shows noncircular motion and they suggest that this could be due to a nuclear CO disk tilted relative to the large stellar disk. Later they discard this scenario and argue that streaming motions along the arms are a more plausible explanation. \citet{fathi2009} studied the pattern speed in late-type barred spirals and derived their ellipticity profiles. Their analysis showed that IC~342 shows a steep increase in ellipticity at a radius of about 2 kpc. This could indicate a different tilt angle of the inner disk.

An argument in favor of the standard geometry \citep[][see also Fig.~\ref{figGeo}, panel a)]{meier2005} is the presence of $\mathrm{HNCO}$ and $\mathrm{CH}_3\mathrm{OH}$ emission, presumably shock excited, at the front side of the trailing arms. However, the spatial resolution of the data ($\sim$ 6\arcsec $\times$5\arcsec) make accurate localization difficult. Comparing the maps of the shock tracers with the various CO isotopologue maps \citep[Fig.~2 in][]{meier2005}, the displacement between the two appears marginal. Nevertheless, any high-resolution data tracing shocked gas or triggered star formation on either side of the spiral arms would be a good test on the underlying geometry and dynamics.

Another scenario  preserving the standard geometrical interpretation would be a very wide-angle SE outflow together with an arm/ring plane that is significantly tilted with respect to the plane of sky. In this case blueshifted emission is expected close to the cluster and redshifted emission in the SE. Comparison of the line profiles of CO and the ionized gas in Fig.~\ref{figObs2} shows a very weak blueshifted component in \cii\ at the (0\arcsec,0\arcsec) position. The \nii\ profile at (0\arcsec,0\arcsec) is wider compared to CO which could be interpreted as an overlay of blue- and redshifted emission. However, the same is also found at almost all other observed positions and is more likely the signature of a larger overall velocity in the ionized gas. Furthermore, the wide-angle outflow scenario would also affect the entire NW lobe and lead to redshifted emission signatures in {\nii}. Again, the larger linewidths in \nii\ and the overall lack of ionized gas emission in the NW inhibit a verification of this scenario.

We conclude that the  \cii\ and \nii\ data are difficult to explain within the current standard geometry of IC~342's nucleus. We suggest a modification of the current image by flipping the ring/arm plane which leads to a leading arm configuration better explaining the kinematics of the molecular gas in the spiral arm and the ring as well as the ionized gas. A disadvantage of the proposed geometry is the leading arm configuration with a strongly tilted axis with respect to the global galactic disk. Theoretical work on orbits in such a configuration as well as high-resolution observations are both required to resolve this  uncertainty.

%-----------------------------
% Section CII/CO ratio
%-----------------------------

\section{The \cii\ to \twco\ (1-0) ratio\label{cii2co}}

\begin{table*}[!pt]
\caption[]{\cii\ and \twco\ (1-0) integrated line intensities at all observed positions.}
\label{tab:3}
{\centering
\begin{tabular*}{1.0\textwidth}{@{\extracolsep{\fill}} c l l l l l}
\hline\hline
\noalign{\smallskip}
($\Delta$RA, $\Delta$Dec)&$I\left(\cii\right)$\tablefootmark{a}&$I\left(\twco{(1-0)}\right)$\tablefootmark{a}&
%I(\cii\ )/I(\twco\ (1-0) )
\multirow{2}{*}{$\dfrac{I\left(\cii\right)}{I\left(\twco{(1-0)}\right)}$}&$\int T_\mathrm{mb}^\mathrm{[CII]} dv/\int T_\mathrm{mb}^\mathrm{CO(1-0)}dv$ \\
(\arcsec,\arcsec)&erg~s$^{-1}$~cm$^{-2}$~sr$^{-1}$&erg~s$^{-1}$~cm$^{-2}$~sr$^{-1}$\\
             \noalign{\smallskip}
             \hline
             \noalign{\smallskip}
             \centering
 (7, 7) &$ 2.29\times 10^{-4} $&$ 4.06\times 10^{-7} $&$ 564 $& 0.13 \\
 (0, 7) &$ 2.74\times 10^{-4} $&$ 5.26\times 10^{-7} $&$ 520 $& 0.12 \\
 (-7, 7) &$ 1.42\times 10^{-4}$ &$ 3.39\times 10^{-7}$ &$ 420$ & 0.09 \\
 (7, 0) &$ 4.62\times 10^{-4} $&$ 3.06\times 10^{-7} $&$ 1508 $& 0.33 \\
 (0, 0) &$ 4.42\times 10^{-4} $&$ 5.07\times 10^{-7} $&$ 873 $& 0.19 \\
 (-7, 0) &$ 3.09\times 10^{-4} $&$ 4.32\times 10^{-7} $&$ 717$ & 0.16 \\
 (7, -7) &$ 3.22\times 10^{-4} $&$ 1.88\times 10^{-7} $&$ 1708$ & 0.38 \\
 (0, -7) &$ 6.55\times 10^{-4} $&$ 3.72\times 10^{-7} $&$ 1760$ & 0.39 \\
 (-7, -7) &$ 1.68\times 10^{-4}$ &$ 4.20\times 10^{-7} $&$ 400$ & 0.09 \\
 (0,-14) &$ 1.48\times 10^{-4}$ &$ 2.39\times 10^{-7}$ &$ 622$ & 0.14 \\
 \noalign{\smallskip}
             \hline
             \noalign{\smallskip} 
average&$3.33\times 10^{-4}$&$3.9\times 10^{-7}$&$855$&$0.19$\\
             \noalign{\smallskip}
             \hline
             \noalign{\smallskip}  
\end{tabular*}
}
\tablefoottext{a}{To convert to temperature units use $10^{-4}$~erg~s$^{-1}$~cm$^{-2}$~sr$^{-1}=14.23$~{\Kkms} for the \cii\ line and $10^{-7}$~erg~s$^{-1}$~cm$^{-2}$~sr$^{-1}=64.16$~{\Kkms} for the \twco\ (1-0) line.} 
\end{table*}

\begin{figure}
\centering
\resizebox{\hsize}{!}{\includegraphics{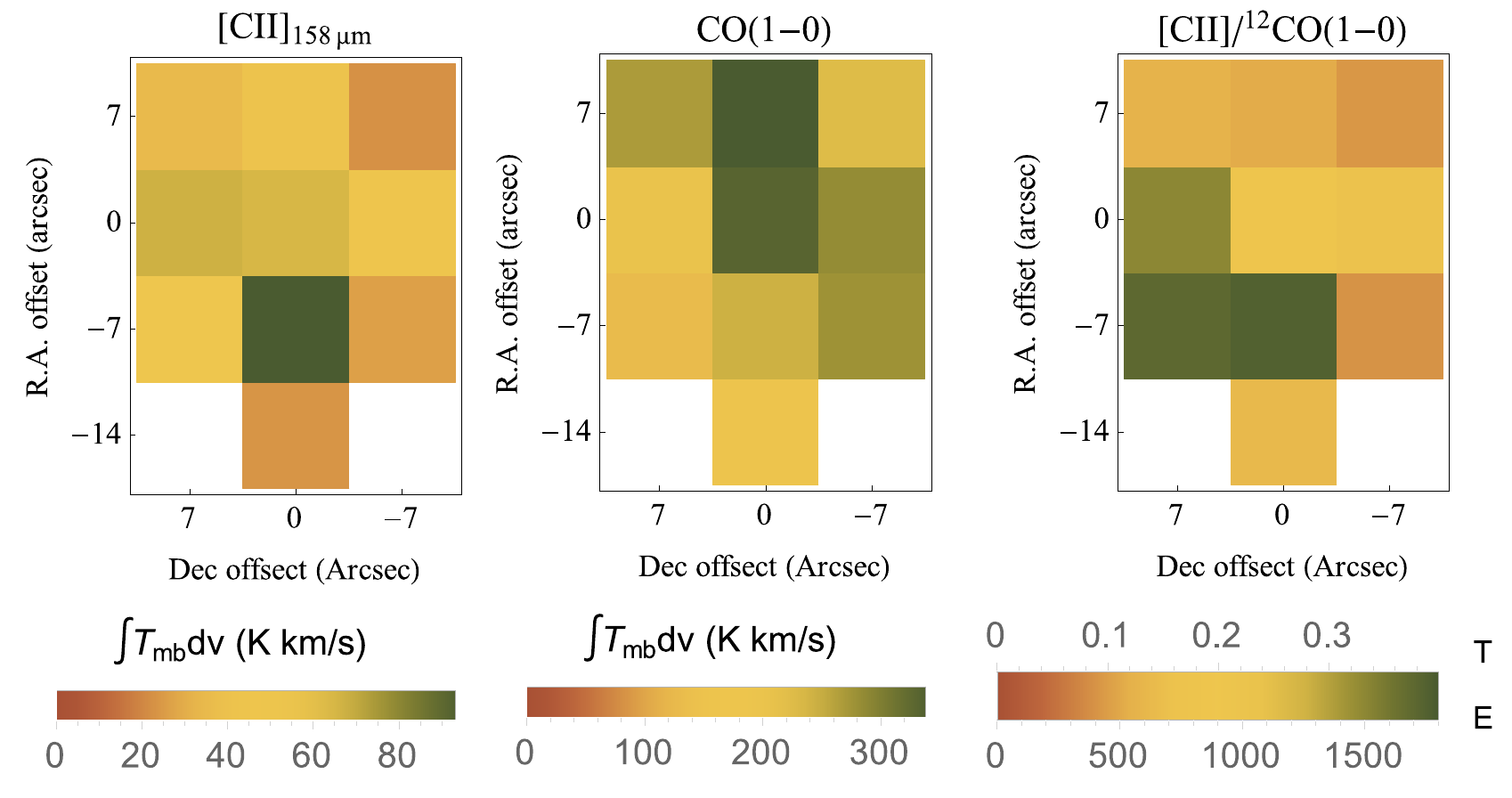}}
\caption{Spatial distribution of the integrated line intensities $\int T_\mathrm{mb}dv$ of \cii\ (left), \twco\ (1-0) (middle), and the $I(\cii\ )/I(\twco\ (1-0))$ line ratio (right). The colorbar of the right panel specifies the values for the line ratio  calculated by using temperature units (T) and energy units (E). To convert T units to E units, multiply T by 4510. The spatial resolution is 14\arcsec and 18\arcsec for \cii\ and {\nii}, respectively.}
\label{figRatio}
\end{figure}
The intensity ratio $I(\cii\ )/I(\twco\ (1-0))$ is often used to characterize the energetic state and star formation activity of the ISM \citep[e.g.][]{stacey1991}. The \cii\ intensity strongly scales with the intensity of the ambient FUV intensity, which is mainly produced by young massive stars. The \twco\ (1-0) line, on the other hand, is emitted from cooler, better shielded material. A high \cii\ /\twco\ (1-0) ratio is indicative of strong PDR emission and star-burst activity. Common values range from a few hundred up to a few $10^4$ in extreme regions, such as 30 Doradus. 
In Table~\ref{tab:3} we compare the integrated line intensity ratio for all observed positions and for the sum spectra, averaged over the central 3x3 grid. The values range from 400 to 1800. In Fig.~\ref{figRatio} we show the spatial distribution of the ratios. The highest values are found along the lower left corner of our 3x3 grid, partly corresponding to the GMC A \citep{meier2001}. The strong star formation in GMC B (and to a lesser degree also GMC E) is most likely causing the slightly increased line ratio at the offsets (0\arcsec,0\arcsec) and (-7\arcsec,0\arcsec) (compare with Fig.~\ref{figObs}). 

Usually, the \cii\ /\twco\ (1-0) ratio is used to deduce the global star formation activity of an object. On global scales, \cii\ emission from \hii\ regions only contributes about 20\% to the total \cii\ intensity \citep{pineda2014}, 
but an increased active star formation gives rise to a higher FUV flux on larger scales and to an enhanced \cii\ /CO line ratio accordingly. The much higher angular resolution of the SOFIA \cii\ data compared to the data available to \citet{stacey1991} reveals significant local variations in the line ratio when pointing on or off sites of active star formation. This is apparent when comparing the position dependent \cii\ /\twco\ (1-0) line ratio with the average value of $\sim 855$ (Table~\ref{tab:3}, last line). From Fig.~\ref{figRatio} we see that the high ratio at (7\arcsec,-7\arcsec) is mostly  driven by the lack of CO emission, while at (0\arcsec,-7), the high ratio is directly caused by the strong \cii\ emission. 

Our data shows only a spatial correlation between the \cii\ and \twco\ (1-0) integrated line intensities (Pearson correlation coefficient $\rho=0.80$). The \cii\ line peaks more around the edges of the CO distribution. 
Our angular resolution is too low to spatially attribute the \cii\ emission to geometrical structures as depicted in Fig.~\ref{figGeo}. 
The  original interpretation of  \citet{meier2005} who 
locate the PDR activity on the inside of the molecular ring 
facing the nuclear cluster remains valid even in the proposed new geometry. The only difference is that the PDR surface, i.e. the cluster facing side of the ring, is facing away from the observer.  

We note that the average line ratio derived in this paper is significantly lower than the value of 3250 \citep[][corrected for a main beam efficiency of 0.65]{stacey1991}. 
Given the much lower area filling factor of the cool CO gas compared to more diffuse and widespread C$^+$, we expect the line ratio to increase with increasing beam sizes. 

%------------------------------------------
% Section NII and CII analysis
%------------------------------------------

\section{\nii\ and \cii\ analysis}
\begin{figure}
 \centering  
    \epsfig{file=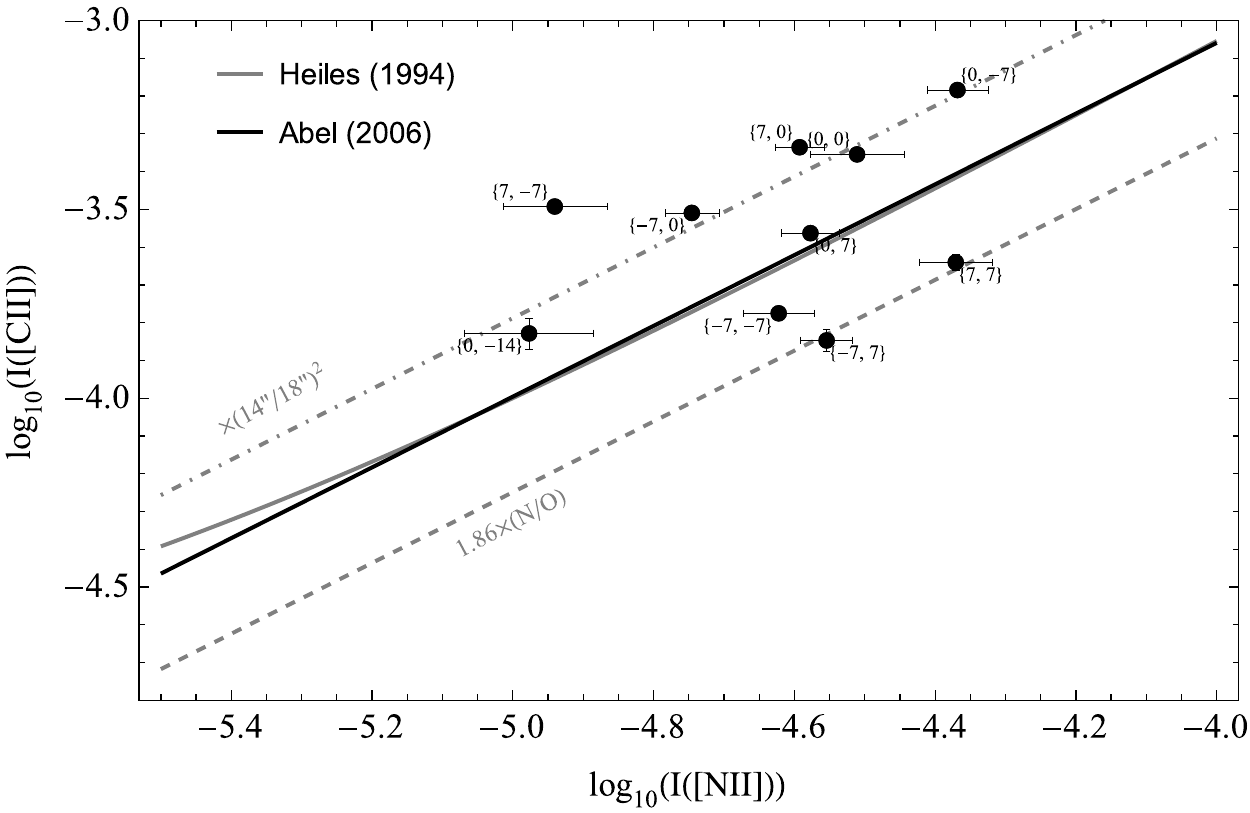,angle=0,width=0.85\linewidth} \\
 \caption{Correlation between $I(\nii)$ (205~{\textmu}m) and $I(\cii)$ (159~{\textmu}m). The data points are labeled with their position (in offsets of \arcsec). The black line corresponds to the best fit line provided by \citet{abel06b} and the solid gray line shows the fit given by \citet{heiles1994}. The dashed and dot-dashed lines show how the relation by \citet{abel06b} is affected by an enhanced elemental (N/O) ratio ($(1.86\times\mathrm{(N/O)}$) and by beam size effects ( $I(\nii)\times(14"/18")^2$), respectively. }\label{figCorrelation}
 \end{figure}

\begin{table*}[!pt]
      \caption[]{\cii\ and \nii\ line intensities at all observed positions. Subscript \textit{H}$^+$ denotes values computed from Eq.~(1) from \citet{abel06b}. Negative results in the difference between observed and theoretical \cii\ intensities in the last column are indicated by a dash.}
      \label{tab:2}
         {\centering
         \begin{tabular*}{1.0\textwidth}{@{\extracolsep{\fill}} l l l l l l l}
           \hline\hline
	    \noalign{\smallskip}
($\Delta$RA, $\Delta$Dec)&$I(\nii)$\tablefootmark{a}&N(N$^+$)&$I$(\cii)\tablefootmark{a}&$I$(\cii)$_\mathrm{H^+}$&\multirow{2}{*}{
$\dfrac{I(\cii)_\mathrm{PDR}}{I(\cii)}$}\tablefootmark{b}&$I$(\cii)-$I$(\cii)$_\mathrm{H^+}$ \\
(\arcsec,\arcsec)&erg~s$^{-1}$~cm$^{-2}$~sr$^{-1}$&cm$^{-2}$&erg~s$^{-1}$~cm$^{-2}$~sr$^{-1}$&erg~s$^{-1}$~cm$^{-2}$~sr$^{-1}$&&erg~s$^{-1}$~cm$^{-2}$~sr$^{-1}$\\
             \noalign{\smallskip}
             \hline
             \noalign{\smallskip}
 7 , 7 &$ 4.26\times 10^{-5} $&$6.86\times 10^{16}$&$ 2.29\times 10^{-4} $&$ 3.92\times 10^{-4}
   $&$ - $&$-  $\\
 0, 7 &$ 2.65\times 10^{-5} $&$4.27\times 10^{16}$&$ 2.74\times 10^{-4} $&$ 2.51\times 10^{-4}
   $&$ 0.08 $&$ 2.22\times 10^{-5} $\\
 -7 ,7 &$ 2.79\times 10^{-5} $&$4.49\times 10^{16}$&$ 1.42\times 10^{-4} $&$ 2.64\times
   10^{-4} $&$ - $&$- $\\
 7 , 0 &$ 2.56\times 10^{-5} $&$4.12\times 10^{16}$&$ 4.62\times 10^{-4} $&$ 2.43\times 10^{-4}
   $&$ 0.47 $&$ 2.19\times 10^{-4} $\\
 0 , 0 &$ 3.08\times 10^{-5} $&$4.95\times 10^{16}$&$ 4.42\times 10^{-4} $&$ 2.90\times 10^{-4}
   $&$ 0.34 $&$ 1.52\times 10^{-4} $\\
 -7 , 0 &$ 1.80\times 10^{-5} $&$2.90\times 10^{16}$&$ 3.09\times 10^{-4} $&$ 1.75\times 10^{-4}
   $&$ 0.44 $&$ 1.35\times 10^{-4} $\\
 7 ,-7 &$ 1.15\times 10^{-5} $&$1.85\times 10^{16}$&$ 3.22\times 10^{-4} $&$ 1.15\times
   10^{-4} $&$ 0.64 $&$ 2.07\times 10^{-4} $\\
 0 , -7 &$ 4.28\times 10^{-5} $&$6.89\times 10^{16}$&$ 6.55\times 10^{-4} $&$ 3.94\times
   10^{-4} $&$ 0.40 $&$ 2.60\times 10^{-4} $\\
 -7 , -7 &$ 2.39\times 10^{-5} $&$3.85\times 10^{16}$&$ 1.68\times 10^{-4} $&$ 2.28\times
   10^{-4} $&$ - $&$-  $\\
 0 , -14 &$ 1.05\times 10^{-5} $&$1.69\times 10^{16}$&$ 1.48\times 10^{-4} $&$ 1.06\times
   10^{-4} $&$ 0.29 $&$ 4.23\times 10^{-5} $\\
\noalign{\smallskip}
\hline
\noalign{\smallskip}
average\tablefootmark{c}&$2.26\times 10^{-5}$&$3.64\times 10^{16}$&$3.03\times 10^{-4}$&$2.16\times
   10^{-4}$&$0.29$&$8.64\times 10^{-5}$\\

 \noalign{\smallskip}
\hline
\end{tabular*}
}
\tablefoottext{a}{To convert to temperature units use $10^{-5}$~erg~s$^{-1}$~cm$^{-2}$~sr$^{-1}=3.13$~{\Kkms} for the \nii\ 205~{\textmu}m line and $10^{-4}$~erg~s$^{-1}$~cm$^{-2}$~sr$^{-1}=14.23$~{\Kkms} for the \cii\ line.} 
\tablefoottext{b}{$I(\cii)_\mathrm{PDR}=I(\cii)-I(\cii)_\mathrm{H^+}$.}
\tablefoottext{c}{The average values are averaged over the central 3x3 grid and correspond to the spectra shown in Fig.~\ref{figAverage}.}
\end{table*}
Because of their different ionization potentials, \nii\ and \cii\ can originate from different environments. While both tracers can be emitted from \hii\ regions, the lack of photons with energies above the Lyman limit in photon-dominated regions prohibits the emission of nitrogen fine-structure lines from PDRs. An interesting question is, to what degree is it possible to use the observed \nii\ emission to disentangle what fraction of \cii\ emission stems from \hii\ regions and from PDRs? Based on numerical models of photo-ionization gas and PDRs using the Cloudy model \citep{ferland2013}, \citet{abel06b} presented the following correlation,
\begin{equation}
\label{eq:1}
\log(I(\cii)_\mathrm{H^+})=0.937\log(I(\nii))+0.689\,\, ,
\end{equation}
where $I(\nii)$ and $I(\cii)_\mathrm{H^+}$ are the integrated line intensities of \nii\ and \cii\ from the \hii\ region (intensities given in energy units). This equation agrees well with a similar expression given by \citet{heiles1994}. In Fig.~\ref{figCorrelation} we compare our data with the expected \cii\ from the \hii\ regions. Points on the theoretical curves correspond to observed \cii\ intensities that are produced exclusively by \hii\ gas. Data points to the left of the lines correspond to a combination of PDR and \hii\ gas; data points to the right of the curves are weaker than is expected from \hii\ regions only. This is the case for three of our ten positions, but the deviation from the theoretical curve is not too strong. The detailed {\cii}-\nii\ correlation is  critically dependent on the overall metallicity and the  elemental carbon-to-nitrogen abundance ratio. Local variations of the elemental abundances will alter the numerical values in Eq.~\ref{eq:1} and shift the theoretical line along the $I(\cii)$ axis. Recently, \citet{florido2015} showed that the nuclei of barred spiral galaxies with star formation have a significantly enhanced (N/O) ratio. They find $\log(\mathrm{N/O})=-0.49$, which is a factor of 1.86 higher than the value assumed by \citet{abel06b}. Assuming a linear scaling between N$^+$ column density and \nii\ emission, this would shift the solid black line in Fig.~\ref{figCorrelation} to the right. Similarly, if \cii\ and \nii\ is emitted from the same local volume then \nii\ needs to be corrected for beam size effects,  $I(\nii)\times(14\arcsec/18\arcsec)^2=I(\nii)\times 0.6$, shifting the relation to the left. In the case of IC~342 these two opposite effects will most likely cancel each other out to some degree. However, this suggests a significantly lower {\cii}/{\nii} ratio in sources with solar or subsolar metallicity  due to the area beam filling
\footnote{There are different beam filling factors that are often confused. Here we use the following: If $A_i$ is the projected area of object $i$ then $\phi_a=A_\mathrm{source}/A_\mathrm{beam}$ measures the coverage of the source extent by the beam.
Thus, factor $\phi_a$ corrects fluxes for source extents larger than the beam (where surface brightness is not affected) as well as surface brightness for source extents smaller than the beam (where flux is not affected). 
}
 $\phi_a\ll 1$. 

Table~\ref{tab:2} lists the \cii\ and \nii\ intensities together with the theoretically expected value from Eq.~\ref{eq:1} and their ratios and differences. It is remarkable that the fraction of \cii\ emission coming from the PDR is only between 10 and 65\%, which means that approximately 35-90\% of all the \cii\ is from the {\hii}. This is a much higher fraction than the often assumed range of 10-60\% \citep{abel06b}. \citet{pineda2014} estimate that in the Milky way the contribution from different gas phases to the \cii\ luminosity is dense PDRs (30\%), cold H{\scriptsize I} (25\%), CO-dark H$_2$ (25\%), and ionized gas (20\%). These are averaged values. They also show that the \cii\ luminosity in the inner few kpc of the Milky Way is dominated by ionized gas and only a minor fraction is contributed by the PDRs. These results are in agreement with our findings from the inner few hundred pc of IC~342 making it  a possible template for our Galactic Center.

Averaged over the central $3\times 3$ we observed a ratio $I(\cii)/I(\nii)=12$ in energy units.
Applying Eq.~\ref{eq:1} to the averaged line intensities instead, we find $\langle I(\cii)_\mathrm{H^+}\rangle=2.62\times 10^{-4}\, \ergcmssr$ corresponding to about 80\% of the total \cii\ intensity that is contributed by the ionized gas.   

 \citet{pineda2014} have also studied the use of the \cii\ luminosity as a tracer for the star formation rate. They give a fit of the form
\begin{equation}\label{eq:2}
\log(SFR)=m \log L_{\cii}+b\,\,.
\end{equation}
If $L_{\cii}$ is the total \cii\ luminosity stemming from all the various contributing phases, then $m=0.98\pm 0.07$ and $b=-39.80\pm 2.94$. However, if most of the \cii\ luminosity is produced in the ionized gas \citet{pineda2014} find $m=0.91\pm 0.06$ and $b=-36.30\pm 2.36$, resulting in a higher SFR for a given $L_{\cii}$. We estimate the total \cii\ luminosity from $L_{\cii}=4\pi\,R^2\,\langle I(\cii)\rangle$, with the radius of the emitting nuclear region $R\approx250$~pc, and the mean \cii\ intensity  $\langle I(\cii)\rangle$ from Table~\ref{tab:2}.  Across our ten positions we find $L_{\cii}=2.7\times 10^{5}-1.3\times 10^{6}\,L_\odot$.  Using Eq.~\ref{eq:2} and assuming an \hii\ dominated \cii\ luminosity this corresponds to star formation rates between 0.16 and 0.65~M$_\odot$yr$^{-1}$ within the central 500~pc, similar to the range of 0.4-0.6~M$_\odot$yr$^{-1}$ derived by \citet{rabidoux2014}. If we assume that the \cii\ luminosity is produced not only in the ionized gas, but stems from all phases, we find  significantly lower star formation rates of 0.03 and 0.13~M$_\odot$yr$^{-1}$. Reversing the argument, we can take the SFR derived by \citet{rabidoux2014} and compute the expected \cii\ luminosities using Eq.~\ref{eq:2}. We already expect the \cii\ emission to be dominated by emission from the ionized phase, thus assuming contributions from all phases in Eq.~\ref{eq:2} we will overestimate the \cii\ luminosities. Putting in the numbers we find $L_{\cii}=4.2-6.4\times 10^{6}\,L_\odot$, significantly more than observed. Assuming that most \cii\ is emitted from \hii\ gas, we find $L_{\cii}=7.4\times 10^{5}-1.2\times 10^{6}\,L_\odot$. This agrees well with our observations and confirms that the \cii\ emission appears to be dominated by emission from the ionized gas.

%------------------------------------------
% SubSection NII and CII analysis from thermal emission
%------------------------------------------

\subsection{\cii\ and \nii\ estimation from thermal emission}\label{sect:thermal}

\begin{table*}[hbt]
\caption[]{Properties of the ionized gas derived from its thermal emission. Boldfaced numbers are the expected intensities when accounting for the frequency dependent coupling of the compact \hii\ regions to the SOFIA beam.  
}
\label{tab:thermal}
{\centering
\begin{tabular*}{1.0\textwidth}{@{\extracolsep{\fill}} l l l l l c c}
\hline\hline
\noalign{\smallskip}
$\theta $
&$\mathcal{L}$\tablefootmark{a}
&$EM$\tablefootmark{b}
&$\langle n_e \rangle$\tablefootmark{c}
&$\langle N_e \rangle$\tablefootmark{d}
&$\tfrac{\theta^2}{14^2}\,\int T^\mathrm{[CII]}_{158\mu\mathrm{m}}\mathrm{d}v$\,\tablefootmark{e}
&$\tfrac{\theta^2}{18^2}\,\int T^\mathrm{[NII]}_{205\mu\mathrm{m}}\mathrm{d}v$\,\tablefootmark{e}\\
$(\arcsec)$&pc& pc cm$^{-6}$&cm$^{-4}$&cm$^{-2}$&\Kkms&\Kkms\\
             \noalign{\smallskip}
             \hline
             \noalign{\smallskip}
             \centering
 21.3 & 269  &$ 1.7\times 10^4 $& 8   &$ 6.59\times 10^{21} $& 758  &85.8 \\
 18   & 227  &$ 2.4\times 10^4 $& 10  &$ 7.17\times 10^{21} $& 856  & 113 \\
 14   & 177  &$ 3.9\times 10^4 $& 15  &$ 8.13\times 10^{21} $& 1013 & 100 \\
 2    & 25.2 &$ 1.9\times 10^6 $& 280 &$ 2.15\times 10^{22} $& 60.1 & 9.5 \\
 1.7  & 21.4 &$ 2.7\times 10^6 $& 350 &$ 2.33\times 10^{22} $& 47.1 & 7.3 \\
 1.61\tablefootmark{f} & 20.3 & &     &                      & 42.3 & \textbf{6.5} \\
 1.55\tablefootmark{g} & 19.5 & &     &                      & \textbf{39.2} & 6.0 \\
 1    & 12.6 &$ 7.7\times 10^6 $& 780 &$ 3.04\times 10^{22} $& 21.3 & 3.0 \\
 \noalign{\smallskip}
\hline
\end{tabular*}
}
\tablefoottext{a}{Scale length, averaged over spherical volume: $\mathcal{L}=\tfrac{4}{3}\pi\mathcal{R}^3/\pi\mathcal{R}^2$, with
$\mathcal{R}=\tfrac{1}{2}\tfrac{\theta}{3600}\tfrac{\pi}{180}D$ and $D=3.9$~Mpc.}\\
\tablefoottext{b}{Emission meassure, averaged over $\theta^2$: $EM=4.85\times 10^3 S(\mathrm{Jy})_\mathrm{5\,GHz} T_e^{0.35}\tfrac{1}{(\theta(\arcsec)/60)^2}$~pc cm$^{-6}$}\\
\tablefoottext{c}{Electron density, averaged over $\theta^2$: $\langle n_e \rangle=\sqrt{{EM}/{\mathcal{L}}}$~cm$^{-3}$}\\
\tablefoottext{d}{Electron column density, averaged over $\theta^2$: $\langle N_e \rangle=\langle n_e \rangle\times\mathcal{L}=\sqrt{{EM}\times{\mathcal{L}}}$~cm$^{-2}$}\\
\tablefoottext{e}{Integrated \cii\ and \nii\ intensity assuming $T_e=8000$~K and an electron density of $\langle n_e\rangle$ for collisional excitation. The intensity is corrected by the area filling factors $\min(1,\theta^2/14^2)$ and $\min(1,\theta^2/18^2)$ for the \cii\ and \nii\ beams, respectively.}\\
\tablefoottext{f}{Effective, aggregated diameter of the compact \hii\ regions in the \nii\ beam. $\langle n_e \rangle$ and $\langle N_e \rangle$ remain unaffected and were taken from the $\theta=1.7\arcsec$ row.}\\
\tablefoottext{g}{Same as \tablefootmark{(f)} but for the \cii\ beam.}\\
\end{table*}

As a comparison we also derive an upper limit to the expected \cii\ and {\nii} emission based on thermal continuum measurements in this section. From \citet{rabidoux2014} we find that in a 21.3{\arcsec} beam IC~342 has a thermal flux of 15.4~mJy at 33~GHz, corresponding to 19~mJy at 5~GHz. 
Based on \citet{mezger1967} we can calculate the emission measure $EM(5\, \mathrm{GHz})=4.85 \times 10^3 S(\mathrm{Jy})_\mathrm{5\,GHz}\times T_e^{0.35} \times \theta(')^{-2}$~pc~cm$^{-6}$. \citet{meier2011} estimated $T_e=8000$~K.
Assuming that the thermal emission comes from a solid angle $\theta\ll21.3\arcsec$, then $S(\mathrm{Jy})_\mathrm{5\,GHz}=const$ for smaller beams. Depending on the angular extent of this ionized gas we can calculate its properties, such as the scale length $\mathcal{L}$, mean electron density $\langle n_e\rangle$, and its mean electron column density $\langle N_e\rangle$. Table~\ref{tab:thermal} summarizes the results for different \hii\ sizes.

We note that the six \hii\ regions given by \citet{tsai2006} (see also Fig.~\ref{figObs}) have a total projected area of $2.26$~sr, equivalent to an aggregated diameter of 1.7\arcsec, which implies an area filling factor $\phi_a(21.3\arcsec)=1.7^2/21.3^2\approx1/160$ in the 5~GHz beam , $\phi_a(18\arcsec)=1.7^2/18^2\approx1/110$ in the SOFIA 205{\textmu}m beam, and $\phi_a(14\arcsec)=1.7^2/14^2\approx1/70$ in the SOFIA 158{\textmu}m beam\footnote{Propagating the errors on the \hii\ size estimates ($\sim$ 30\%) gives an uncertainty of $\Delta \phi_a/\phi_a=42$\%. The errors on $S(5\mathrm{GHz})$ of $\sim$1/30 result in $\Delta \langle n_e\rangle/\langle n_e\rangle=15$\% and $\Delta \langle N_e\rangle/\langle N_e\rangle\approx 15$\%. Because $\int T dv=const\times \phi_a\times N$, we find $\Delta T_\mathrm{obs}/T_\mathrm{obs}=45$\%.}. This is an upper limit for the total \hii\ region area because of the varying coupling to the beam  depending on the position of the \hii\ regions. Accounting for the beam coupling at the (0\arcsec,0\arcsec) position we find effective areas of 1.89 and 2.02~sr in a 14\arcsec\ and 18\arcsec\ beam, respectively. This corresponds to effective diameters $\theta=1.55\arcsec$ and $\theta=1.61\arcsec$ for \cii\ and for {\nii}, respectively (see Table.~\ref{tab:thermal}). We note that to determine the local physical properties, such as $\langle n_e\rangle$ and $\langle N_e\rangle$, $\theta=1.7\arcsec$ should be used instead of the effective diameters.
We also note that for $\theta=1.7\arcsec$, we find $\langle n_e\rangle\approx350$~cm$^{-3}$, smaller than the value of 700~cm$^{-3}$ found by  \citet{meier2011}.

 We assume in the center of IC~342 a metallicity twice solar $(12+\log(\mathrm{[O]/[H]})_\odot = 8.5)$ and a carbon depletion factor of 0.4. However, in an \hii\ region the gas-phase abundance equals the elemental abundance because all dust is destroyed. We also assume that the elemental abundance of nitrogen scales linearly with the metallicity, even though there is some evidence that [N]/[H] increases more quickly under high-metallicity conditions \citep[see e.g.][]{liang2006}. Thus,  gas-phase = elemental [C]/[H] = $2 \times 3.16 \times 10^{-4}$ \citep{simon-diaz2011}. For nitrogen we assume [N]/[H] = $2 \times 8.32 \times 10^{-5}$ \citep{simon-diaz2011}. The total column density then is $\langle N_e\rangle\times[X]/[H]$ assuming that all carbon and nitrogen are in singly ionized form.

Based on updated electron collision strengths for N$^+$ \citep{tayal2011} \citet{goldsmith2015} analyzed the \nii\ fine structure emission in the Galactic plane. Using their expression for the level population ratios we can derive the optically thin \nii\ emission assuming LTE conditions. For the 205{\textmu}m line we find
\begin{equation}
\int T^\mathrm{[CII]}_{205\mu\mathrm{m}}\mathrm{d}v=2.1\times 10^{-16}  N(\mathrm{N}^+)\,\,\mathrm{K\,km\,s^{-1}}
\end{equation}
with the total N$^+$ column density $N(\mathrm{N}^+)$ and assuming $T_e=8000$~K and taking $n_e=350$~cm$^{-3}$ from Table~\ref{tab:thermal} (see the Appendix~\ref{niiemission} for details and Table~\ref{tabTmb} for different values of $n_e$ and $T_e$).
Table~\ref{tab:thermal} also gives the expected intensities for different sizes $\theta$. 
Coupling the spatial distribution of \hii\ regions to the \nii\ beam gives $\theta=1.61\arcsec$ 
leading to $\int T^\mathrm{[NII]}_{205\mu\mathrm{m}}\mathrm{d}v=6.5$~\Kkms, significantly lower than the observed value of 9.4~\Kkms. However, here we assumed [N]/[O]$\propto$[O]/[H]. If the elemental nitrogen abundance scales super-linearly, then the N$^+$ column density and intensity is enhanced accordingly and could, at least partially, explain the discrepency. 

Again assuming $T=8000$~K and $n=350$~cm$^{-3}$  and a collisional de-excitation coefficient with electrons at 8000~K of $4.9~\times10^{-7}$ \citep{wilson2002}, we find $\int T^\mathrm{[CII]}_{158\mu\mathrm{m}}\mathrm{d}v=2.17 \times 10^{-16}\times N(\mathrm{C}^+)$~{\Kkms}  using Eq.~1 from \citet{pineda2013}. Accounting for the effective extent of the \hii\ regions in the \cii\ beam due to coupling to a Gaussian beam gives $\theta=1.55\arcsec$. Then the net result is the predicted contribution to the \cii\ emission from the \hii\ regions $\int T^\mathrm{[CII]}_{158\mu\mathrm{m}}\mathrm{d}v=39.2$~\Kkms. Comparison with the observed value at the central position of 62.9~{\Kkms} then shows that a fraction of about 62\% of the \cii\ intensity is contributed by the compact \hii\ regions consistent with our earlier estimates (see Table~\ref{tab:2}).

The assumption that all the thermal emission comes from the compact \hii\ regions might be wrong. A fraction $a<1$ of $S(\mathrm{Jy})_\mathrm{5 GHz}$ could be contributed by large-scale, extended $(\phi_a=1)$ ionized gas. By adding this second component we can estimate how much it would contribute to the fine-structure emission for a given value of $a$. We find that $a$ cannot exceed the percent level. Otherwise such an extended contribution to the \cii\ and \nii\ emission would be much too high because it would not suffer from any area filling effect.  Therefore we do not expect the extended ionized gas to contribute significantly to the thermal radio emission, but it might still contribute to the fine-structure emission. However, when doing the same analysis for the other positions, we find that an additional component, possibly extended and clumpy, is required to explain the observed \nii\ intensities because of the even weaker coupling of the compact \hii\ emission to the beam at the off-center positions. 

Another uncertainty is the distance to IC~342. Assuming a smaller distance would lower our estimates for the \cii\ and \nii\ intensities because the same angular extent would correspond to lower values of $\mathcal{L}$ and therefore to higher $\langle n_e\rangle$ but lower $\langle N_e\rangle$. 
Summarizing, we find a high fraction of the \cii\ emission expected from the compact \hii\ regions in the center of IC~342 based on its thermal emission.
%------------------------------
% Kinematic correlation section
%------------------------------
\subsection{Kinematic {\cii}-\nii\ correlation\label{sect:kin}}
The \cii\ /\twco\ (1-0) line ratio and the {\cii}-\nii\ correlation in Eq.~\ref{eq:1} are both based on integrated line intensities, discarding any additional kinematic information, but \cii\ emission originating from the \hii\ region should carry the same kinematic signature as the pure \hii\ tracer, the \nii\ line. The SOFIA/GREAT data has sufficient spectral resolution and signal-to-noise ratio to allow a more detailed analysis of the {\cii}-\nii\ correlation.

The \nii\ emission shows different peak velocities and FWHM linewidths. We quantify the \cii\ emission coming from the \hii\ region using the following approach: \textbf{1)} Assuming that Eq.~\ref{eq:1} correctly predicts the amount of \cii\ being emitted from the ionized gas, we simulate a \cii\ spectral line assuming a Gaussian with line center velocity and FWHM line width taken from the Gaussian fit of the corresponding \nii\ spectrum\footnote{We did not scale the \nii\ spectrum directly to avoid the effects of noise amplification.}  (Table~\ref{tab:1}) and with an integrated line intensity corresponding to $I(\cii)_\mathrm{H^+}$ from Table~\ref{tab:3}. This simulated \cii\ line is  subtracted from our observed \cii\ for each velocity channel and gives the residual \cii\ intensity, cleaned of \hii\ contributions.\footnote{The baseline RMS is conserved during subtraction.} \textbf{2)} A Gaussian is fitted to the residual \cii\ line.  The line parameters are then correlated to the \twco\ (1-0) line parameters.

Figure~\ref{figKinCorr} compares the observed data from the center position (0\arcsec,0\arcsec) with the simulated \cii\ data.  The  line shapes of the modeled \cii\ and the residual \cii\ lines are significantly different. The Gaussian line parameters for the {\cii\}$_\mathrm{res}$ are $T_\mathrm{pk}=0.46\pm 0.02$~K, $v_0=29.9\pm 1.2$~{\kms}, and $\sigma_\mathrm{FWHM}=54.3\pm 2.9$~{\kms}. The line shape and position is close to the CO and C line shapes at this position (compare with Table~\ref{tab:1}). 
\begin{figure}
\centering
\resizebox{\hsize}{!}{\includegraphics{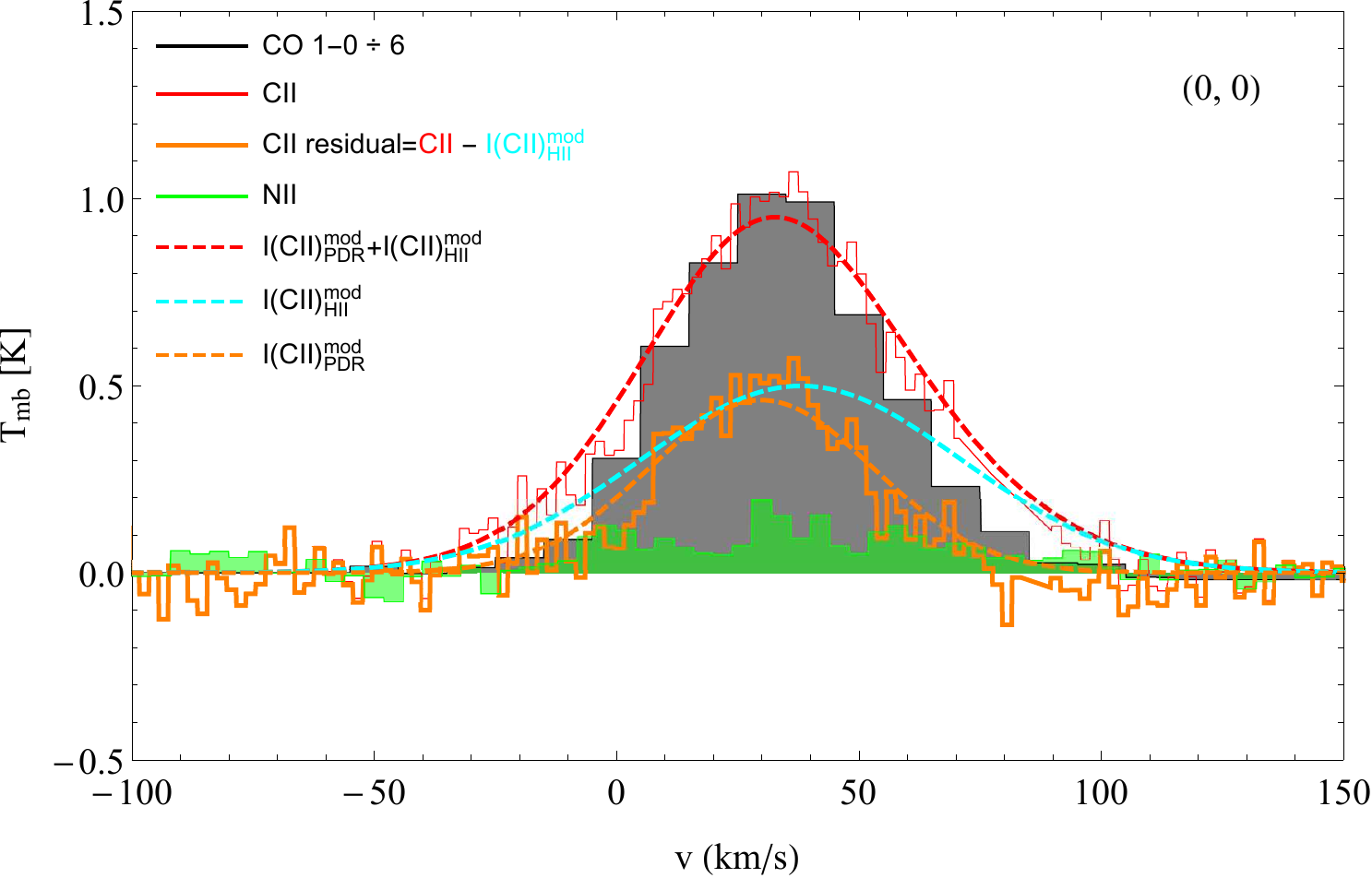}}
\caption{Comparison of the observed \cii\ (red), \nii\ (green), and \twco\ (1-0) (black, gray filling, suppressed by a factor 6) lines at the center position (0\arcsec,0\arcsec) together with the simulated \cii\ spectrum (cyan) derived from the \nii\ line and the residual \cii\ line (orange) corresponding to ``pure'' PDR emission. The dashed lines give the respective Gaussian lines.}
\label{figKinCorr}
\end{figure}

We performed the above analysis for all ten positions. Figure~\ref{figRes} shows the spatial distribution of the derived line properties of the residual \cii\ line. The left panel shows the integrated line intensity of {\cii}$_\mathrm{res}$. The intensities are stronger to the SE than to the NW and we find no spatial correlation to the \twco\ (1-0) data. We note that scaling up the \nii\ emission also increases the error of $I(\cii)_\mathrm{H^+ }$ and consequently also of  {\cii}$_\mathrm{res}$. It is also possible that the {\cii}-\nii\ correlation from Eq.~\ref{eq:1} varies spatially.  We attribute the remaining \cii\ emission in the southeast to pick up from PDRs in the ring regions within the beam. 

The three diagonally hatched positions in Fig.~\ref{figRes} show no  residual \cii\ emission, i.e. \cii\ emitted from PDRs. For the NW position the most likely reason is that the simple scenario of an \hii\ region neighboring a PDR is probably not applicable. The majority of the \nii\ observed there can be attributed to the expanding lobe of ionized gas that is not associated with a transition to a PDR/GMC. Hence, we do not expect strong {\cii}$_\mathrm{res}$ emission. For the other two positions the explanation are less obvious. At the NE position we find the lowest signal-to-noise ratio for the {\nii}, i.e. the largest error in the computation of the {\cii}$_\mathrm{res}$. Nevertheless, the \nii\ intensity is surprisingly strong compared to the total \cii\ intensity. Obviously, we observe strong \nii\ emission from gas that is not spatially associated with PDR gas, in contrast to the standard scenario of an \hii\ region situated in close proximity to molecular clouds. Its origin is unclear. Perhaps it is a strong contribution from the inter-arm gas east of the mini-spiral or from the expanding southeastern \hii\ lobe. The (-7\arcsec,7\arcsec) position is similar in that it shows strong \nii\ emission with a much broader linewidth than that of {\cii}. We cannot find these additional kinematic components in the (-7\arcsec,0\arcsec) position, which would be case if the northwestern \hii\ lobe was contributing to the \nii\ line profile. The origin of this additional kinematic component is unclear.

The middle panel shows the velocity shift between the {\cii}$_\mathrm{res}$ and \twco\ (1-0) line center velocities. All residual spectra are redshifted with the exception of the (0\arcsec,0\arcsec) position, which shows a blue-shift of $\sim 3$~{\kms}. In the right panel we compare the line width of the residual \cii\ and the CO line. We note a significant trend in the spatial distribution. The {\cii}$_\mathrm{res}$ FWHM linewidth related to northern arm is much narrower than the southern arm. In the south, the simulated \cii\ residua are 30-40\% wider than \twco\ (1-0), while they are of comparable line width in the north. We conclude that, assuming Eq.~\ref{eq:1} is valid, the kinematic correlation of the residual \cii\ emission is different between the northern and southern arms of the mini-spiral. In the norther arm we find comparable line widths, indicating that the \cii\ emission is produced in the hot PDR gas layer around the GMCs in the arm. In the southern arm, the much wider lines of the residual \cii\ indicates a much stronger contribution from the diffuse gas between the GMC and from the inter-arm regions.
It is difficult to form a consistent picture from the analysis of the simulated \cii\ to \nii\ properties. The SE quadrant shows the strongest residual \cii\ emission, consistent with the highest values of {\cii}/\twco\ (1-0) indicating the influence of intense FUV radiation from massive stars. Overall, the {\cii}/\twco\ (1-0) shows a spatial correlation with the residual {\cii}. We should add that the broader the \nii\ lines are compared to {\cii} the more unreliable this method becomes because subtracting the scaled {\cii}$_\mathrm{th}$ will result in negative features in the wings of the {\cii}$_\mathrm{res}$ profiles. In other words, \nii\ emission is observed that is kinematically not associated with \cii\ and thus in violation of the assumptions underlying to Eq.~\ref{eq:1}, namely an \hii\ region transitioning into a PDR and then a molecular cloud along one dimension \citep{abel06b,ferland2013}.

\begin{figure}
\centering
\resizebox{\hsize}{!}{\includegraphics{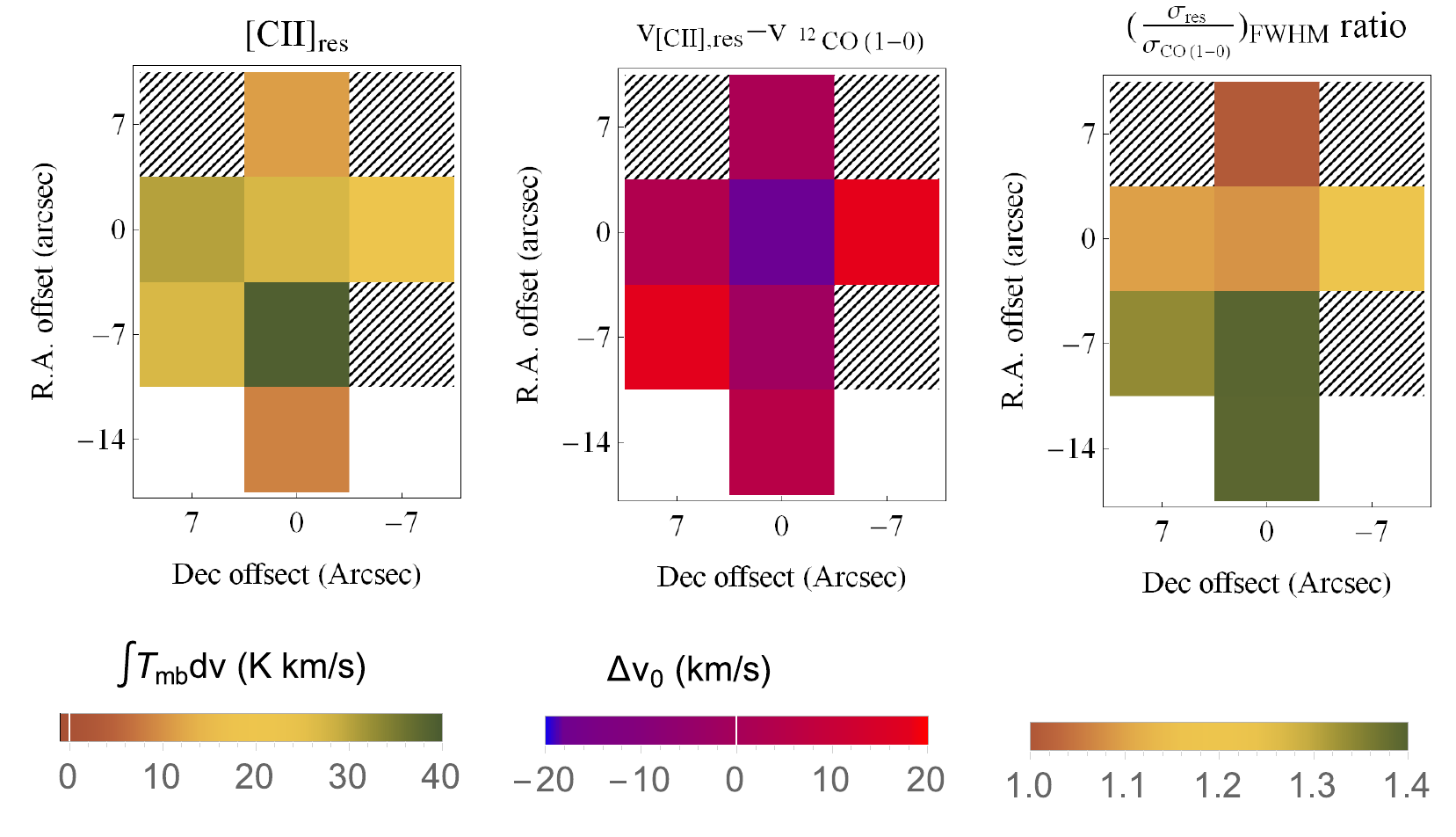}}
\caption{Spatial distribution of the properties of residual \cii\ line as described in Sect.~\ref{sect:kin}: integrated line intensities $\int T_\mathrm{mb,res}dv$ (left), line center shift with respect to \twco\ (1-0) (middle), and line width ratio between the residual \cii\ and \twco\ (1-0)  (right). The positions with no residual \cii\ are diagonally hatched. The baseline RMS of the residual \cii\ line is between 42 and 98~mK. The error of the line intensity integrated over $\Delta v$ is computed as $T_\mathrm{RMS} \Delta v/\sqrt{N}$ with the number of frequency channels $N$. With channel widths of 2~{\Kkms}, $N=15$ for a line width of 30~{\kms} and the total error is 0.3-0.8~{\Kkms}. The spatial resolution is 14{\arcsec} and 18{\arcsec} for \cii\ and {\nii}, respectively.}
\label{figRes}
\end{figure}       

\subsection{Super-resolution \cii\ composition \label{sect:super}}
In the sections above we took the {\cii}-\nii\ correlation as given and used it to estimate the fraction of \cii\ emission coming from the ionized and the PDR gas. In Sect.~\ref{sect:kin} we additionally made use of the spectrally resolved line shapes of \cii\ and {\nii}. In the following we describe an alternative technique to derive additional conclusions based on the kinematic information at hand. In addition to the high spectral resolution of the SOFIA data, we also have a data set with high angular resolution, the \twco\ (1-0) data with a resolution of 5.5$\arcsec$. If there is a correlation between C$^+$ and CO, as discussed above and as shown by \citet[e.g.][]{stacey1991}, then this correlation should also be visible in the kinematic signature of the \cii\ lines. The spectral line profile of \cii\ emitted at a certain position should therefore be the result of the Gaussian  smoothing of the unresolved structures within the beam. 
We propose the following algorithm:
 
\begin{enumerate}
\item  We use the \twco\ (1-0) data with a higher angular resolution and a higher spatial sampling to simulate an artificial, high-resolution set of {\cii}$_i^{hi}$ spectra by constructing a Gaussian with centroid velocity and FWHM line width of the CO spectrum and a peak intensity $f_i$ for each position $i$. In other words,  $f_i$ is the peak intensity of {\cii}$_i^{hi}$ at the position $i$.
\item The artificial, high-resolution spectra {\cii}$_i^{hi}$ are convolved with a Gaussian beam corresponding to the SOFIA \cii\ resolution to create an artificial low-resolution spectrum {\cii}$^{lo}$. 
\item The convolved {\cii}$^{lo}$ spectrum is subtracted from the observed {\cii}$_\mathrm{obs}$. The resulting residuum is minimized by varying the  peak intensities $f_i$, using a simulated annealing algorithm.
\item The result of the nonlinear model fit is a set of peak intensities $f_i$ specifying the spatial variations of the strength of the {\cii} emission, and therefore also the {\cii}-CO relationship, in a sub-beam resolution. 
\end{enumerate}      
\begin{figure*}
\centering
\includegraphics[width=17cm]{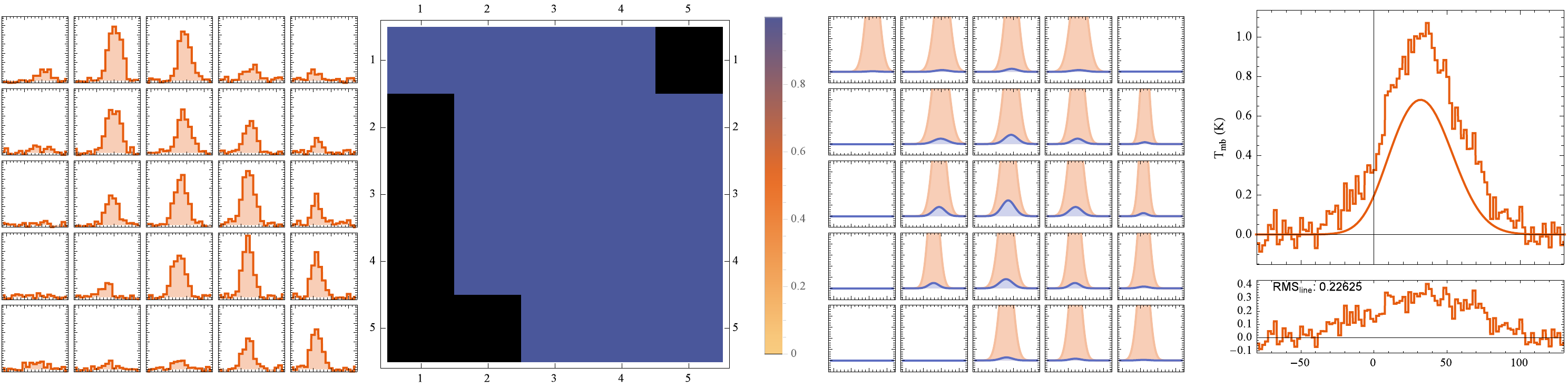}
\caption{Result of the super-resolution line composition as described in Sect.~\ref{sect:super} for the central \cii\ spectrum (at (0\arcsec,0\arcsec) offset).  The array on the left shows the high-resolution Berkeley Illinois Maryland Association (BIMA) \twco\ (1-0) data with a 5{\arcsec} spatial sampling. The \cii\ beam FWHM covers the central 3x3 pixel. The second array from the left shows the spatial distribution of the peak intensities $f_i$.
Darker pixels contribute more to the overall emission. The black pixels show masked positions that are not used.  The third panel shows the array of artificial {\cii}$_i^{hi}$ spectra in light red, and already weighted with the Gaussian weights applied during beam convolution in blue. The final beam convolved spectrum is a simple sum of all blue spectra in this panel. The plot on the right compares the final {\cii}$^{lo}$ spectrum (line) with the observed {\cii}$_\mathrm{obs}$. The residuum is shown below the spectrum. The peak intensities is limited to $f_i\le1$. The line RMS is 199~mK.}
\label{figSuper1}
\end{figure*}
\begin{figure*}
\centering
\includegraphics[width=17cm]{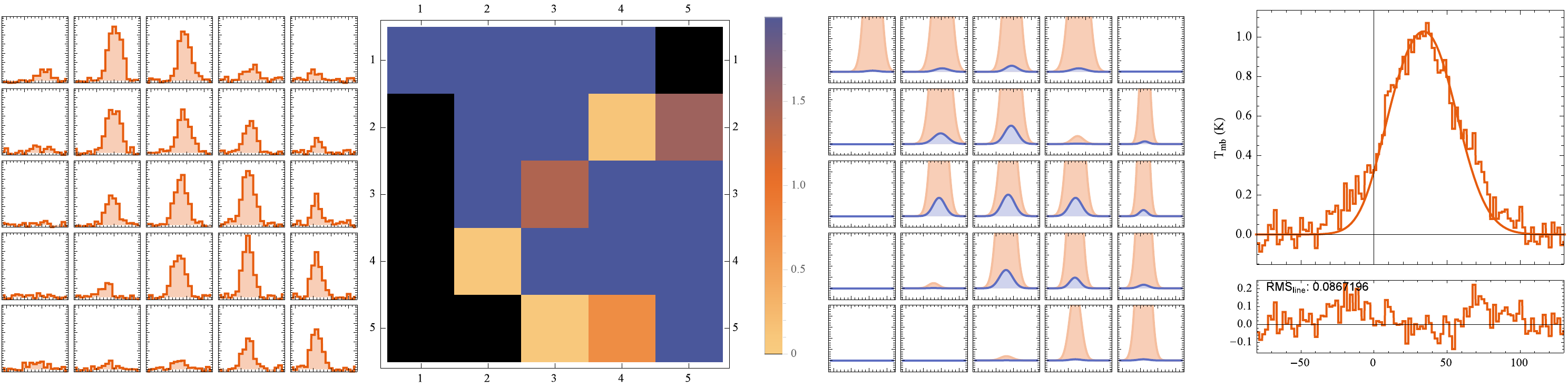}
\caption{Same as Fig.~\ref{figSuper1}, but assuming $f_i\le2$. The line RMS is 84~mK.}
\label{figSuper2}
\end{figure*}
\begin{figure*}
\centering
\includegraphics[width=17cm]{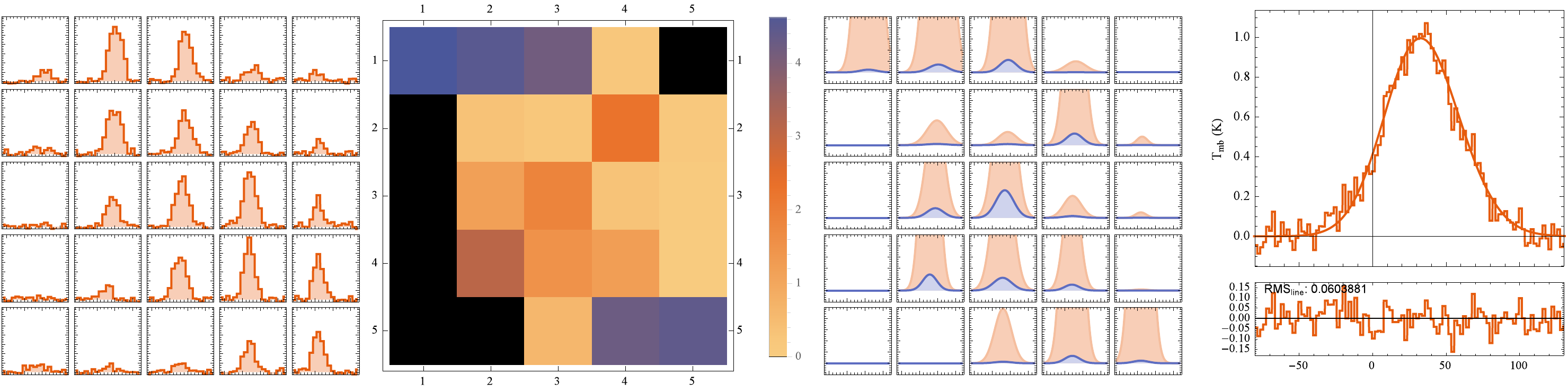}
\caption{Same as Fig.~\ref{figSuper1}, but assuming $f_i\le2$ and also allowing the fitting to adapt the FWHM line width of the {\cii}$_i^{hi}$ spectra. The {\cii}$_i^{hi}$ spectra are 37\% wider than their \twco\ (1-0) counterparts. The line RMS is 60~mK.}
\label{figSuper3}
\end{figure*}
\begin{figure*}
\centering
\includegraphics[width=17cm]{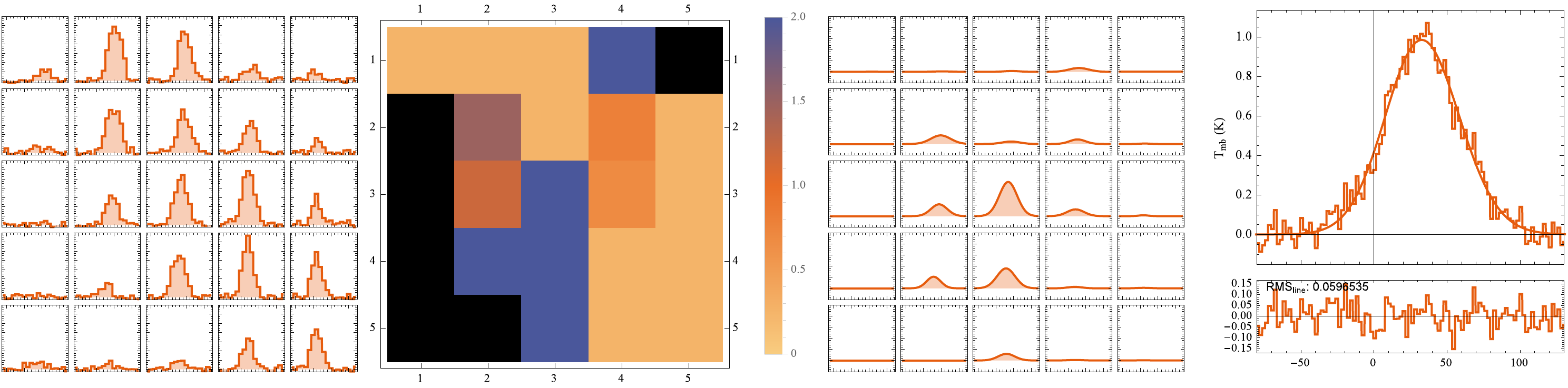}
\caption{Same as Fig.~\ref{figSuper3}, but assuming $f_i\le5$. The {\cii}$_i^{hi}$ spectra are 40\% wider than their \twco\ (1-0) counterparts. The line RMS is 59~mK.}
\label{figSuper4}
\end{figure*}

Figure~\ref{figSuper1} shows the work-flow of the above algorithm together with the final result. The starting point is the high-resolution \twco\ (1-0) data. We chose a set of 25 positions centered on our (0\arcsec,0\arcsec) position sampled every 5\arcsec. The spectra are shown in the left panel in Fig.~\ref{figSuper1}. The central bend of the mini-spiral is visible. We also note a velocity gradient from the NE to the SW. The success of our suggested super-resolution method critically depends on the velocity gradient across the 5x5 map not being too small. The second panel in Fig.~\ref{figSuper1} shows the final result of the numerical fitting of the 25 $f_i$. In this example case, we limited $\max(f_i)\le1$. Additionally, to only focus on the apparent spiral-structure and to remove CO spectra with signal-to-noise ratios that are too low, we applied a zero weighting factor to the spectra at the SE and NW edge of the map (diagonally hatched pixel in the array). We note that all pixels require a maximum value of $f_i=1$, which indicates that we did not reach a real minimum in the fitting. The third panel shows the artificial grid of {\cii}$_i^{hi}$ spectra, basically the Gaussian version of the \twco\ (1-0) spectra with peak intensities $f_i$ (light red spectra) and additionally already multiplied with the weights applied during the Gaussian beam convolution as blue spectra (FWHM=15\arcsec). This way it is possible to directly compare the relative spatial contributions to the final spectrum {\cii}$^{lo}$. The final panel on the right compares the observed {\cii}$_\mathrm{obs}$ spectrum at position (0\arcsec,0\arcsec) with the result from the super-resolution simulation {\cii}$^{lo}$ (line). Below the spectrum we give the residuum {\cii}$_\mathrm{obs}-{\cii}^{lo}$. The residuum RMS across the line is given in the panel.
First of all, we note that the resulting {\cii}$^{lo}$ fails to reproduce the observed line. This is a direct result of limiting $f_i\le1$. The algorithm is not able to reduce the RMS any further, even by including the maximum allowed emission. This highlights the strong dependence of the method on the details of the applied fitting method.

Figure~\ref{figSuper2} shows the result of the fit, if we allow a peak intensity limit $f_i\le2$. This maximum peak intensity is employed over large fraction of the beam area as visible from the second panel. We note that a diagonal stripe (NW to SE) of the ring contributes only marginally. The remainder of the ring and the regions where ring and arms join are the major contributors to the composite {\cii}$^{lo}$ spectrum. The right panel shows that the spectral shape of the {\cii}$^\mathrm{obs}$ is roughly met, but the setup fails to reproduce kinematic features in the red and blue wings, also visible from the non-negligible structures in the residuum plotted below.  

In Fig.~\ref{figSuper3} we set again the peak intensity limit to $f_i\le2$. Additionally, we assume that the {\cii}$_i^{hi}$  lines are wider than the \twco\ (1-0)  by a fixed factor that is assumed to be the same for all positions $i$ and is determined by the numerical fitting to be 1.37, i.e. the  {\cii}$_i^{hi}$ spectra are wider than CO by 37\% consistent with the fact that \cii\ is often observed to have wider lines than CO because it traces warmer, more turbulent material at cloud/clump edges. The second panel now shows significant variations in the spatial contributions  to the \cii\ emission. The arms of the mini-spiral have very little kinematic impact on the central \cii\ spectrum. The ring is dominantly contributing, and here it is mainly the lower left quadrant that is dominant. We also note a contribution from the more diffuse material off the arms to the S and to the NW. The composite {\cii}$^{lo}$ spectrum in the panel to the right matches the observed {\cii}$^\mathrm{obs}$ very well. The residuum does not show significant structure.

The same behavior is confirmed even in cases where we relax the peak intensity limit of the {\cii}$_i^{hi}$ even further. In Fig.~\ref{figSuper4} we show the results for  $f_i\le5$. Again the agreement between observed and simulated \cii\ is very good. The major difference to the case  $f_i\le2$ (Fig.~\ref{figSuper3}) is that the central spectrum at the position of the nuclear cluster is now contributing much less to the overall emission. This is compensated by emission pick-up from the more distant spectra to the north (with a much lower Gaussian weight). Again the residuum plot confirms that no major kinematic component remains unaccounted for.

The strong influence of the numerical details of this method makes it difficult to draw quantitative conclusion. The velocity gradient across our 5x5 field is not strong enough to prevent kinematic degeneracies to occur. Spectral shapes from the center can be replaced by spectra from other positions given a sufficiently large $f_{i,max}$. To make sure that the influence of these degeneracies is not influencing our conclusions, we performed a  series of simulations with progressively increasing peak intensity limits. The qualitative behavior was always the same: The kinematic structure of the ionized carbon on the center position (0\arcsec,0\arcsec) visible in its spectral shape is consistent with a scenario where the largest contribution to the {\cii}$^\mathrm{obs}$ with strong emission from beam-unresolved structures (PDRs/\hii) comes from the SE part of the ring complemented by additional but weaker PDR emission from clouds along the rest of the ring and possibly parts of arms. This includes emission coming from the ionized gas expanding  out of the nucleus. This picture is in agreement with what had been found earlier by other authors \citep[e.g.][and references therein]{meier2005}. 

This demonstrates that the suggested kinematic super-resolution method is working. So far we have constrained ourselves to modeling only the central position. In the following section, we will apply the super-resolution analysis to a more interesting position in IC~342.

\subsubsection{Super-resolution analysis of the southeast lobe.}

\begin{figure*}
\centering
\includegraphics[width=\linewidth]{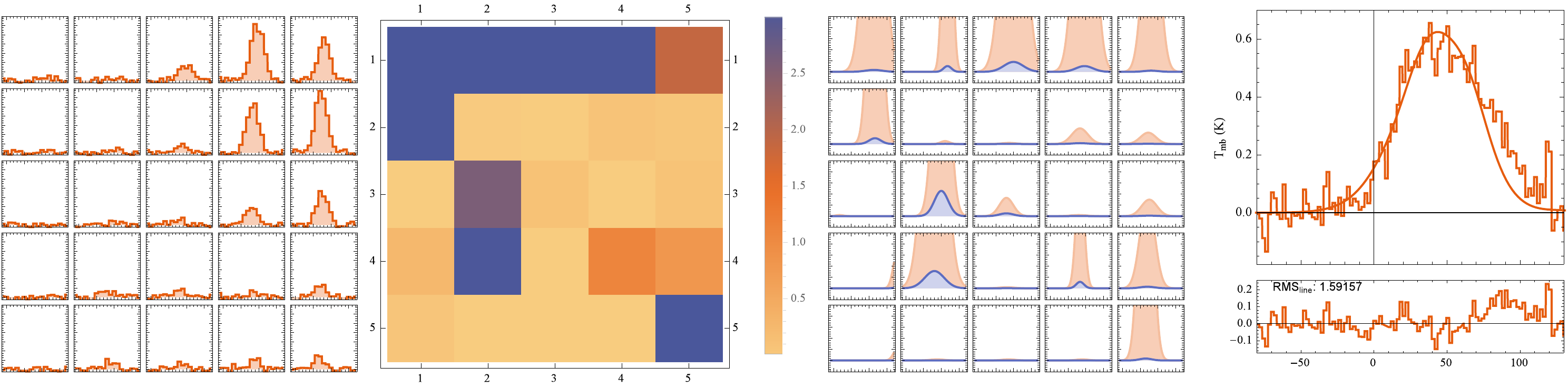}
\vfill
\includegraphics[width=\linewidth]{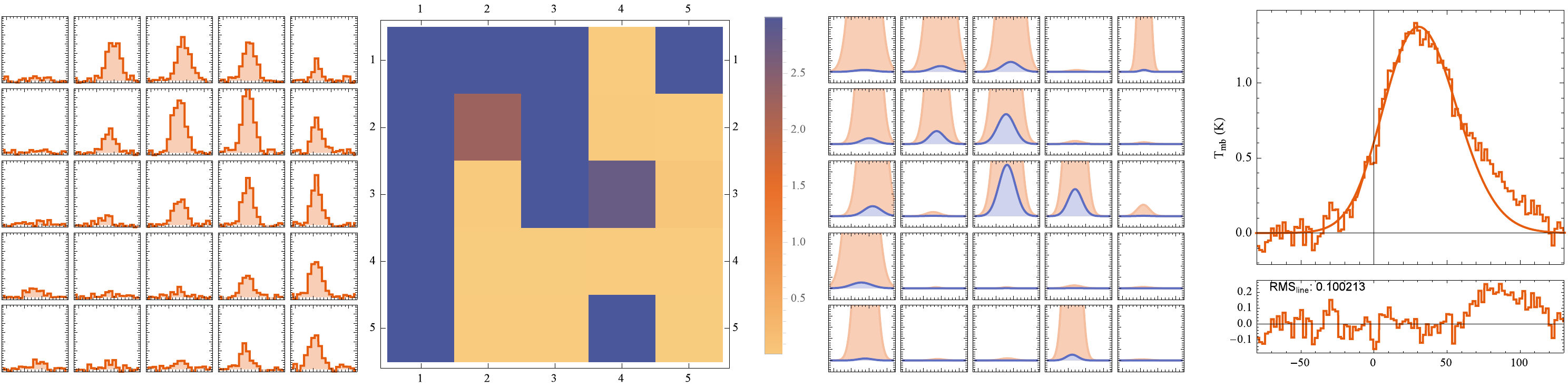}
\caption{Same as Fig.~\ref{figSuper3} assuming $f_i\le3$. The {\cii}$_i^{hi}$ spectra are 30\% wider than their \twco\ (1-0) counterparts. \textbf{Top panel)} position  (7\arcsec,-7\arcsec). \textbf{Bottom panel)} position  (0\arcsec,-7\arcsec).}
\label{figSuper7-7}
\end{figure*}

In the previous section we presented the super-resolution analysis and applied it to our center position. The results show that basically all the observed \cii\ emission at the center position can be explained as CO-correlated \cii . The residuum shows very little structure. There might be a marginal blueshifted component visible in Fig.~\ref{figSuper4}. To further demonstrate the usefulness of the approach we  apply the technique to  the SE lobe of ionized gas.

Figure~\ref{figSuper7-7} shows the results of the decomposition at (7\arcsec,-7\arcsec) (top panel) and (0\arcsec,-7\arcsec) (bottom panel). We notice several points: the comparison between composed and observed \cii\ line profiles shows a significant redshifted residuum that cannot be explained by any \cii\ emission related to \twco (1-0). This is visible at both positions. Furthermore, at both positions we need to assume factors $f_i=3$ at many unresolved positions in order to account for the observed intensity. This is a relatively high ratio given that $f_i=1$ in temperature units is equivalent to a ratio \cii\ /\twco (1-0)=4509 in energy units, already indicating strong starburst activity \citep{stacey1991}. However, this line ratio strongly depends on the different beam filling ratio of \cii\ /\twco. With increasing angular resolution and remaining larger beam filling of \cii\ compared to cold \twco\ we expect larger values of $f_i$.

Comparing the top and bottom panels of Fig.~\ref{figSuper7-7} we note an inconsistency resulting from the independent fitting. The 7\arcsec position offset between the two is equivalent to a 1 1/2 grid-shift of the underlying high-resolution data. Consequently we would expect to see a similar {\cii}$_i^{hi}$ pattern in the bottom panel but shifted 1 1/2 boxes with respect to the top panel. However, the strong {\cii}$_i^{hi}$ contribution at the central position in the bottom panel has no corresponding counterpart in the top panel. One reason might be the different line strengths and signal-to-noise ratio at both positions. The super-resolution in the top panel is applied to much noisier data and relies on weak CO emission in its convolution. The residual RMS is more than a factor of 10 higher than the analysis at position (0\arcsec,7\arcsec) in the bottom panel. A possible strategy to improve this behavior could be a two-step coupling between the two positions, starting with the higher quality data and using the results as a starting point in the analysis of the positions with lower signal-to-noise ratio, but this is beyond the scope of this paper. Nevertheless we also note qualitative consistencies between the {\cii}$_i^{hi}$ in both panels, i.e. {\cii}$_i^{hi}$ patterns shifted by 1-2 boxes\footnote{We chose a 5\arcsec spacing for the {\cii}$_i^{hi}$  (and CO) data because this is the spatial resolution of the BIMA CO data. Since the \cii\ data is spaced 7\arcsec apart, this means that the 5\arcsec spaced {\cii}$_i^{hi}$ grid of two neighboring \cii\ positions does not overlap, which makes a quantitative comparison between neighbor pixels difficult.}.

 In  Appendix \ref{AppSR} we applied the same analysis to all other positions. We do not discuss all the positions in detail, but would like to point out that our super-resolution approach is able to reproduce the observed \cii\ line profiles with the exception of the contribution from the off-plane emission of the expanding lobes of ionized gas at positions (7\arcsec,-7\arcsec) and (0\arcsec,-7\arcsec), as shown in Fig.~\ref{figAppendixSR2}. This is the expected behavior because of the underlying assumption of a kinematic correlation between \twco (1-0) and {\cii}.
 
 It is also possible to simultaneously compose all ten {\cii}$^\mathrm{obs}$ spectra with a single underlying field of \twco\ (1-0) spectra and peak intensities $f_i$. Since the \cii\ spectra are spatially almost fully sampled this would add additional constraints to the fitted values of $f_i$ and somewhat reduce the kinematic degeneracies. However, this would also increase the dimensionality of the numeric fit significantly. The same would be true if we allow the \cii\ /CO line width ratio to vary from position to position. This analysis is beyond the scope of this paper. A spatial super-resolution analysis requires CII (and CO) data on more extended scales, with heterodyne resolution. With the upcoming 14-pixel upGREAT receiver this will become possible within reasonable observing time requests. 

The super-resolution analysis, applied to all ten observed positions, shows that only a small fraction of the observed \cii\ emission cannot be correlated kinematically to the molecular gas. At first this seems to be in conflict with our previous result of the \hii\ regions contributing significantly to the overall \cii\ emission. However, this contradiction is due to the design of the method, which makes the assumption that all \cii\ is kinematically related to the CO emission. The algorithm searches CO-to-\cii\ scaling factors such that the \cii\ residuum is minimized,
but it cannot distinguish between emission from PDRs and from their nearby \hii\ regions as long as they share the same kinematic signature.  Therefore, by choosing high-resolution molecular emission data, we make the algorithm insensitive to any emission from ionized gas that is not related to a nearby PDR and force it to attribute as much \cii\ emission as possible to PDRs. 
\citet{israel1978} presented a blister model, where the ionized gas in the \hii\ regions is streaming away from its associated molecular cloud resulting in velocity differences of $\pm 10-12$~{\kms} times $\cos(i)$, where $i$ is the streaming angle. Because $i$ is random, the average velocity difference for a number of \hii\ regions/molecular clouds comes out close to zero. 
Even a high spectral resolution does not guarantee resolving the degeneracy in attributing the \cii\ to either PDRs or \hii\ regions.
If we had high-resolution (spatially and spectroscopically) data tracing the ionized gas we could turn the analysis around, but we would still face the same problem that the assumedly uncorrelated gas will be neglected.
One way to resolve this limitation is to clean the lower-resolution \cii\ emission from any contribution by the expanding lobes of ionized gas as described in Sect.~\ref{sect:kin} and perform the super-resolution analysis on the residual \cii\ emission. This may be problematic owing to the higher noise of the \nii\ lines and the different center velocities, which might introduce negative features in the residual \cii\ lines. Another possibility is to limit the CO-to-\cii\ scaling factors $f_i$ to lower values as shown in Fig.~\ref{figSuper1}. This limits the capability of the algorithm to fully explain the \cii\ line by correlation to the CO gas. Here the problem is a reasonable choice of $f_{i,max}$ and the fact that the {\cii}/CO ratio depends on the respective beam filling factors (see Sect.~\ref{cii2co}).

\section{Conclusions}
We observed spectroscopically resolved data of the nucleus of IC~342. Using the GREAT receiver on SOFIA we mapped the \cii\ 158{\textmu}m and  \nii\ 205{\textmu}m emission lines in a small map centered on the nuclear star cluster. In the following analysis we demonstrate how the high angular and spectral resolution of the data leads us to revise some earlier conclusions on the general understanding of IC 342's central structure.  

Comparing the line center velocities of the fine-structure lines with the \twco\ (1-0) data we find additional kinematic components not visible in the molecular gas and most likely tracing the ionized material. We attribute this kinematic signature to the ionized gas surrounding the nuclear cluster and expanding in two lobes out of the plane of the mini-spiral/molecular ring. The Doppler-shift of the two lobes makes it necessary to modify the established geometric image of the center of IC~342 and we present two possible alternatives that are kinematically consistent with the observations. 

We examine the ratio of \cii\ over \twco\ (1-0) and find a value of $I(\cii)/I(\twco\ (1-0))\sim800$ when using the averaged intensities. Spatially, we find significant variations of the ratio between 400 and 1800 indicating a mixture of quiescent and more active star forming conditions along the arms and the molecular ring. We find the highest ratio, most likely indicative of the strongest star formation activity, in the southwestern quadrant of our small map.

Assuming a theoretically predicted  and observationally confirmed correlation between the 205{\textmu}m line and the amount of \cii\ being emitted by corresponding \hii\ region \citep{heiles1994,abel06b}, we find significant spatial variations in how much the ionized gas in IC~342 contributes to the total line intensities. Averaged over the central few hundred parsec we find a {\hii}-to-PDR ratio of 70:30. The spatial distribution of this ratio can deviate significantly from this value. We find that the northern edge of our small map is mostly dominated by \hii\ contributions. For the remainder of the map, the fraction of \cii\ emission from PDRs varies between 30 and 65\%.     

We present various methods for estimating the amount of \cii\ emission coming from the ionized gas and from PDRs. The first method uses a theoretically predicted {\nii}-{\cii} correlation to simulate the \cii\ spectrum coming from the \hii\ region for each positions. These predicted spectra are subtracted from the observed \cii\ spectra and the residual emission is examined. Similar to the results of the analysis of the line-integrated intensities, we find a residual \cii\ emission, i.e. \cii\ coming from PDRs between 24\% and 58\% if both phases contribute. Three observed positions show no residual \cii\ emission, suggesting that the emission is dominated by \hii\ regions exclusively. We also find a significant trend in the linewidths of the residual \cii\ emission. The lines in the northern part of our map show linewidths similar to the \twco\ (1-0) lines while the lines in the southern half are 30-40\% wider than CO. This is consistent with the scenario of  much more active star formation in the south and southeastern part of the map leading to a more clumpy and turbulent composition of the PDRs. Using information on the the thermal emission of the embedded \hii\ regions we model the optically thin emission of \cii\ and \nii\ under LTE conditions and find that the known compact \hii\ regions at the center of IC~342 account for about 2/3 of the observed intensities, which is similar to our other results.

We also present a more complex method for assembling the observed \cii\ line profiles from simulated, unresolved structures within the beam. This super-resolution technique is combined with a numerical fitting scheme to find the unresolved kinematic structure of the gas that explains the observed line profiles the best. Our findings are consistent with the other results presented in this paper and with earlier findings of other authors. The southwestern quadrant of the ring/arm is the dominant contributor kinematically. The emission of C$^+$ gas in both arms and the rest of the ring is not important in order to explain the \cii\ emission observed toward the nucleus of IC~342.

\acknowledgements{

We thank the SOFIA engineering and operations teams, whose tireless  
support and excellent teamwork was essential for the GREAT  
accomplishments during Early Science, and say Herzlichen Dank to the
DSI telescope engineering team.

Based in part on observations made with the NASA/DLR Stratospheric Observatory for Infrared Astronomy. SOFIA Science Mission Operations are conducted jointly by the Universities Space Research Association, Inc., under NASA contract NAS2-97001, and the Deutsches SOFIA Institut under DLR contract 50 OK 0901.  

The research presented here was supported by the {\it Deutsche Forschungsgemeinschaft, DFG} through project number SFB956C.
}

\bibliographystyle{aa}
\setlength{\bibsep}{-2.1pt}
\bibliography{ic342}

\appendix
\section{Data summary}
\begin{table}[ht]
\caption{Summary of angular resolution and spatial sampling of the complementary data. } % title of Table
\label{tabResolution} % is used to refer this table in the text
\centering % used for centering table
\begin{tabular}{l l l l} % centered columns (4 columns)
\hline\hline % inserts double horizontal lines
line& frequency &  orig. resol.& grid \\ % table heading
&GHz& (\arcsec)& (\arcsec)\\
\hline % inserts single horizontal line
\twco\ (1-0) & 115 & 5.5  & 1 \\
\thco\ (2-1) & 220 & 23  & 15 \\ 
\twco\ (2-1) & 230 & 22  & 15 \\ 
\thco\ (3-2) & 330 & 15  & 7.5 \\ 
\twco\ (3-2) & 330 & 15  & 7.5 \\ 
\twco\ (4-3) & 461 & 11  & 8 \\
\cilo        & 492 & 10  & 8 \\  
\hline %inserts single line
\end{tabular}
\end{table}

\section{Velocity shift of $I(\cii)_\mathrm{res}$}
\begin{figure}[h]
\centering
\resizebox{\hsize}{!}{\includegraphics{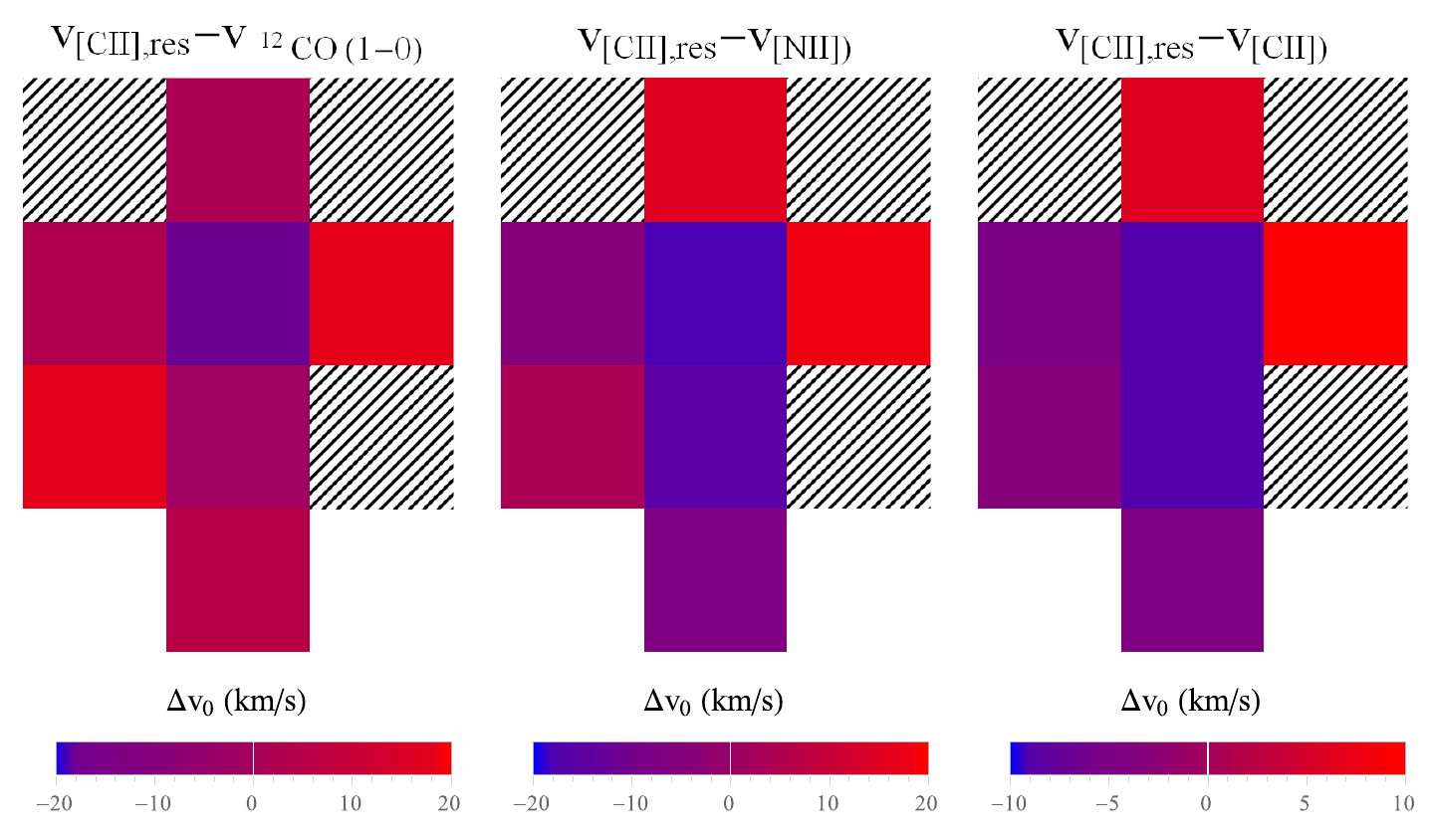}}
\caption{Spatial distribution of the properties of the line center shift of the residual \cii\ line with respect to \twco\ (1-0) (left), to \nii\ (middle), and to \cii\ (right), as described in Sect.~\ref{sect:kin}. The positions with no residual \cii\ are diagonally hatched. The baseline RMS of the residual \cii\ line is between 42 and 98~mK. The spatial resolution is 14\arcsec and 18\arcsec for \cii\ and {\nii}, respectively.}
\label{figRes2}
\end{figure}      

\section{Individual spectra overlayed with complementary data \label{app1}}
\begin{figure*}
\centering
\begin{minipage}{8.5cm}
\centering
\includegraphics[width=8.5cm]{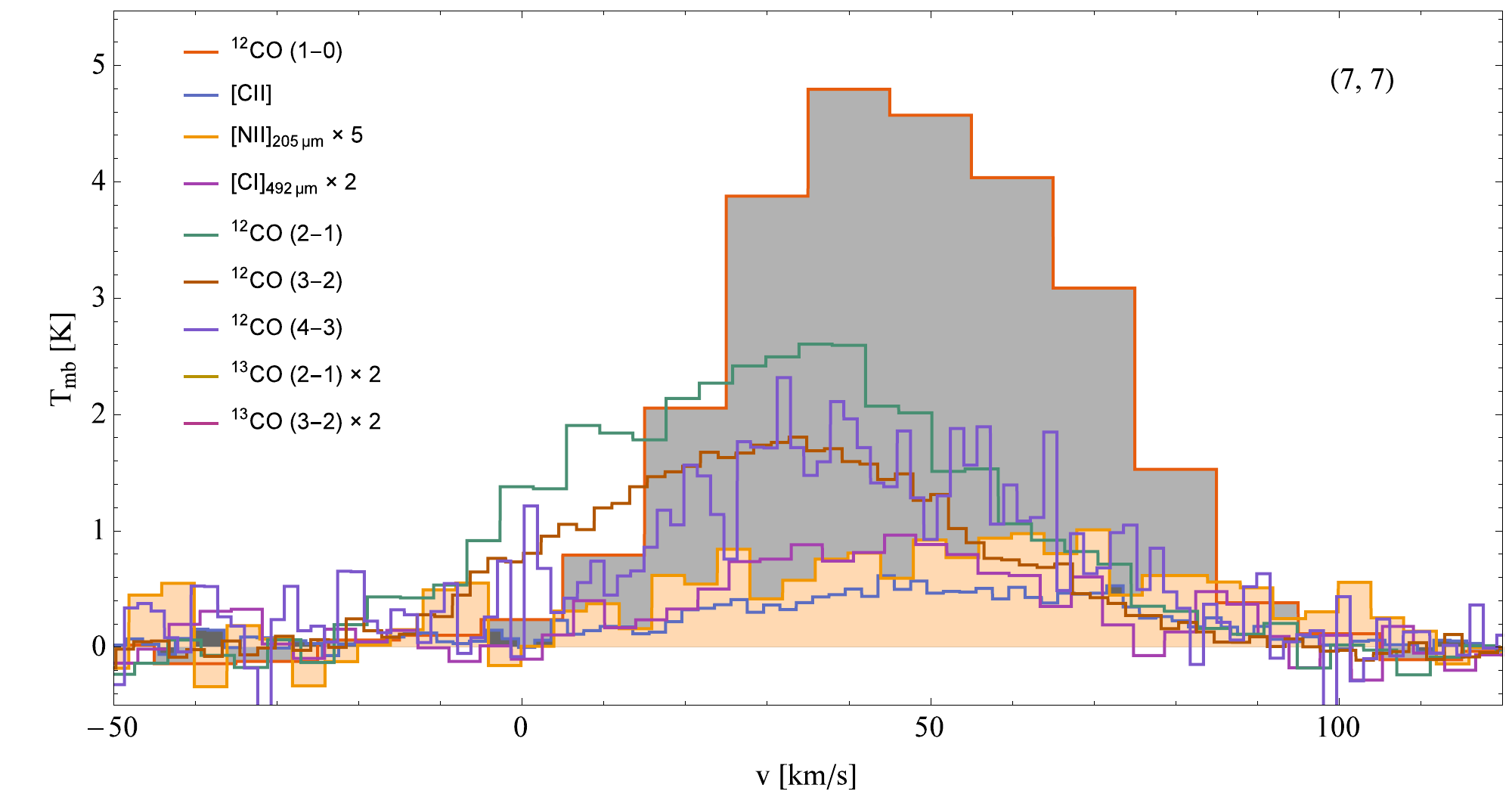}
\end{minipage}
\begin{minipage}{8.5cm}
\centering
\includegraphics[width=8.5cm]{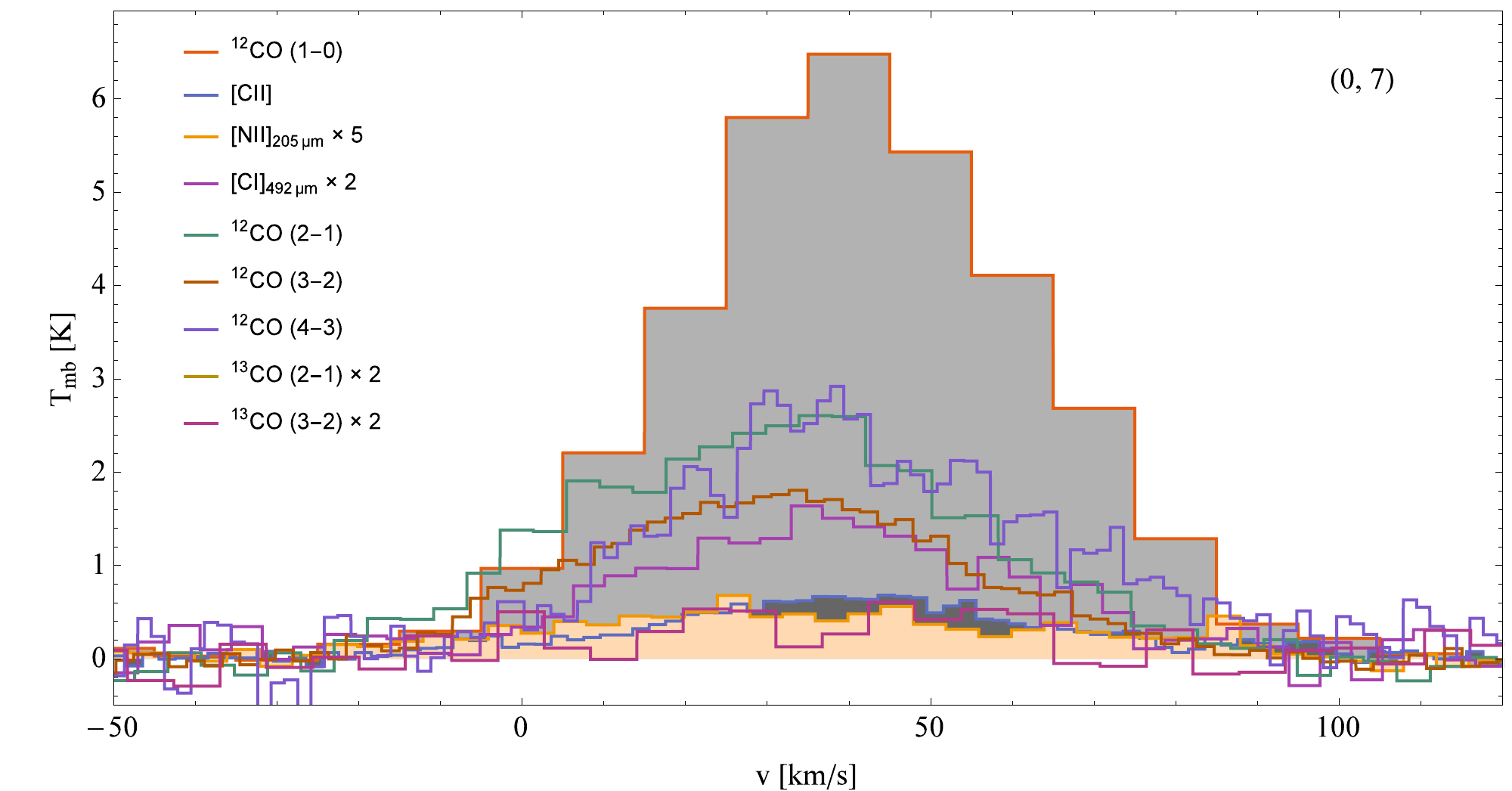}
\end{minipage}
\vfill
\centering
\begin{minipage}{8.5cm}
\centering
\includegraphics[width=8.5cm]{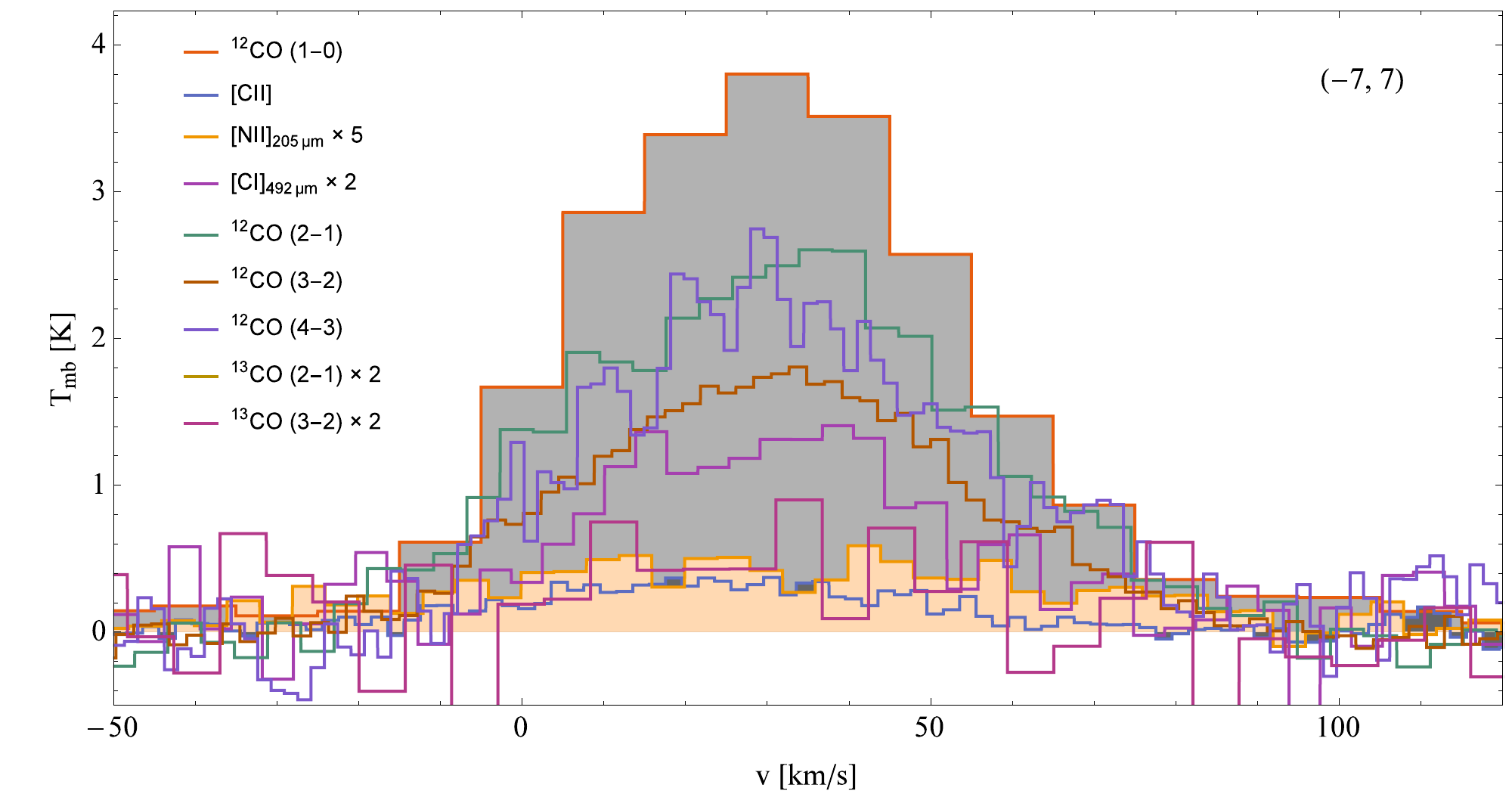}
\end{minipage}
\begin{minipage}{8.5cm}
\centering
\includegraphics[width=8.5cm]{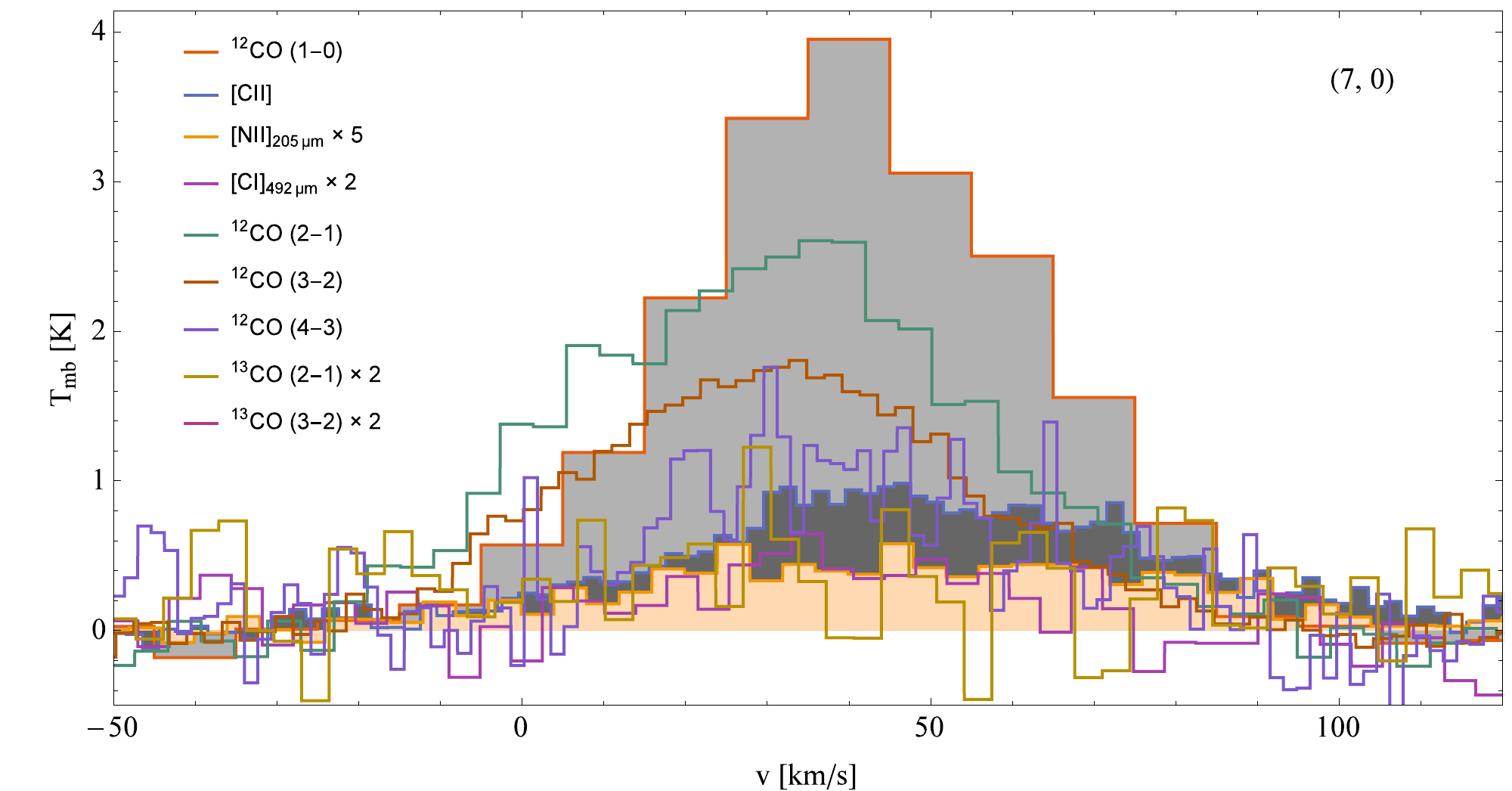}
\end{minipage}
\vfill
\begin{minipage}{8.5cm}
\centering
\includegraphics[width=8.5cm]{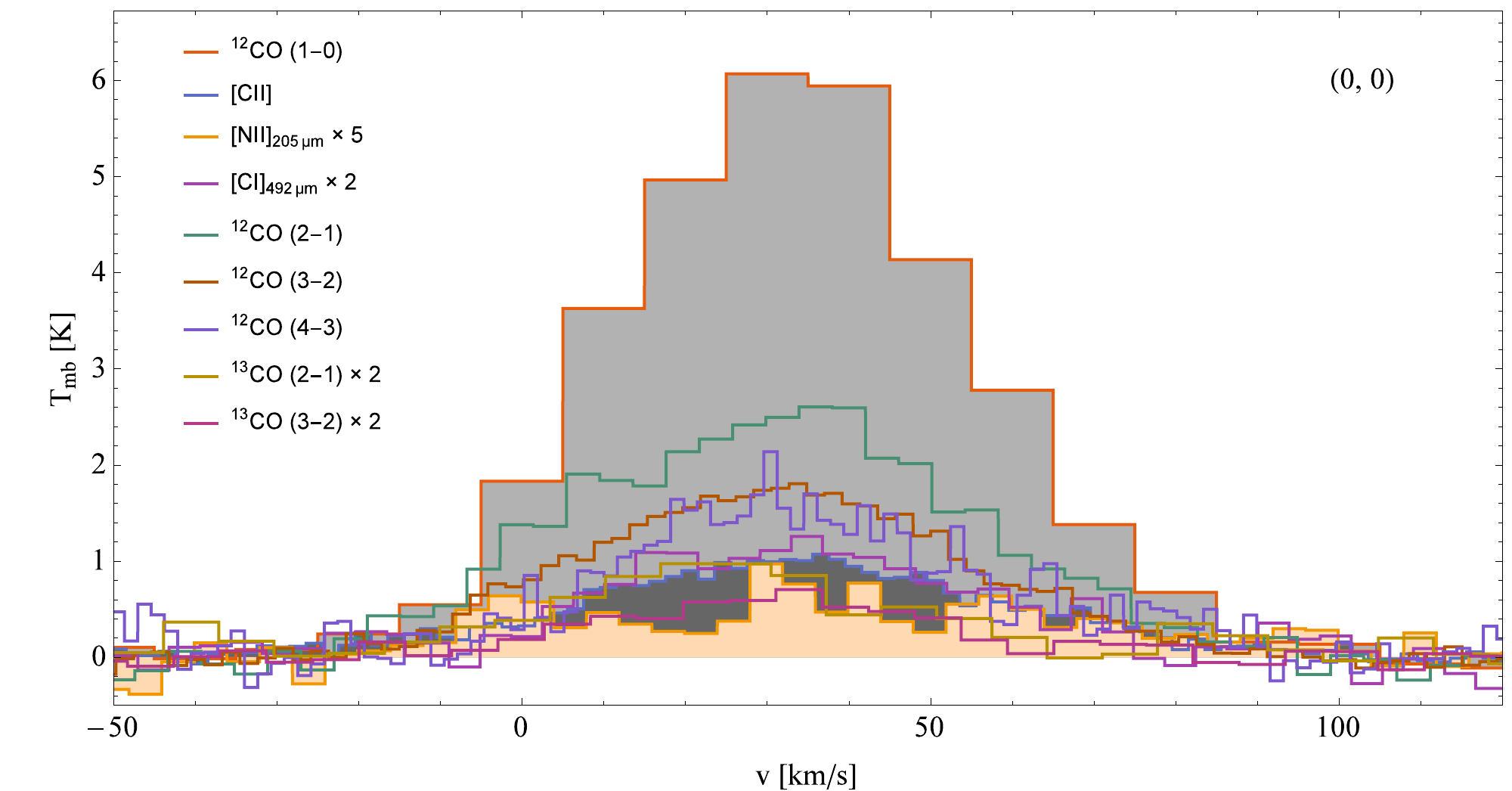}
\end{minipage}
\begin{minipage}{8.5cm}
\centering
\includegraphics[width=8.5cm]{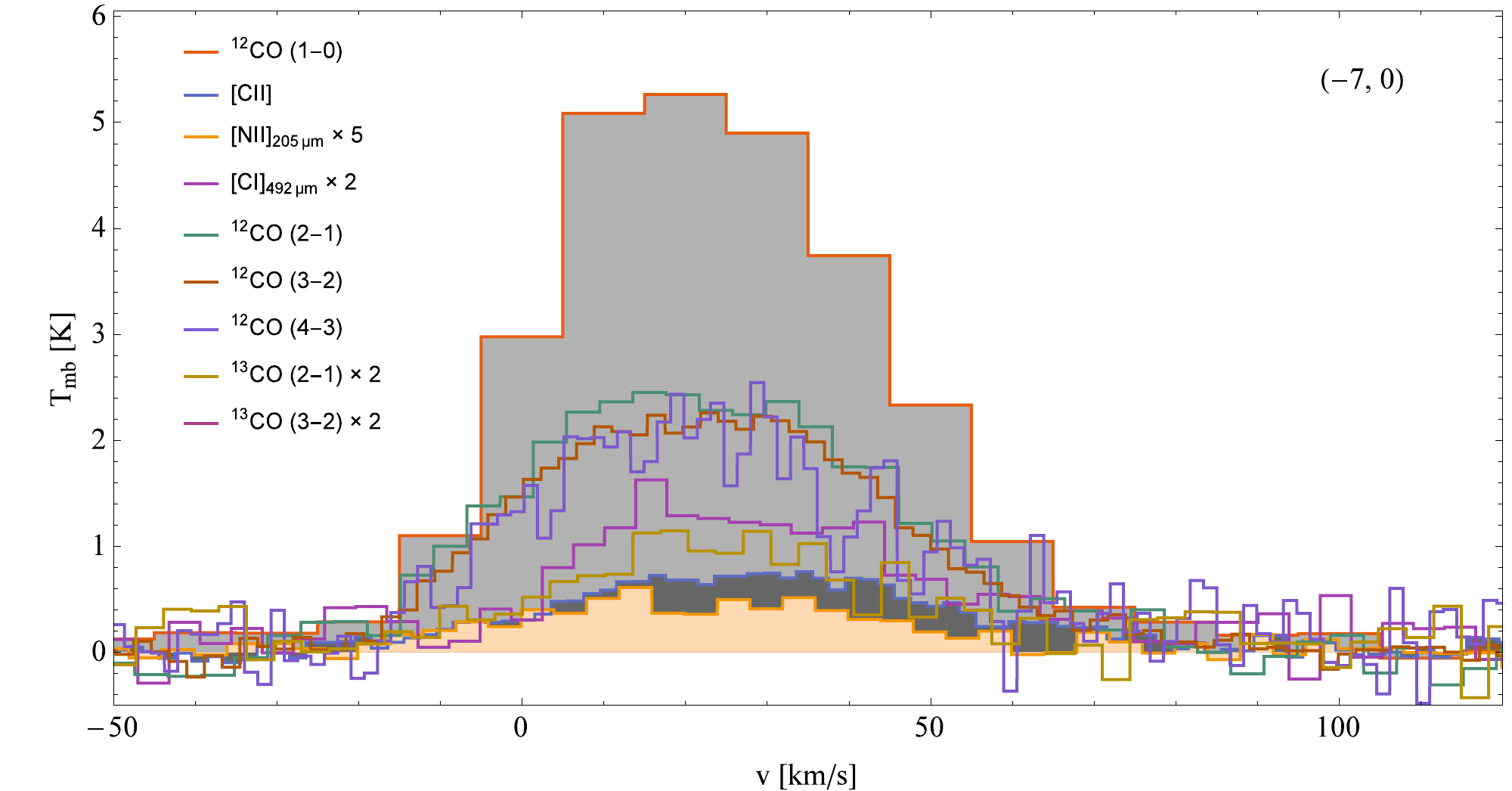}
\end{minipage}
\vfill
\centering
\begin{minipage}{8.5cm}
\centering
\includegraphics[width=8.5cm]{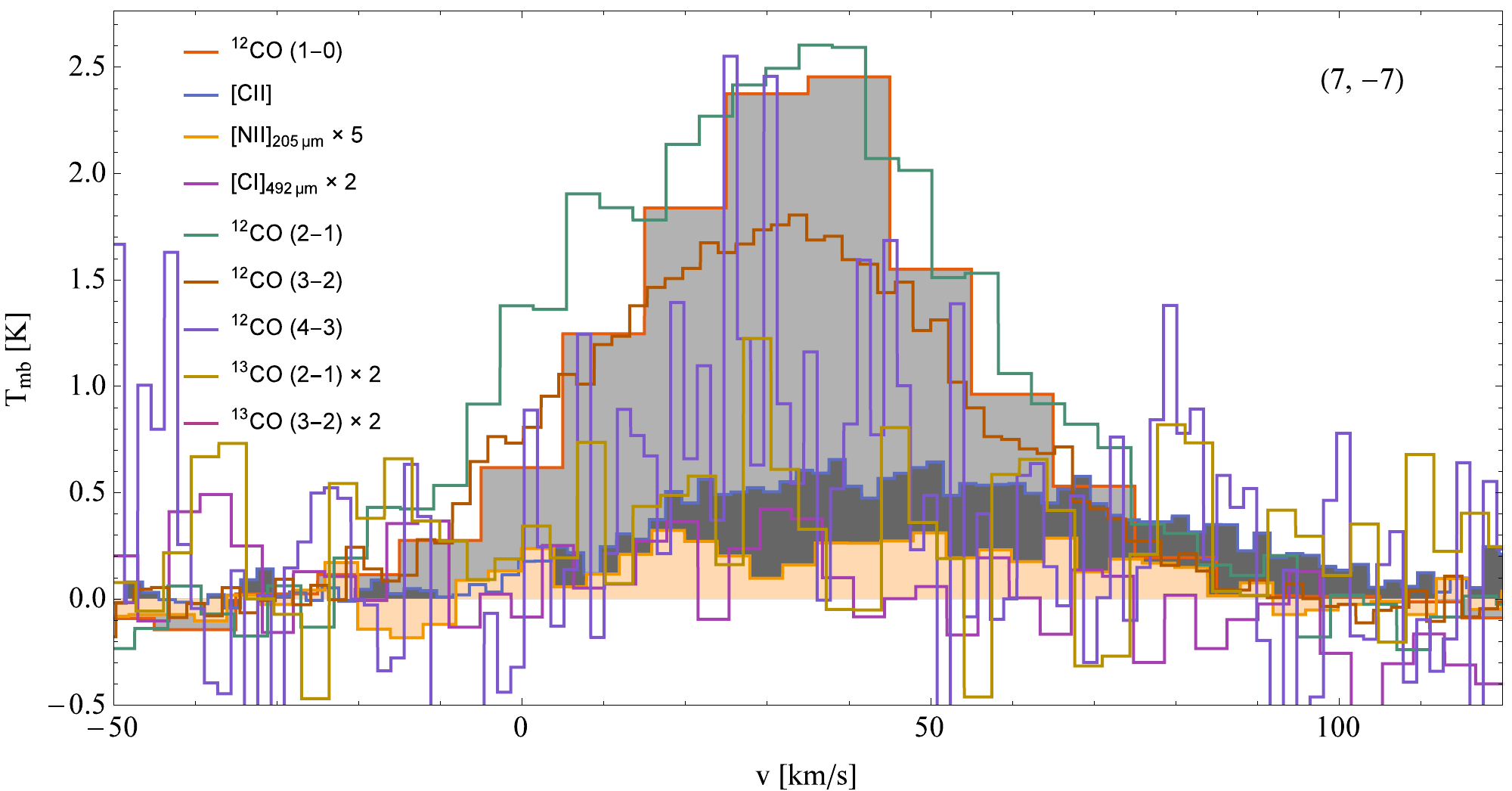}
\end{minipage}
\begin{minipage}{8.5cm}
\centering
\includegraphics[width=8.5cm]{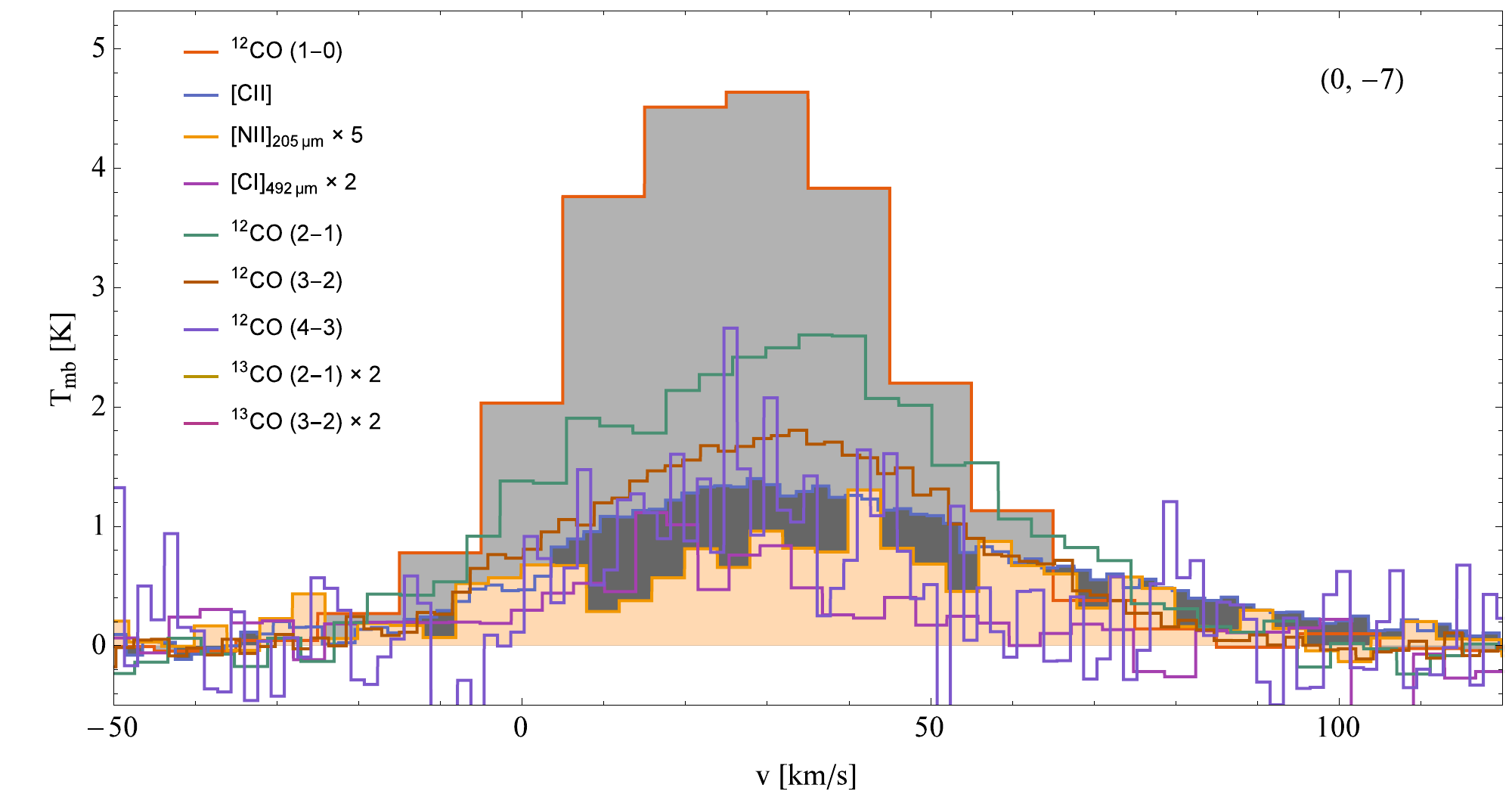}
\end{minipage}
\vfill
\centering
\begin{minipage}{8.5cm}
\centering
\includegraphics[width=8.5cm]{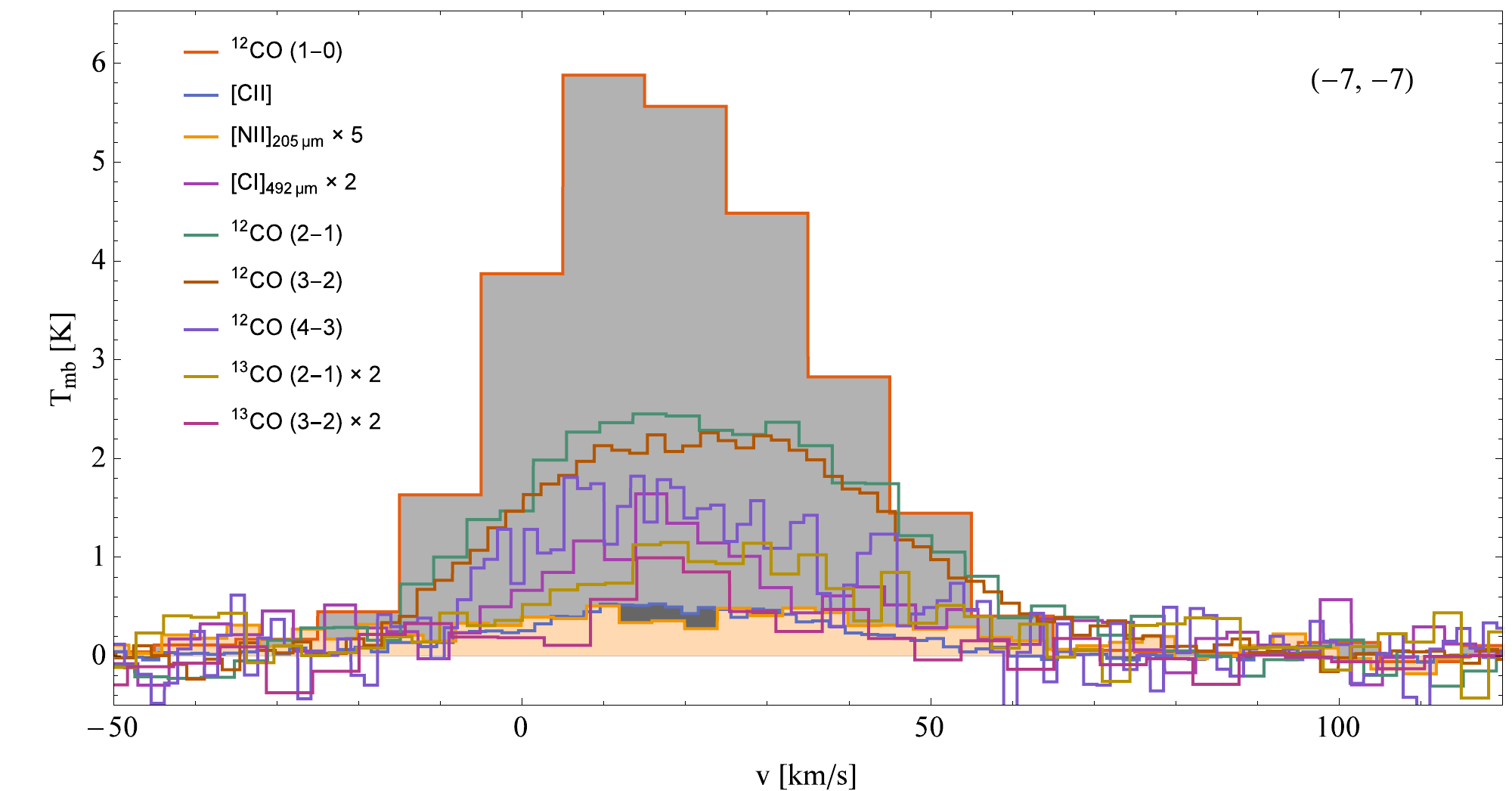}
\end{minipage}
\begin{minipage}{8.5cm}
\centering
\includegraphics[width=8.5cm]{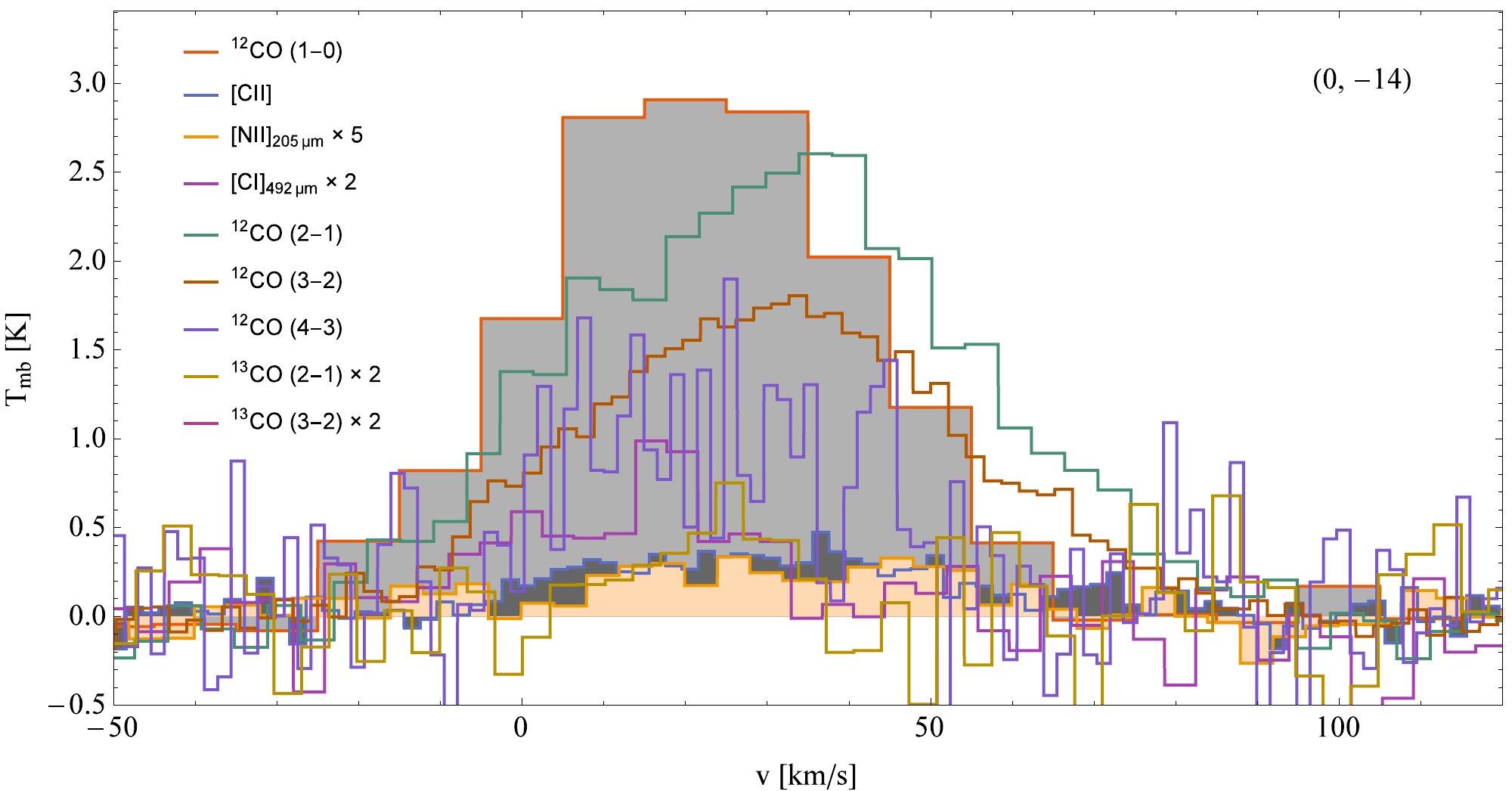}
\end{minipage}
\caption{Comparison of the SOFIA data with complementary data. For easier recognition, \twco\ (1-0), {\nii}, and  \cii\ spectra are shaded ({\cii}: dark gray, {\nii}: light orange, CO: light gray). The observed offset position is indicated in the top right corner of each panel. Note the varying scales of the ordinates.}\label{figAppendix}
\end{figure*}
In Fig.~\ref{figAppendix} we show a comparison between our SOFIA data and a selection of complementary data. On each spectrum we overlay  $^{12}$CO(1-0) data from BIMA-SONG\footnote{Berkeley Illinois Maryland
Association, \url{http://ned.ipac.caltech.edu/level5/March02/SONG/SONG.html}}\citep{bimasong}, as well as $^{12}$CO(2-1),  $^{12}$CO(3-2),  $^{12}$CO(4-3),  $^{13}$CO(2-1),  $^{13}$CO(3-2), and [CI]~$^3P_1-^3P_0$ spectra from \citet{israel2003}. 
 Because of the different beam sizes, gridding, and map coverage it was not always possible to re-grid and convolve all spectra to the positions and beam sizes of the GREAT observations. We kept the \twco\ (2–1) and \thco\ (2–1) spectra on their native resolution of 22\arcsec and 23\arcsec. We kept \twco\ (3–2) and \thco\ (3–2) on their native resolution of 15\arcsec. \twco\ (1–0), \twco\ (4–3), and \cilo\ spectra were smoothed to the resolution of our [C II] data.

\section{Optical thin \nii\ emission under LTE conditions}\label{niiemission}
Assuming local thermal equilibrium (LTE) the specific intensity $I_{ul}$ for an optically thin spectral line of frequency $\nu_{ul}$, Einstein-A value $A_{ul}$ and upper level column density $N_u$ is given by
\begin{equation}
\int I_{ul} d\nu=\frac{h \nu_{ul}}{4\pi}A_{ul} N_u \,\,\mathrm{erg\,s^{-1}\,cm^{-2}\,sr^{-1}} .
\end{equation}
With $T_{mb}=I_{ul}(h \nu_{ul})/(2 k \nu_{ul}^2)$ and $\Delta \nu/\nu=\Delta v/c$ it follows that
\begin{equation}\label{eqnTmb}
\int T_\mathrm{mb} dv=\frac{h c^3}{8\pi k \nu_{ul}^2} A_{ul} N_u \,\Kkms 
\end{equation}
and $N_u=n_u/n_\mathrm{tot} N_\mathrm{tot}$, where $n_u/n_\mathrm{tot}$ is the relative population of the upper energy level. 
Following \citet{goldsmith2015} and denoting the three fine-structure levels $^3P_0$, $^3P_1$, and $^3P_2$ as 0, 1, and 2, respectively, the relative population can be written as
\begin{gather}
\frac{n_2}{n_\mathrm{tot}}=\frac{R(1/0)R(2/1)}{1+R(1/0)+R(1/0)R(2/1)}\\
\frac{n_1}{n_\mathrm{tot}}=\frac{R(1/0)}{1+R(1/0)+R(1/0)R(2/1)}\\
\frac{n_0}{n_\mathrm{tot}}=\frac{1}{1+R(1/0)+R(1/0)R(2/1)}
\end{gather}
where $R(u/l)=n_u/n_l$ is given by \citet{goldsmith2015}. 
Using their numerical values for the collisions with electrons and for spontaneous decay we find that for the 205{\textmu}m line 
\begin{equation}
\int T^\mathrm{[NII]}_{205\mu\mathrm{m}} dv=5.06\times 10^{-16} n_u/n_\mathrm{tot} N_\mathrm{tot}
\end{equation}
 and for the 122{\textmu}m line 
 \begin{equation}
 \int T^\mathrm{[NII]}_{122\mu\mathrm{m}} dv=6.38\times 10^{-16} n_u/n_\mathrm{tot} N_\mathrm{tot}\,\,.  
 \end{equation}
The relative population is a function of the electron density and temperature, but for a given density and temperature the line integrated intensity scales linearly with the total N$^+$ column density $\int T_\mathrm{mb} dv= c(n_e,T_e)\times N_\mathrm{N+}$~\Kkms. In Tab.~\ref{tabTmb} we provide the scaling factors $c(n_e,T_e)$ for common values of $n_e$ and $T_e$.
\begin{table}[ht]
\caption{Scaling constants for the \nii\ fine-structure emission   $\int T_\mathrm{mb} dv= c(n_e,T_e)\times N_\mathrm{N+}$~{\Kkms} for common values of $n_e$ and $T_e$. The form $A(B)$ corresponds to $A\times 10^{B}$.} % title of Table
\label{tabTmb} % is used to refer this table in the text
\centering % used for centering table
\begin{tabular}{l l l l l} % centered columns (4 columns)
\hline\hline % inserts double horizontal lines
\noalign{\smallskip}
\backslashbox{$n_e/\mathrm{cm^{-3}}$}{$T_e/\mathrm{K}$}& 1000& 5000& 8000&10000 \\ 
             \noalign{\smallskip}
            \hline
            \noalign{\smallskip}
\multicolumn{5}{c}{\niilo\   205{\textmu}m}\\
            \noalign{\smallskip}
            \hline
            \noalign{\smallskip}
10 &8.66(-17)&9.24(-17)&9.30(-17)&9.32(-17)\\
100 &2.28(-16)&2.29(-16)&2.29(-16)&2.30(-16)\\
500 &2.08(-16)&2.02(-16)&2.02(-16)&2.01(-16)\\
700 &2.02(-16)&1.95(-16)&1.94(-16)&1.94(-16)\\
1000 &1.96(-16)&1.89(-16)&1.88(-16)&1.88(-16)\\
3000 &1.85(-16)&1.77(-16)&1.77(-16)&1.76(-16)\\
            \noalign{\smallskip}
            \hline
            \noalign{\smallskip}
\multicolumn{5}{c}{\niiup\ 122{\textmu}m }\\
            \noalign{\smallskip}
            \hline
            \noalign{\smallskip}
10 &1.48(-17)&1.71(-17)&1.73(-17)&1.74(-17)\\
100 &1.39(-16)&1.53(-16)&1.55(-16)&1.55(-16)\\
500 &2.67(-16)&2.85(-16)&2.86(-16)&2.87(-16)\\
700 &2.83(-16)&3.01(-16)&3.03(-16)&3.03(-16)\\
1000 &2.97(-16)&3.15(-16)&3.16(-16)&3.17(-16)\\
3000 &3.20(-16)&3.38(-16)&3.39(-16)&3.40(-16)\\
            \noalign{\smallskip}
            \hline
            \noalign{\smallskip}
\end{tabular}
\end{table}

\section{Super-resolution fit to all observed positions}\label{AppSR}
\begin{figure*}
\centering
\includegraphics[width=\textwidth]{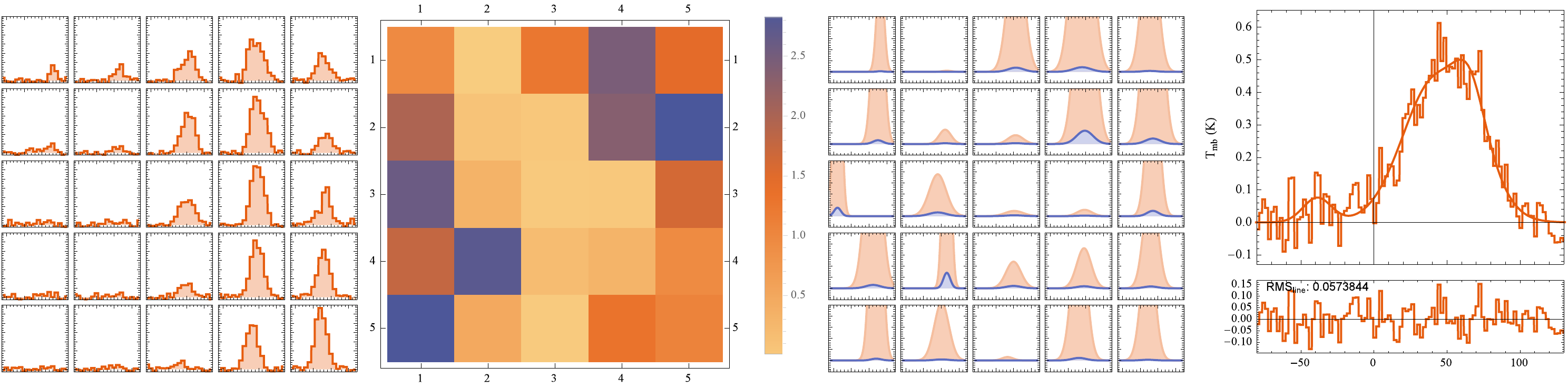}
\vfill
\includegraphics[width=\textwidth]{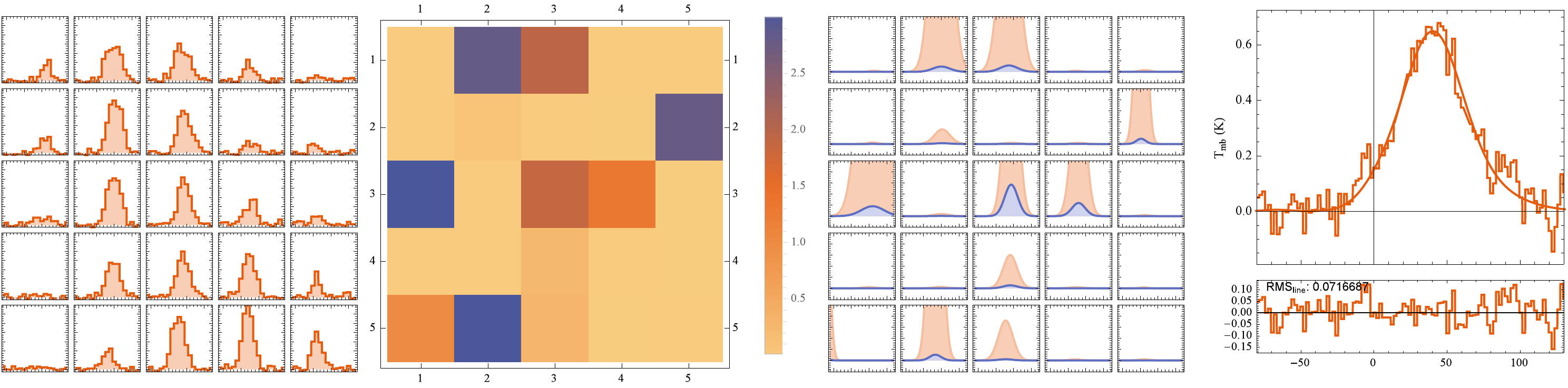}
\vfill
\includegraphics[width=\textwidth]{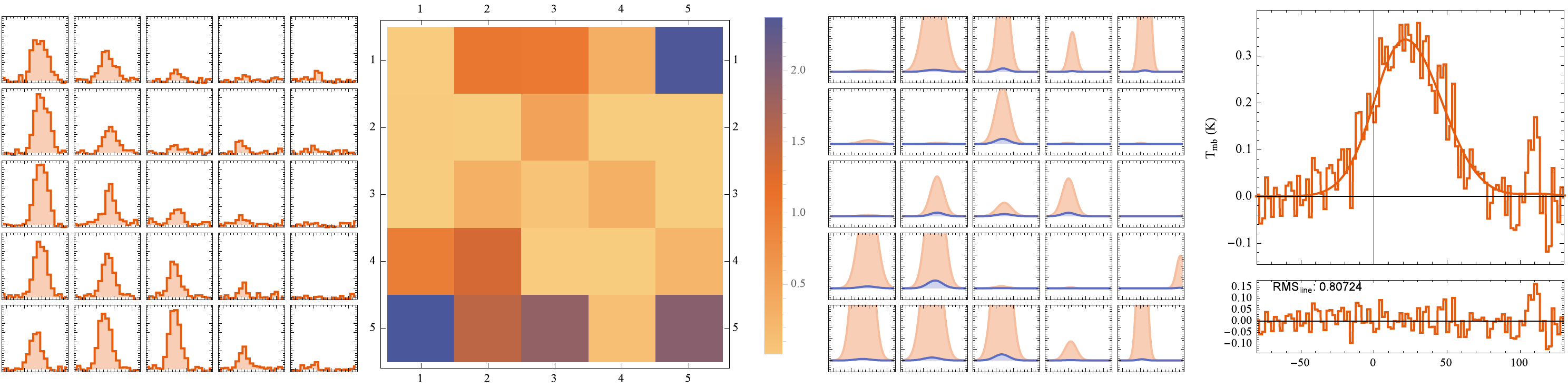}
\vfill
\includegraphics[width=\textwidth]{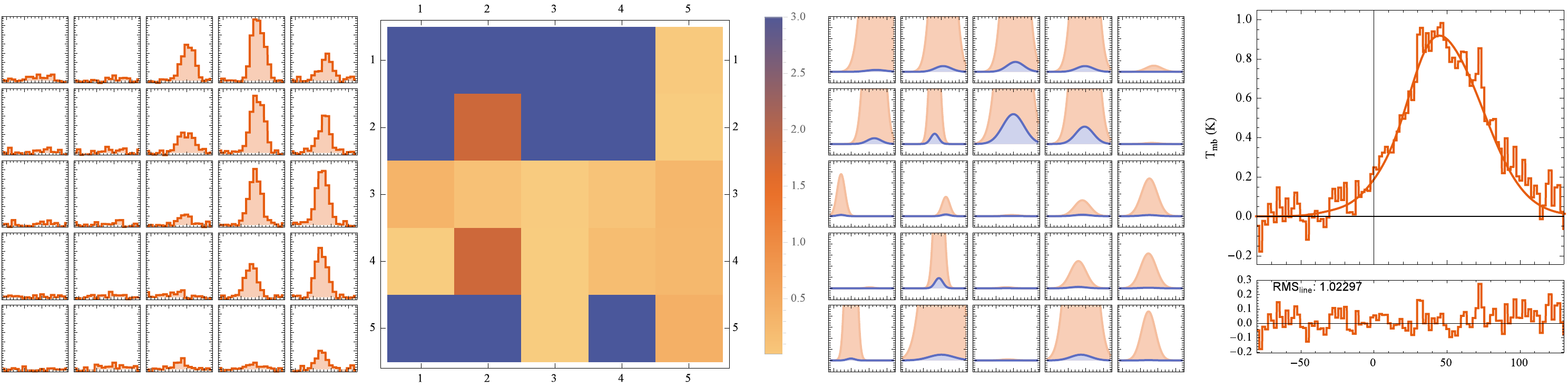}
\vfill
\includegraphics[width=\textwidth]{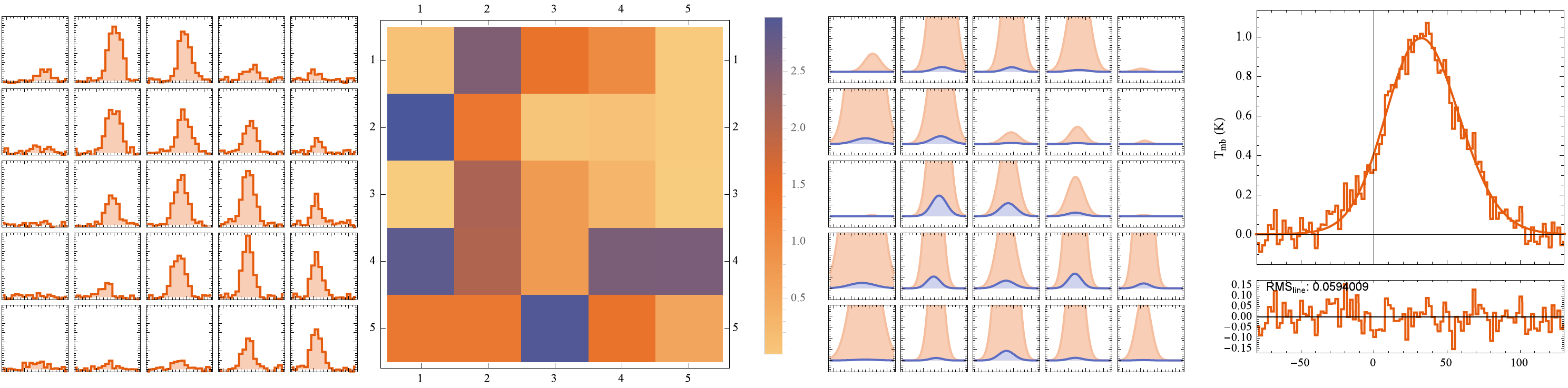}
\caption{\footnotesize{Same as Fig.~\ref{figSuper1}. The peak intensities are limited to $f_i\le3$. \cii\ line widths are up to 30\% wider than \twco\ (1-0). The various panels show the results for the positions (7\arcsec,7\arcsec), (0\arcsec,7\arcsec), (-7\arcsec,7\arcsec), (7\arcsec,0\arcsec), (0\arcsec,0\arcsec) from top to bottom. }}\label{figAppendixSR1}
\end{figure*}
\clearpage
\begin{figure*}
\centering
\includegraphics[width=\textwidth]{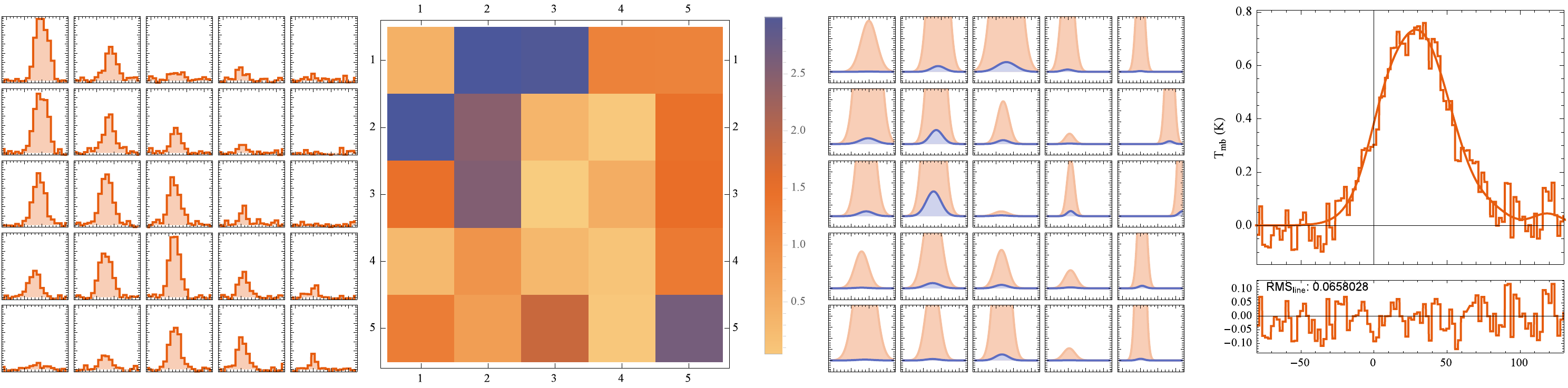}
\vfill
\includegraphics[width=\textwidth]{{SR-7_-7.max3}.pdf}
\vfill
\includegraphics[width=\textwidth]{{SR-0_-7.max3}.pdf}
\vfill
\includegraphics[width=\textwidth]{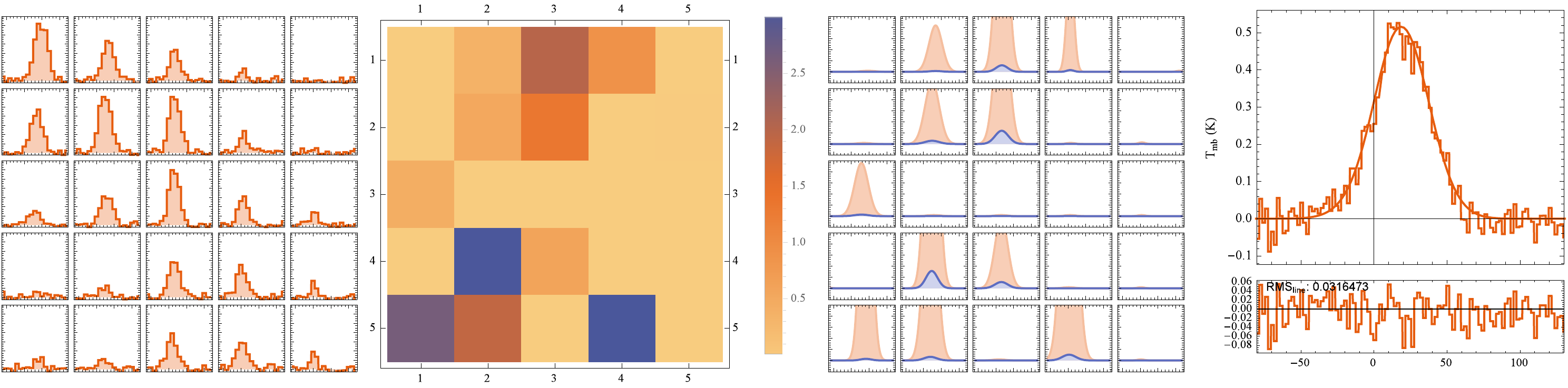}
\vfill
\includegraphics[width=\textwidth]{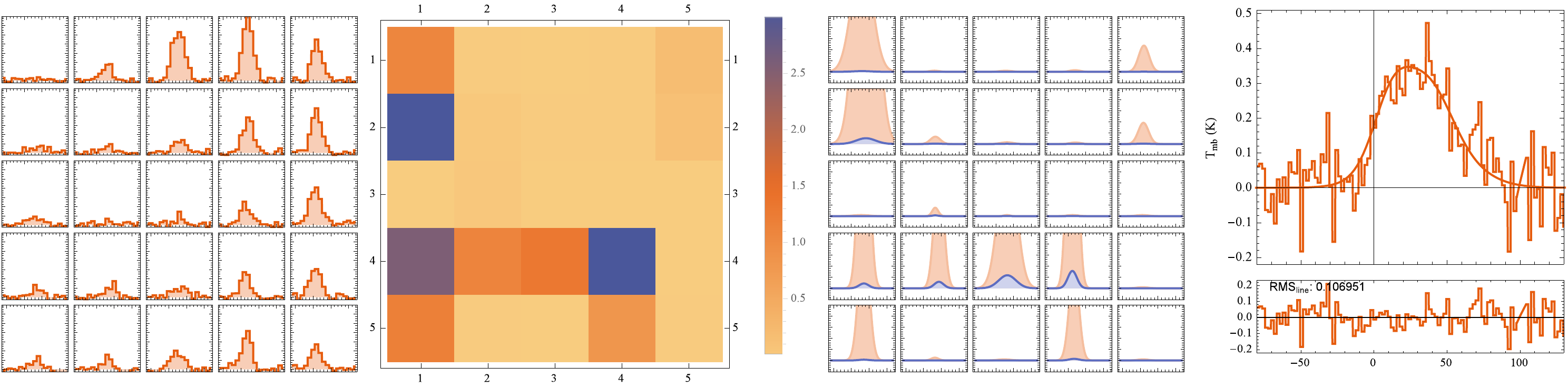}
\caption{Same as Fig.~\ref{figAppendixSR1}. The various panels show the results for the positions  (-7\arcsec,0\arcsec), (7\arcsec,-7\arcsec), (0\arcsec,-7\arcsec), (-7\arcsec,-7\arcsec), (0\arcsec,-14\arcsec) from top to bottom}\label{figAppendixSR2}
\end{figure*}

\end{document}